\documentclass{statsoc}
\usepackage{amsmath, amssymb}
\usepackage{graphicx}
\usepackage{natbib}
\usepackage{algorithm, algorithmicx, algpseudocode}
\usepackage{url} 
\usepackage[a4paper]{geometry} 
\newcommand\StateX{\Statex\hspace{\algorithmicindent}}
\newcommand{\invGam}{\textrm{inv-$\Gamma$}}
\newcommand{\pacc}{p_{\textrm{acc}}}
\newcommand{\boldbeta}{\boldsymbol{\beta}}
\newcommand{\boldx}{\boldsymbol{x}}

\newcommand{\boldlambda}{\boldsymbol{\lambda}}

\title[The Bayesian Spatial Bradley--Terry Model]{The Bayesian Spatial Bradley--Terry Model: Urban Deprivation Modelling in Tanzania}
\author[Seymour {\it et al.}]{Rowland G. Seymour, David Sirl, Simon P. Preston, Ian L. Dryden}
\address{School of Mathematical Sciences, University of Nottingham, UK}
\author[Seymour {\it et al.}]{Madeleine J. A. Ellis, Bertrand Perrat, James Goulding}
\address{N/LAB, University of Nottingham, UK}

\begin{document}

\begin{abstract}
Identifying the most deprived regions of any country or city is key if policy makers are to design successful interventions. However, locating areas with the greatest need is often surprisingly challenging in developing countries. Due to the logistical challenges of traditional household surveying, official statistics can be slow to be updated; estimates that exist can be coarse, a consequence of prohibitive costs and poor infrastructures; and mass urbanisation can render manually surveyed figures rapidly out-of-date. Comparative judgement models, such as the Bradley--Terry model, offer a promising solution. Leveraging local knowledge, elicited via comparisons of different areas' affluence, such models can both simplify logistics and circumvent biases inherent to house-hold surveys. Yet widespread adoption remains limited, due to the large amount of data existing approaches still require. We address this via development of a novel Bayesian Spatial Bradley--Terry model, which substantially decreases the number of comparisons required for effective inference. This model integrates a network representation of the city or country, along with assumptions of spatial smoothness that allow deprivation in one area to be informed by neighbouring areas. We demonstrate the practical effectiveness of this method, through a novel comparative judgement data set collected in Dar es Salaam, Tanzania.
\end{abstract}


\section{Introduction}
Deprivation statistics are used by governmental and non-governmental organisations to describe the standard of living in a small administrative areas \citep{MHCLG19}. Yet assessment of deprivation depends not only on the financial situation of those living in an area, but also factors such as health, housing, commercial activity,  and access to education. If correctly estimated, such statistics can be central to the design of successful policy interventions \citep[see, e.g.,][]{USAID, Williams2021}, supporting citizens and guiding decision makers in local government, non-governmental organisations and the business sector alike. However, obtaining deprivation estimates is often a surprisingly challenging task, particularly in developing countries. In such contexts traditional household surveys are often prohibitively expensive or logistically intractable. Data collection efforts are impaired by poor physical infrastructures restricting sample sizes. Mass urbanization can render estimates rapidly out-of-date; and a lack of financial transparency in the face of vast informal economies exacerbates the well-established response biases inherent to household surveying \citep{randall2015poverty,lynn2002separating}.  

In Africa, according to the World Bank's chief economist, such issues have generated a ``statistical tragedy'' \citep{Devarajan2013}. Dar es Salaam, the largest city in Tanzania, is a case in point. With a population of over 6 million the city has doubled in size in just a decade, leaving official statistics generated but 5 years ago broadly inapplicable. The United Nations has estimated that the annual growth rate of the city will continue to be 4.8\%, and by 2030 Dar es Salaam will be home to at least 10 million people \citep{UN18}. Such rapid growth means citizens lack resources, with poor physical infrastructures and absent public services resulting in a low quality of living. Over 70\% of citizens in Dar es Salaam live in unplanned settlements and slums \citep{Lim16}, water sources in the city are polluted \citep{Nap10} and outbreaks of diseases are common \citep{McCrick17}. Determining the level of deprivation in each part of this rapidly changing city is key to designing policies and strategies to alleviate these issues, especially in the face of limited resources, yet traditional household surveys are simply not viable \citep{randall2015poverty}.

Citizen science and comparative judgement offer a way to address the lack of official data and the rapid changes in the city, providing access to informed and up-to-date opinions from local citizens. Comparative judgement methods contrast sharply with traditional surveying approaches, in which a respondent might be asked to indicate the affluence level of an area, or their own household income, based upon some arbitrary scale. Instead, individuals are shown pairs of areas and asked which is the more affluent of the two. Making pairwise comparisons is preferable to making absolute judgements, which are well-evidenced as subject to strong biases and inconsistencies \citep{kalton1982rss}. With household income levels often being highly volatile in developing world contexts, and respondents often reticent to provide accurate responses due to the scale of the informal economy \citep{randall2015poverty}, this also provides scope to reduce response bias and logistical costs.

To achieve this one might fit a Bradley--Terry (BT) model \citep{Brad52} to pairwise comparative judgement data. This allows not only areas to be ranked, but deprivation levels in each neighbourhood or region to be estimated. However, existing models still require an obstructively large number of individual comparisons to be provided in order to produce accurate estimates. With data collection infrastructures remaining poor in developing countries \citep[see, e.g.,][]{Etten19,engelmann2018unbanked}, comparative judgement solutions can only become viable in practice if the amount of data required can be reduced. We address this key issue via development of a novel Bayesian Spatial Bradley--Terry (BSBT) model, which substantially decreases the amount of data required for reliable estimates of the parameters of interest. This model integrates a network representation of the city or country, along with assumptions of spatial smoothness that allow deprivation in one area to be informed by neighbouring areas.

Adding structure by including covariates in the standard BT model has only generally been achieved in a parametric framework with linear predictors \citep[see, e.g.,][]{Spring73,Stern11}. Nonparametric methods have received comparatively little attention. For example, a more flexible spline-based approach for explanatory variables has been proposed by \cite{Des93}. A semi-parametric approach, which allows for subgroups within the set of objects being compared, has also been developed \citep{Str11}. However, these methods are unsuitable for spatial explanatory variables, as it is difficult to propose covariates which can describe complex spatial structures. We instead avoid specifying any parametric functions and use a multivariate normal prior distribution to model the spatial structure. This novel treatment allows for far more flexibility as we do not need to propose strict parametric models, which often do not describe well the latent structure. 
We also extend the BSBT model to include ways to examine if different groups of judges (participants in the study who make the comparative judgements) hold different opinions. In developing countries, we are particularly interested in the differing opinions of men and women, as women can face starkly different health, social and economic difficulties to men. The BSBT model with judge information allows us to locate areas where men and women hold notably different opinions about the deprivation level.

\subsection{Empirical Background} \label{sec:empirical}

To demonstrate the practical effectiveness of this new method, we have additionally collected a large, novel comparative judgement data set to infer deprivation in Dar es Salaam. Ethical approval for this part of the study was obtained from the Nottingham University Business School ethical review committee, application reference No. 201819072. We include the resulting data set in the \texttt{BSBT} R package that accompanies the paper, as well as in the supplementary material. The Dar es Salaam comparative judgement data set contains 75,078 comparisons made by 224 local participants, whom we refer to henceforth as judges, as well as the gender of each judge. Dar es Salaam is divided into 452 administrative areas called \emph{subwards}, which are the lowest level of administrative division in the city.

\begin{figure}[htbp]
    \centering
    \includegraphics[width = 0.5 \textwidth]{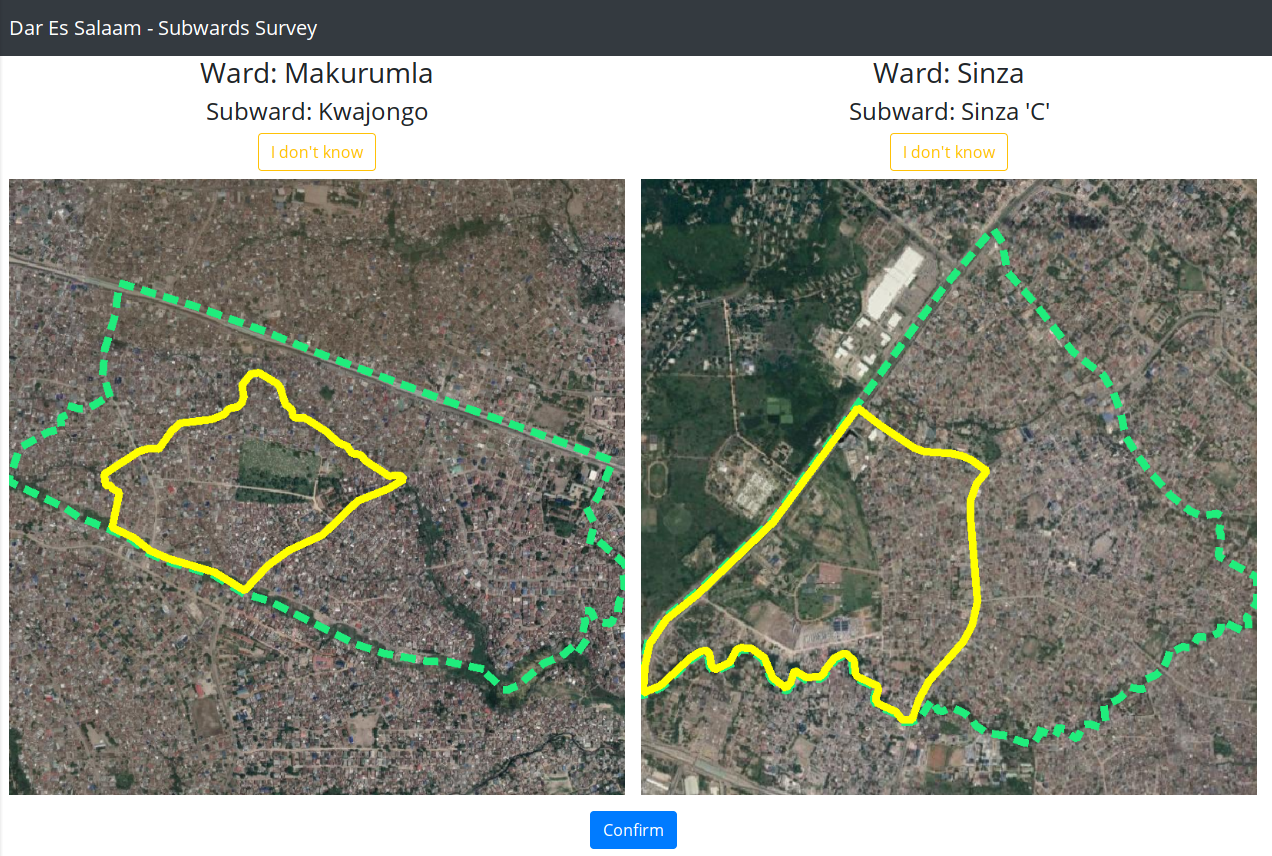}
        \caption{A screenshot of the software designed to carry out the comparative judgement study. In this example a user is asked to choose the most affluent between two subwards, \textit{Kwajongo} and \textit{Sinza C}. Images were zoomable, with both the subward and ward named directly in order to contextualise the user.}
    \label{fig: dar comparison software}
\end{figure}

To carry out the judgements, we designed a web interface (see Figure \ref{fig: dar comparison software}) so that judges could be shown images of pairs of subwards and asked to compare the affluence. The interface relied on a Python back end alongside a relational database (PostgreSQL was used for the study) to collect and store comparative judgements. At the start of the study, judges were asked to identify areas of the city they were familiar with. Then, during the judging process, judges had the option to indicate either i) which of the two subwards they felt was more affluent, ii) that the subwards were roughly equal in affluence, or iii) that they were unfamiliar with at least one of the two subwards. Comparisons corresponding to ii) were recorded as a tie, and outcomes corresponding to iii) were discarded and the judges were not asked about the subwards they were unfamiliar with again. Pairs of subwards for each judge were chosen uniformly at random from the list of all possible pairs of subwards which the judge was familiar with.  

Judges were recruited through word of mouth by students at local universities, NGOs and also via a local taxi driver association. The rationale for recruiting these judges was that they were all citizens of Dar es Salaam with a wide working knowledge of different subwards in the city. Of the judges, by occupation 37\% were students, 19\% were unemployed, 13\% had white-collar jobs (e.g. teacher, accountant), and the remaining 31\% either had a job not in those preceding categories, or chose not to disclose this information. By gender, 40\% of the judges were female, and 60\% were male although males judges made 72\% of the comparisons in the data set. This is because, on average, the women took longer to complete each comparison. Data was collected \emph{in situ} over two weeks in August 2018 via 17 data collection sessions each lasting two hours. Sessions were run in the morning, early and late afternoon, and evening to ensure as many judges as possible could attend. Judges were only allowed to attend one session. At the start of each session, the judges received a 15 minute training session in English and Swahili explaining how to make judgements and guidance on how to judge areas based on affluence and deprivation. Accompanying written instructions for the judges were provided in English and Swahili. One judge, who made over 2,000 comparisons, was excluded from the study as the comparisons seem spurious -- they are not included in any of the data we report. 

\begin{figure}
    \centering
    \includegraphics[width = 0.49 \textwidth]{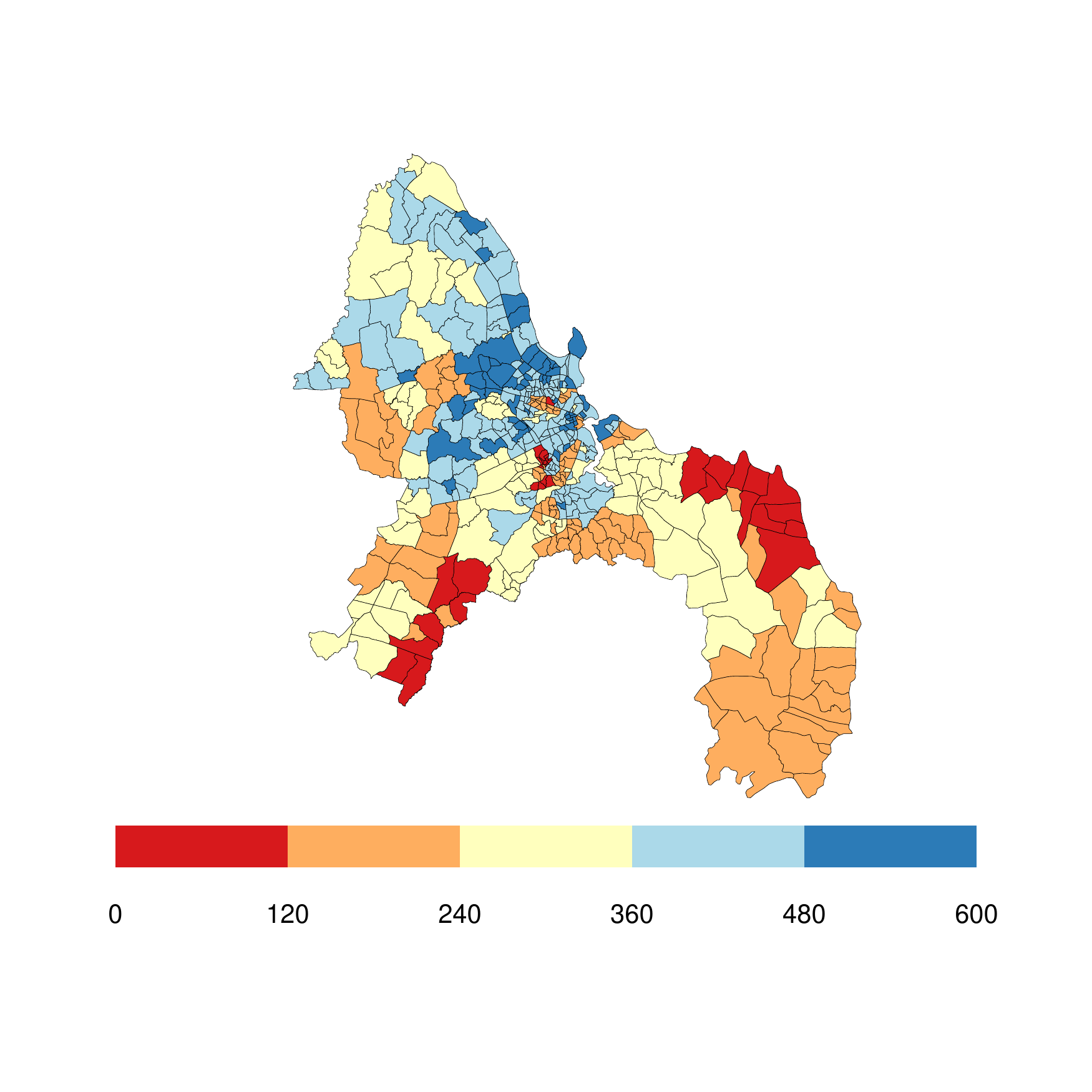}
    \includegraphics[width = 0.49 \textwidth]{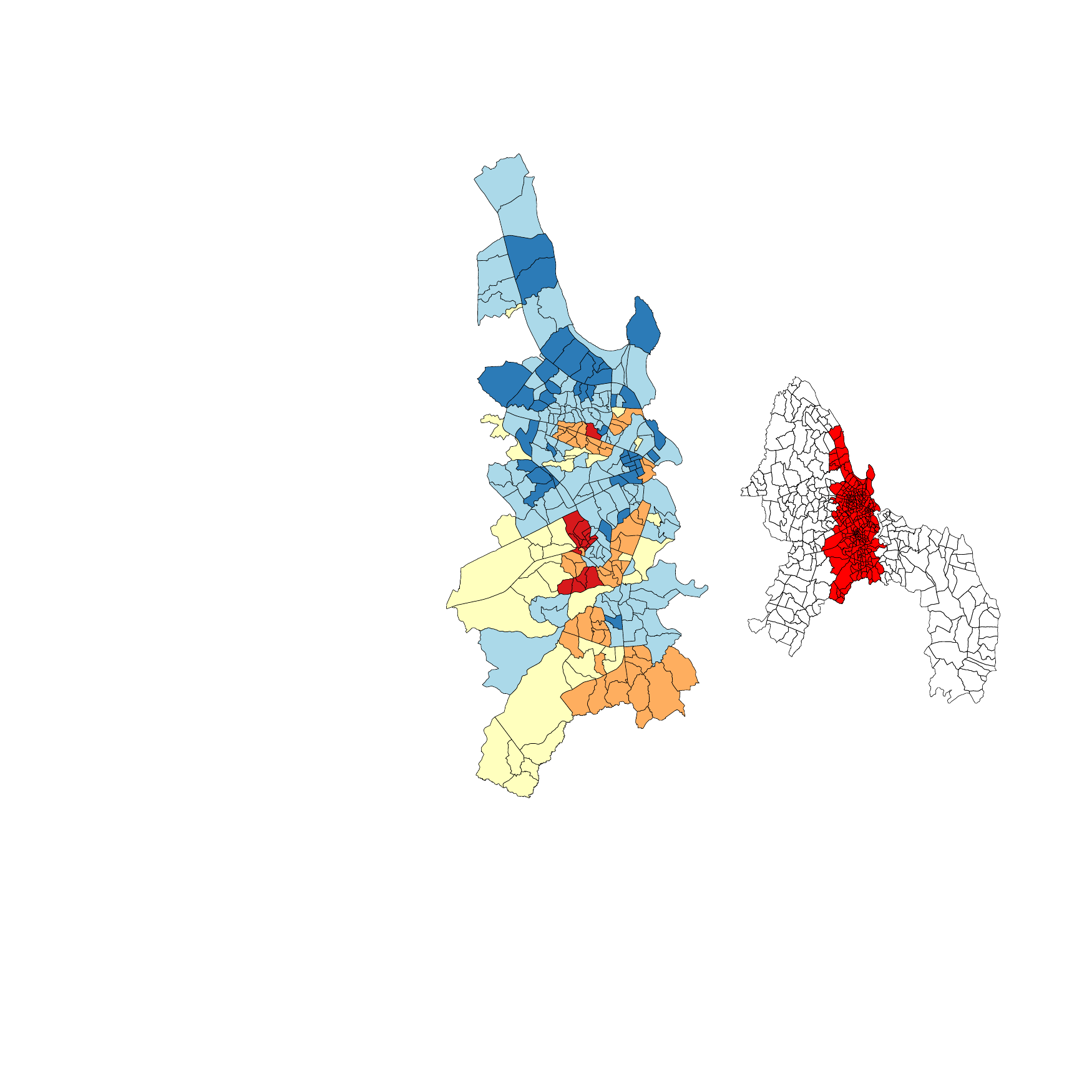}
        \caption{A map showing the number of times each subward featured in a comparison. The figure on the right shows a magnified section of the centre of the city, together with this central area highlighted on a map of the whole city.}
    \label{fig: comparison historgram}
\end{figure}

Figure \ref{fig: comparison historgram} shows how many comparisons each subward was featured in, which ranged between 65 and 588 comparisons, with mean 150. The affluent areas in the city and central business district were the most well known areas.  A total of 14.6\% of the comparisons made in the study were tied comparisons. There are several ways of dealing with tied comparisons \citep[see, e.g.,][]{Rao67, Dav70, Turner2012} and we discuss these in Section~\ref{sec:darResults}. We chose to randomly assign one of the pair to be the more deprived subward. 

An important aim of this work is to develop methodology that enables more efficient data collection, able to overcome the organisational challenges faced in the field. Two weeks were invested in collecting this large data set, and organisation and recruitment of participants prior to the study took several months. A key aim of the scale of the data collection process undertaken was to conclusively evidence, with similar future efforts in mind, that the first two days could have been sufficient if improved modelling procedures are employed. This would save considerable time and resources in both collecting the data and reducing the number of participants needed to recruit, train and organise.


\section{A spatial framework for the Bradley--Terry model}

\subsection{The standard Bradley--Terry model} \label{sec: standard BT}
Consider a comparative judgement data set consisting of $K$ pairwise comparisons of $N$ areas.   We assign to each area what we call a {\it relative deprivation parameter}  $\lambda_i \in \mathbb{R}$ ($i=1,\ldots,N$) and infer the value of each parameter using a comparative judgement model. We use the term ``deprivation parameter'' because identifying deprivation is the primary focus of the paper, but note that, in keeping with most measures of this kind, larger (respectively, smaller) values are associated with more affluent (respectively, deprived) areas.

We begin by outlining the standard BT model, a commonly used comparative judgement model. If areas $i$ and $j$ are compared $n_{ij}$ times, the number of times area $i$ is judged to be more affluent than area $j$ is modelled as 
$$
Y_{ij} \sim \hbox{Bin}(n_{ij}, \pi_{ij}),
$$
and we assume $Y_{ij}$ are independent. Here the probability $\pi_{ij}$ that area $i$ is judged to be more affluent than area $j$ depends on the difference in relative deprivation of $i$ and $j$ and is
\begin{equation}
    \hbox{logit}(\pi_{ij}) = \lambda_i - \lambda_j
    \iff
    \pi_{ij} = \frac{\exp(\lambda_i)}{\exp(\lambda_i) + \exp(\lambda_j)}
    \qquad
    (i \neq j, 1\leq i, j \leq N).
    \label{eq: logit difference}
\end{equation}
Model (\ref{eq: logit difference}) is invariant to translations $\lambda_i \longrightarrow \lambda_i + c$ (for any $c\in \mathbb{R}$), so an identifiability constraint is needed. A common choice is $\sum_{i=1}^N \lambda_i = 0$, which means that relatively deprived areas will have negative parameters, relatively affluent areas will have positive parameters and areas with middling levels of relative deprivation will have parameters near zero.

We write $y_{ij}$ for the number of times area $i$ was judged to be more affluent than area $j$, and denote by $\boldsymbol{y}$ the set containing these outcomes for all pairs of areas. The likelihood function for the model is given by
\begin{equation}
  \pi(\boldsymbol{y}\mid \lambda_1, \ldots, \lambda_N) = \prod_{i=1}^N\prod_{j<i} \begin{pmatrix} n_{ij} \\ y_{ij}
\end{pmatrix} \pi_{ij}^{y_{ij}} (1-\pi_{ij})^{n_{ij} - y_{ij}}. \label{eq: BT likelihood}
\end{equation}
We will compare our model to the standard BT model and the implementation provided in the \texttt{BradleyTerry2} R package \citep{Turner2012}, as this is a popular implementation of the method \citep[see, e.g.,][]{Cat12b, Van16, Grin18}. This package computes MLEs for the model parameters. We follow \cite{Turner2012} and \cite{Firth2004} and construct 95\% confidence intervals using the quasi-variance for the estimates in the standard BT model. This is done using the \texttt{qvcalc} package \citep{Firth20}, as this allows us to readily compare the relative deprivation levels.



\subsection{The Bayesian Spatial Bradley--Terry model} \label{sec: BSBT}
In the BSBT model, we assume the relative deprivation parameters $\lambda_i$ to be random and dependent on one another, with a higher level of dependence between nearby areas than areas which are further apart. To avoid making specific parametric assumptions about the level of deprivation in each area, we model the relative deprivation parameters using a multivariate normal prior distribution. We use a zero-mean multivariate normal prior distribution for the deprivation level parameters $\boldlambda = \{\lambda_1, \ldots, \lambda_N\}$ subject to the  constraint $\boldsymbol{1}^T\boldlambda = 0$, where $\boldsymbol{1} = (1, \ldots, 1)^T$ is a vector of ones. This matches the condition in the standard BT model, that the sum of the deprivation levels is 0.  Conditional on this constraint 
\begin{equation}
(\boldlambda \mid \boldsymbol{1}^T\boldlambda = 0) \sim \textrm{MVN}\Big(\textbf{0}, \, \Sigma - \Sigma\boldsymbol{1}(\boldsymbol{1}^T\Sigma \boldsymbol{1})^{-1}\boldsymbol{1}^T\Sigma\Big).
\label{eq: constrained prior}   
\end{equation}

\subsubsection{Modelling spatial covariance}
The structure of the covariance matrix $\Sigma$ is a modelling choice and there are number of options to choose from. In the simplest terms, we want to assign high covariance between deprivation levels in nearby subwards and low covariance between levels in distant subwards. A widely used option in Euclidean spatial domains is to use the squared-exponential covariance function \citep{Ras06}. Using this function, the covariance between the deprivation levels in subwards $i$ and $j$ is
\begin{equation}
    \hbox{cov}(\lambda_i, \lambda_j) = \Sigma_{i,j} = k(i, j;\, \alpha, l) = \alpha^2 \exp\left(- \, \frac{d_{ij}^2}{l^2} \right),
    \label{eq: sq exp function}
\end{equation}
where $d_{ij}$ is the Euclidean distance between areas $i$ and $j$, $\alpha^2$ is the prior variance hyperparameter and $l$ is the characteristic length scale, which describes what is meant by nearby and distant. However, using a function which is stationary in Euclidean space may not capture the change in deprivation in all parts of the city. City centres are typically densely packed with small areas, with peri-urban and rural areas being larger. Modelling the spatial structure using a Euclidean metric is therefore unsuitable since, for example, two points 1km apart in a rural area are likely similar, but two points 1km apart near a city centre may be very different.

\begin{figure}
    \centering
    \includegraphics[width = 0.49\textwidth]{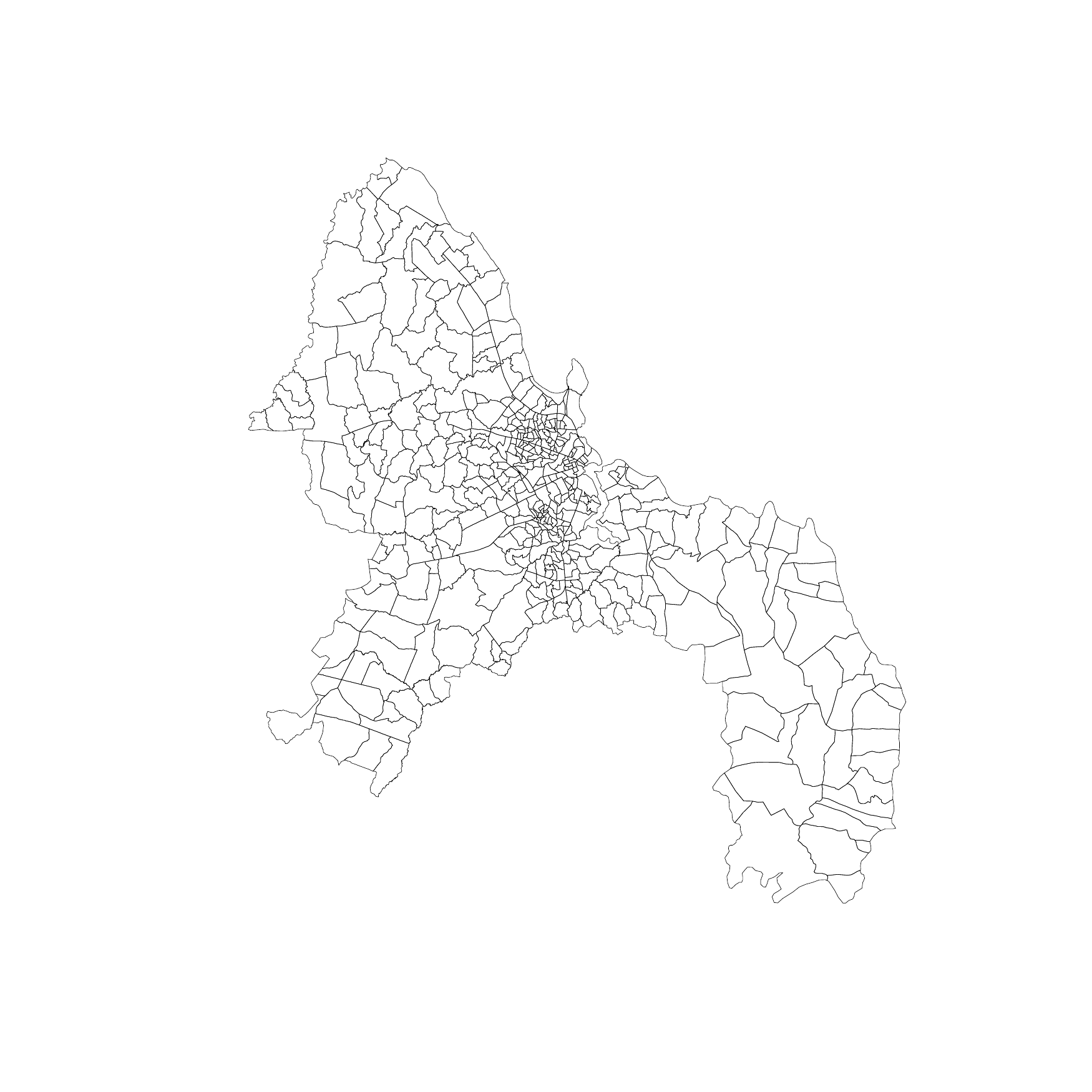}
    \includegraphics[width = 0.49\textwidth]{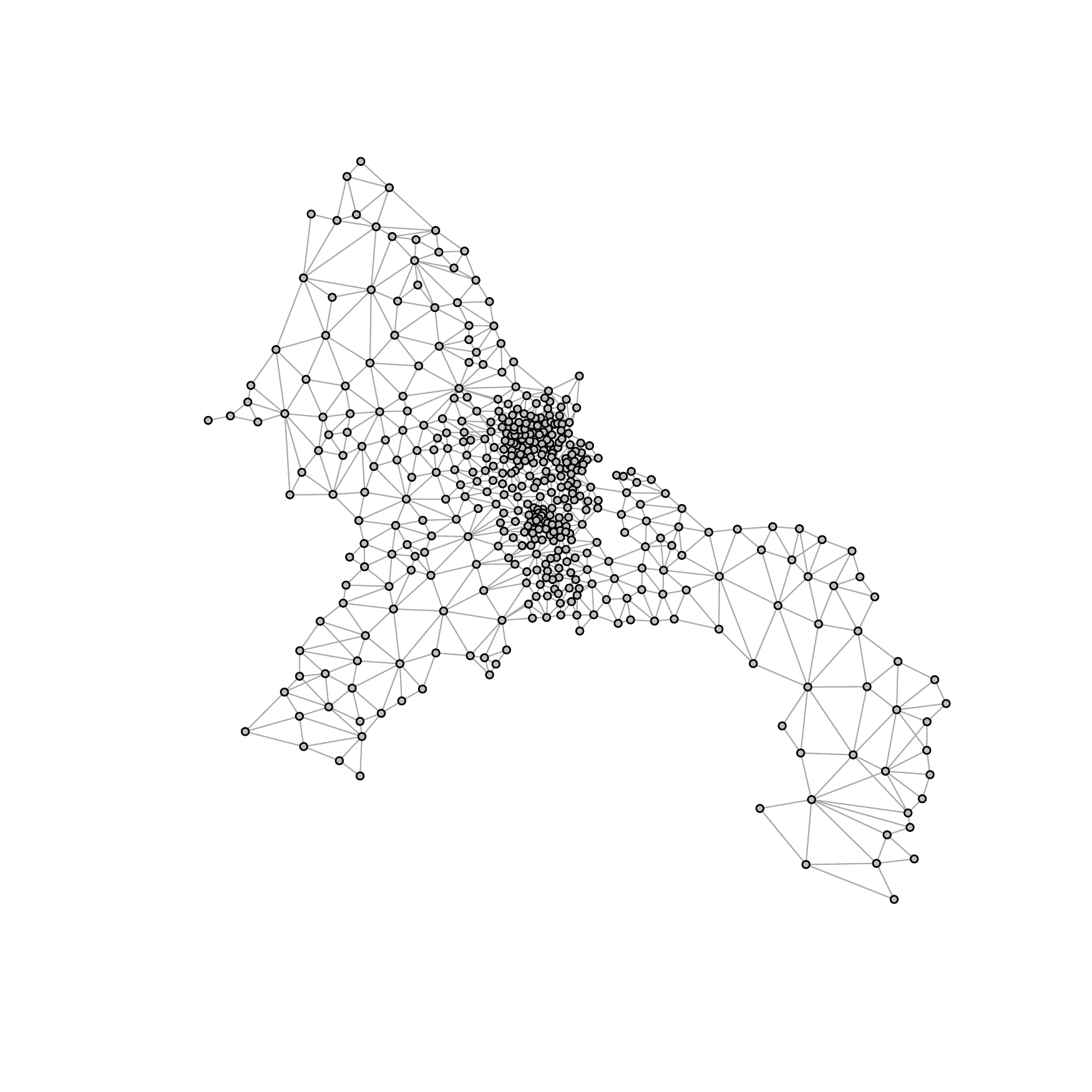}
    \caption{A map and the network representation of subwards in Dar es Salaam.}
    \label{fig: dar map and network}
\end{figure}

Urban regions are typically divided into sub-areas for administrative purposes, and these neighbourhoods often provide natural units over which to quantify deprivation. While spatially connected, such areas often vary greatly in size. In this paper, we model an urban region as a network, whereby these low-level areas are represented as nodes with edges joining neighbouring areas, such that we can use a network-based (i.e.\ non-Euclidean) distance to define spatial `closeness' between pairs of areas when defining prior assumptions of spatial smoothness. Using a network metric allows us to model nonstationary structures. We therefore begin by transforming the set of areas into a network by treating each area as a node and placing edges between adjacent areas; some modelling choices are required when dealing with noncontiguous areas or islands. In the Dar es Salaam network, we add two additional edges over the Kurasini creek to reflect the high-volume road and ferry connections. Figure \ref{fig: dar map and network} shows a map of Dar es Salaam and the corresponding network.

We can adapt the squared exponential covariance function in (\ref{eq: sq exp function}) for use with a network by letting $d_{ij}$ be the distance of the shortest path between subwards $i$ and $j$. The shortest distance can be computed using Dijkstra's algorithm \citep[see, e.g, ][]{Cor01}. Although using a network reduces the issue of stationarity, specifying the value of the length scale still may be challenging or restrictive; instead, when using the rational quadratic covariance function, which is is a mixture of squared exponential covariance functions with different length scale values, we can specify the relative importance of long and short scale variation in deprivation. Another option is to use the Mat\'ern covariance function, which would remove the assumption that the spatial structure is smooth. However, when using shortest-path distances in (\ref{eq: sq exp function}), the resulting matrix is not guaranteed to be positive semi-definite and we may need to project the matrix into the space of covariance matrices. This can be done in a number of ways, including setting the negative eigenvalues to 0 or modifying the polar decomposition \citep{Higham1988}.

Instead of using a distance based approach, we can construct the covariance matrix directly from the structure of the network. \citet{Estrada2010} describe several options for quantifying the `communicability' between two nodes of a network in terms of functions of the adjacency matrix of the network. The option we choose is based on the matrix exponential of the adjacency matrix as this measure emphasises connectedness over short distances rather than long distances to a greater extent than the alternatives described in \citet{Estrada2010}. Let $\Lambda = e^A$, where $A$ is the network's adjacency matrix, and let $D$ be a diagonal matrix containing the elements on the diagonal of $\Lambda$. The covariance matrix is given by
\begin{equation}
 \Sigma = \alpha^2 D^{-\frac{1}{2}}\Lambda D^{-\frac{1}{2}},
 \label{eq: cov function}
\end{equation}
where $\alpha^2$ is a hyperparameter describing the variance in the deprivation levels. The matrix $\Sigma$ therefore has diagonal entries $\alpha^2$ and off-diagonal entries proportional to the communicability of each pair of subwards in the network. We thus achieve our aim of assigning higher covariance between better-connected pairs of subwards, using a natural characterisation of the network. Although we use the matrix exponential covariance matrix in the paper, we find no discernible difference in the results of Sections \ref{sec: sim studies} and \ref{sec: dar study} when using the (network-adapted) squared-exponential covariance function.  

\subsection{Incorporating judge information} \label{sec: judge info}
We now incorporate judge covariates into the model as this avoids the assumption that the judges act homogeneously. Suppose there are $G$ groups of judges and let $\boldx_g$ be the vector of length $P$ containing the covariates for group $g$. We assume judges in the same group act homogeneously. The vector $\boldx_g$ may contain categorical,  discrete or continuous covariates or a mixture of all three; for a categorical covariate we represent the $q$ levels of the covariate by $q$ indicator functions. If $\boldx_g$ contains categorical covariates the number of groups may be small, but if $\boldx_g$ contains a continuous covariate each judge may be its own group. 

We model the deprivation in area $i$, as perceived by judges in group $g$, to be 
$$
\lambda_i^g = \lambda_i + \sum_{p = 1}^P x^g_p\beta_{pi},
$$
where $x^g_p$ is the $p^{th}$ element of $\boldx_g$ and $\beta_{pi}$ is the parameter corresponding to $x^g_p$ and area $i$, where $i = 1,\ldots,N$. Modifying the likelihood function in equation (\ref{eq: BT likelihood}) to take account of the contributions from each group of judges, we now have the likelihood function
\begin{equation}
  \pi(\boldsymbol{y}\mid \boldlambda, \boldbeta_1, \ldots, \boldbeta_P) = \prod_{g=1}^G\prod_{i=1}^N\prod_{j<i} \begin{pmatrix} n_{ijg} \\ y_{ijg}
\end{pmatrix} \pi_{ijg}^{y_{ijg}} (1-\pi_{ijg})^{n_{ijg} - y_{ijg}} \label{eq: judge likelihood},
\end{equation}
where $n_{ijg}$ is the number of times judges in group $g$ compared areas $i$ and $j$, $y_{ijg}$ is the number of times judges in group $g$ judged area $i$ to be more affluent than area $j$, $\pi_{ijg}$ is the probability judges in group $g$ judge area $i$ to be more affluent than area $j$ and is given by $\textrm{logit}(\pi_{ijg}) = \lambda_i^g - \lambda_j^g$, and $\boldbeta_p = \{\beta_{p1}, \ldots, \beta_{pN}\}$ is the set of parameters corresponding to $p^{th}$ element of the set of judge covariates. We recover the model and likelihood of Section \ref{sec: BSBT} by taking $G=1$ and $P=0$ in this formulation.

As in the BSBT model with no judge covariates, we place a constrained multivariate normal prior distribution on the spatial parameters $\boldlambda$, shown in (\ref{eq: constrained prior}). We also place an independent, constrained, multivariate normal prior distribution on each $\boldbeta_p$ which allows us to model the effect of each covariate spatially. So that the deprivation parameters, $\boldsymbol{\lambda}$, represent the grand mean of the deprivation for all judges, we enforce a second constraint amongst the set of parameters $\boldbeta_p$, which correspond to a given categorical covariate, as this allows us to treat each category symmetrically, i.e.\ we avoid fixing one category as a reference category and then not having any uncertainty associated with it. For a group of $q$ covariates representing the $q$ categories of covariate $p$, the corresponding parameters $\beta_{p_1i}, \ldots, \beta_{p_qi}$ need a constraint to ensure identifiability. We use $\beta_{p_1i} + \ldots +\beta_{p_qi} = 0$ for each area $i = 1,\ldots,N$. 

An example of including judge information is investigating how judges of different genders view different subwards. In less developed countries, women may be more vulnerable to different forms of exploitation than men (e.g.\ female genital mutilation, modern slavery, and forced marriage) and finding areas women view as more deprived than men may indirectly give information about where these practices are happening. We sort the judges into two groups (i.e.\ $G=2$), men and women. We let $\boldx_1^T = (1\ 0)$ for male judges and $\boldx_2^T = (0\ 1)$ for female judges (i.e.\ $P=2$). The appropriate constraint to ensure identifiability is then $\beta_{1i} + \beta_{2i} = 0$ for each area $i$. 

\subsection{Fitting the model} \label{sec: MCMC}
Now we have described the BSBT model, we develop a Markov chain Monte Carlo (MCMC) algorithm to infer the model parameters given the observed comparative judgements $\boldsymbol{y}$, and the judge covariates $\boldsymbol{x}$. The model parameters are: the deprivation parameters $\boldlambda$, any covariate parameters $\boldbeta_p$, and the covariance function variance hyperparameters $\alpha^2_\lambda$ and $\alpha^2_1, \ldots, \alpha^2_P$. By Bayes' theorem, the posterior distribution is 
\begin{align}
    \pi(\boldlambda, \boldbeta_1, \ldots, \boldbeta_P, \alpha^2_\lambda, \alpha^2_1, \ldots, \alpha^2_P \mid \boldsymbol{x}, \boldsymbol{y})\propto\ &\pi(\boldsymbol{y}\mid \boldlambda, \boldbeta_1, \ldots, \boldbeta_P) \, \pi(\boldlambda \mid \alpha^2_\lambda,\ \boldsymbol{1}^T\boldlambda = 0) \, \pi(\alpha^2_\lambda) \nonumber  \\
    &\times  \prod_{p=1}^P\pi(\boldbeta_p \mid \alpha^2_p,\ \boldsymbol{1}^T\boldbeta_p = 0) \, \pi(\alpha^2_p).\label{eq:posterior}
\end{align}
The first term on the right hand side is the likelihood function (\ref{eq: judge likelihood}) and the second term is the prior density for the spatial component $\boldlambda$, for which we use the constrained prior distribution (\ref{eq: constrained prior}). We place an independent prior distribution on the variance hyperparameter $\alpha^2_\lambda$, which is the third term on the right hand side. The product term contains the prior distributions for the covariate parameters $\boldbeta_1, \ldots, \boldbeta_P$ and the variance hyperparameters $\alpha^2_1, \ldots, \alpha^2_P$ for these distributions. 

The posterior density cannot be computed explicitly, but it can be sampled from using Algorithm \ref{alg: BSBT}. This MCMC algorithm involves iterating Gibbs updates for the variance hyperparameters, $\alpha^2_\lambda$, $\alpha^2_1, \ldots, \alpha^2_P$, and  Metropolis-Hastings updates for the spatial components, $\boldlambda$ and $\boldbeta_1, \ldots, \boldbeta_P$. For analytical convenience, we place a conjugate inverse-Gamma prior distribution on the variance hyperparameters, the density function of which is
$$
\pi(x; \chi, \omega) = \frac{\omega^\chi}{\Gamma(\chi)} \frac{1}{x^{\chi + 1}}\exp\left(-\frac{\omega}{x}\right) \qquad (x>0; \, \chi>0,\, \omega>0).
$$
The Gibbs updates are possible because the full conditional distribution for $\alpha^2_\lambda$ has a closed form. It is given by
$$
\alpha^2_\lambda \mid \boldlambda \sim \invGam\left(\chi + \frac{N}{2}, \omega + \frac{1}{2}\boldlambda\bar{\Sigma}^{-1}\boldlambda^T\right),
 $$
where $\bar{\Sigma}$ is the covariance matrix of the constrained prior with $\alpha^2=1$ in (\ref{eq: cov function}). Analogously, the full conditional distribution for $\alpha^2_p$ is 
$$
\alpha^2_p \mid \boldbeta_p \sim \invGam\left(\chi + \frac{N}{2}, \omega + \frac{1}{2}\boldbeta_p\bar{\Sigma}^{-1}\boldbeta_p^T\right),
$$
To update the deprivation parameters, $\boldlambda$, we use a Metropolis-Hastings sampler with an underrelaxed proposal mechanism \citep{Neal1998}. This allows us to update the parameters as a block and reduces the computational complexity compared to updating each deprivation parameter individually. Given the current deprivation parameters $\boldlambda$, we propose new values by 
$$
\boldlambda' = \sqrt{1- \delta^2}\boldlambda + \delta\boldsymbol{\nu},
$$
where $\delta \in (0, 1]$ is a tuning parameter and $\boldsymbol{\nu}$ is a draw from the constrained prior distribution in equation (\ref{eq: constrained prior}). We accept this proposal with probability
$$
\pacc = \min\left(\frac{\pi(\boldsymbol{y}\mid \boldlambda', \boldbeta_1, \ldots, \boldbeta_P)}{\pi(\boldsymbol{y}\mid \boldlambda, \boldbeta_1, \ldots, \boldbeta_P)}, 1\right).
$$
The proposal ratio using the underrelaxed proposal mechanism is the inverse of the prior ratio, meaning the acceptance probability is the ratio of the likelihood function with the proposed and current deprivation parameters. We follow an analogous process for the covariate parameters $\boldbeta_1, \ldots, \boldbeta_P$.

\begin{algorithm}
\caption{MCMC Algorithm for the BSBT Model}
	\begin{algorithmic}[1]
		\State Choose initial values for $\boldlambda$, $\boldbeta_1, \ldots, \boldbeta_P$ and ${\alpha^2_\lambda, \alpha^2_1, \ldots, \alpha^2_P}$. 
		\StateX \textit{On iteration $j $ of the MCMC algorithm do} 
		\State Update $\boldlambda_i$ using a Metropolis-Hastings step with underrelaxed proposal mechanism;
		 \State Update $\boldbeta_1, \ldots, \boldbeta_P$ using a Metropolis-Hastings step with underrelaxed proposal mechanism;
		 \State Update $\alpha^2_\lambda$ using a Gibbs step;
		 \For{$i$ in 1 to $P$} 
		 \State Update $\alpha^2_i$ using a Gibbs step;
		 \EndFor
	\end{algorithmic}
	\label{alg: BSBT}
\end{algorithm}

\subsection{Implementing the Model}
We have developed an R package to allow any user to implement this method on a comparative judgement data set. The package \texttt{BSBT} is available on CRAN \citep{BSBT}. It includes the novel comparative judgement data set on deprivation in Dar es Salaam, Tanzania, shapefiles for the 452 subwards in the city and a vignette containing instructions on how to reproduce the analysis in section \ref{sec: dar study}. The package allows users to place a constrained multivariate normal prior distribution for deprivation parameters over a predetermined network (it also facilitates constructing the network) and fit the model using the MCMC algorithm in Algorithm \ref{alg: BSBT}.  We provide a number of covariance functions, including the squared-exponential, Mat\'ern and matrix exponential functions. The MCMC functions included in the package can be used to fit either a spatial model, or a spatial model with a covariate for judge information. Due to our formulation of the likelihood function, the computational time for the BSBT implementation scales according to the number of areas, whereas the implementation provided in the \texttt{BradleyTerry2} package scales according to the number of comparisons.

\section{Simulation Study} \label{sec: sim studies}
To assess the model's ability to infer deprivation levels in a realistic scenario, we simulate deprivation levels for the subwards in Dar es Salaam by drawing a sample from the prior distribution, then seek to infer these from simulated comparative judgements. A map of the city and the corresponding network are shown in Figure \ref{fig: dar map and network}. We simulate the comparisons according to the model in equation (\ref{eq: BT likelihood}) and choose pairs of areas uniformly at random to compare. We simulate data sets of various sizes to mimic real data collection. The sizes of simulated data sets used in this paper are shown in Table \ref{tab:simStudySizes11}. We use `judge hours' to quantify the number of comparisons by the total judging time required, assuming 20 seconds per comparison or 180 comparisons per judge hour. 

\begin{table}
\caption{Data set sizes used in the simulation studies, using 180 comparisons per judge hour.\label{tab:simStudySizes11}}
\centering
\begin{tabular}{c|*{10}{c}}
Judge hours & 1 & 2 & 5 & 10 & 25 & 50 & 100 & 250 & 500 & 1,000\\
\hline
Comparisons & 180 & 360 & 900 & 1,800 & 4,500 & 9,000 & 18,000 & 45,000 & 90,000 & 180,000
\end{tabular}
\end{table}

We fit the model to each data set, running the MCMC algorithm for 1,500,000 iterations and removing the first 500,000 iterations as a burn-in period. We fix the tuning parameter $\delta = 0.01$, based on initial runs of the algorithm. For the prior distribution on $\alpha^2_\lambda$, we fix $\chi = \omega = 0.1$ which results in a somewhat noninformative distribution \citep{Gel06}. To assess the model fit, we compute the Mean Absolute Error (MAE) for the result of each set of comparisons, which is given by 
$$
\textrm{MAE} = \frac{1}{N}\sum_{i=1}^N |\lambda_i - \hat{\lambda}_i|,
$$
where $\hat{\lambda}_i$ is the estimate corresponding to the MLE or posterior mean for area $i$. 

\begin{figure}[ht]
    \centering
    \includegraphics[width=0.49\linewidth]{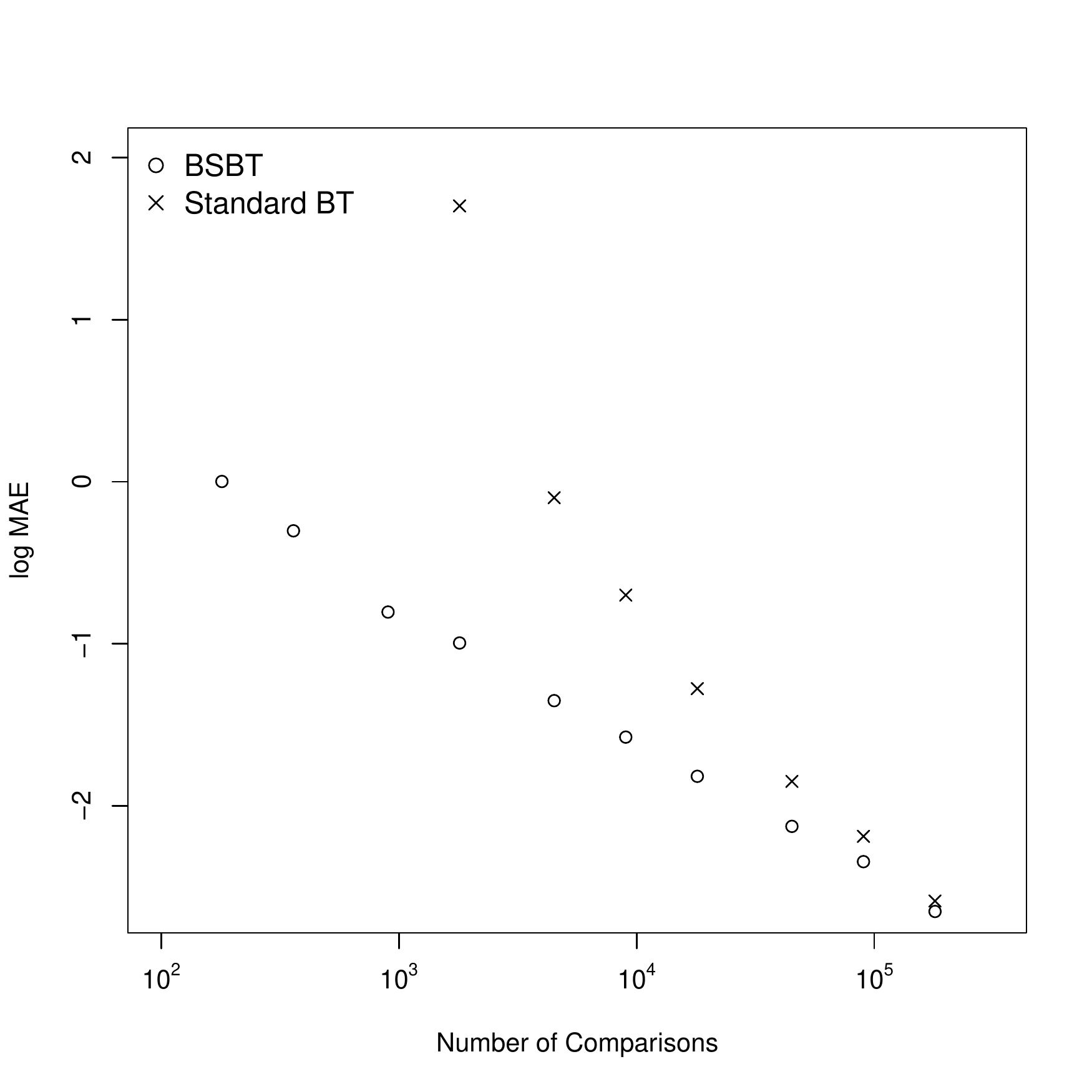} 
    \caption{Log MAE for the simulation study comparing performance of the standard BT and the BSBT models in terms of error as a function of the number of comparisons.
    }
    \label{fig: sim studies MAE}
\end{figure}

Figure \ref{fig: sim studies MAE} shows the log MAE for each data set. The BSBT model outperforms the standard model for all sizes of data set used. For a fixed number of comparisons, the BSBT model has smaller error than the standard model. For example, when using 1,800 comparisons (10 judge hours), the MAE using the BSBT model (0.260) is less than a third of the error in the standard model (0.907). Figure \ref{fig: sim studies MAE} also shows that we can substantially reduce the number of comparisons required to achieve a given level of error by using the BSBT instead of the standard BT model. For example, MAE in the BSBT with 5 judge hours is similar to that in the standard model with 50 judge hours, a decrease in judge hours of 90\%; and 250 judge hours with the standard model yields similar MAE to 100 judge hours with BSBT, a still substantial reduction of 60\% in terms of the data required to give a similar level of performance. For small data sets we are unable to compute the MLE for all areas and so the corresponding MAE is undefined for the standard BT model. Here we see one of the main advantages of the BSBT model: including weak prior assumptions about spatial correlations allows it to learn about areas featured in very few, or even no, comparisons from information about nearby areas.

We observe that the performance of the BT and BSBT models are very similar when the number of judgements is large. This is to be expected from the Bernstein--von Mises theorem \citep{Kleijn12} whereby the posterior distribution of finite dimensional parameters and the MLEs tend to the same asymptotic multivariate normal distribution for large samples, subject to smoothness and identifiability conditions on the prior distribution and a positivity condition on the prior at the true value.

We also present a simulation study on a synthetic 1-d `city' in Section 1 of the supplementary material. Although less realistic than the 2-d study above, it has the significant advantage of allowing much easier visualisation of the synthetic ground truth, the simulated data and the results of fitting our model; aiding interpretation of what our methods achieve. 

\section{Deprivation in Dar es Salaam, Tanzania} \label{sec: dar study}

\subsection{Bayesian Spatial Bradley--Terry model}
\label{sec:darResults}
We fit the BSBT model to the data and run the MCMC algorithm shown in Algorithm \ref{alg: BSBT} for 1,500,000 iterations, removing the first 500,000 iteration as a burn-in period. This took around 3 hours on a 2019 iMac with a 3 GHz CPU. We examined trace plots to ensure adequate mixing of the Markov chain and to choose the length of the burn-in period. These are given in Section 2 of the supplementary material. We fix the tuning parameter $\delta = 0.01$, based on initial runs of the algorithm, and the inverse gamma prior distribution parameters $\chi = \omega = 0.1$.

\begin{figure}[ht]
    \centering
    \includegraphics[width=0.49\textwidth]{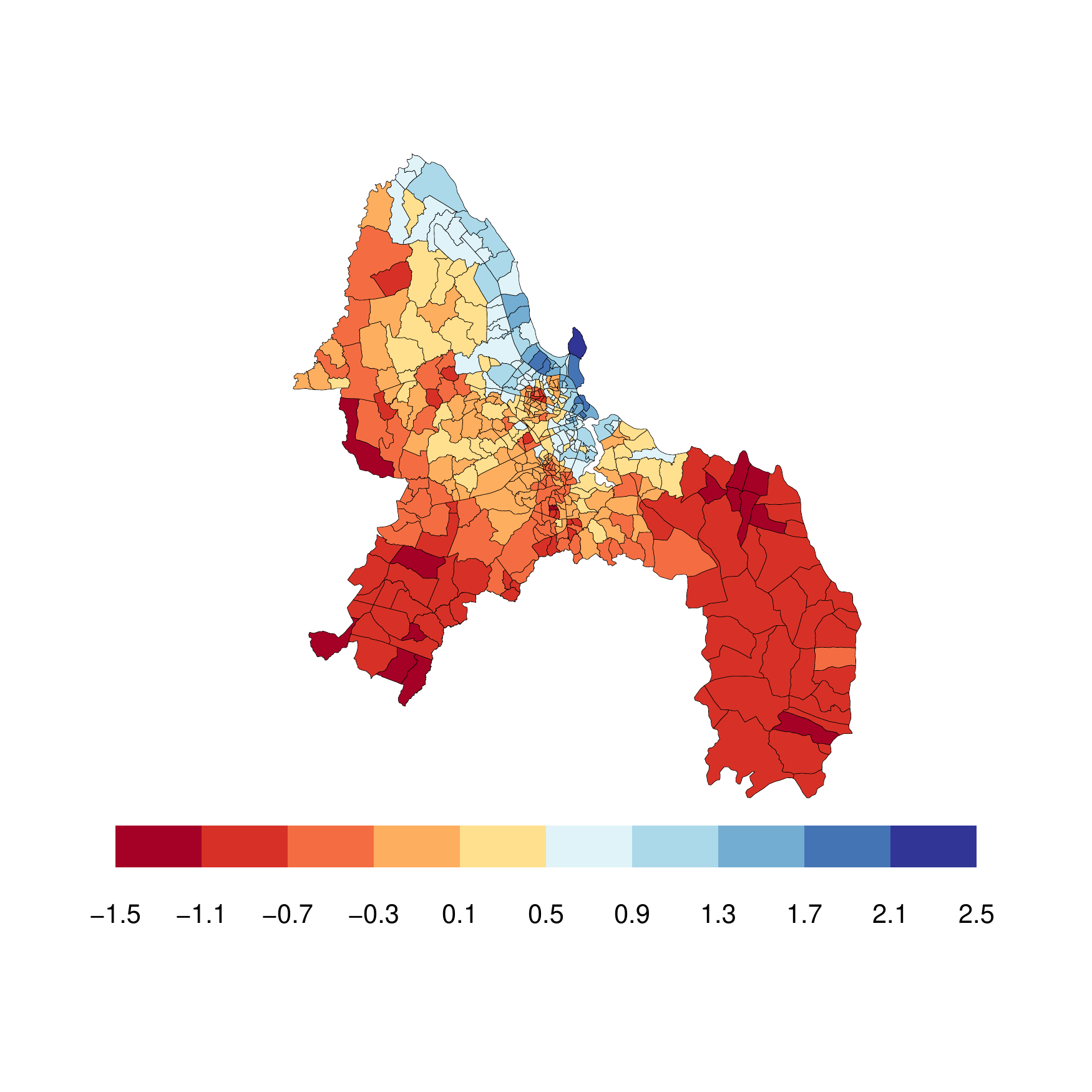}
    \includegraphics[width=0.49 \textwidth]{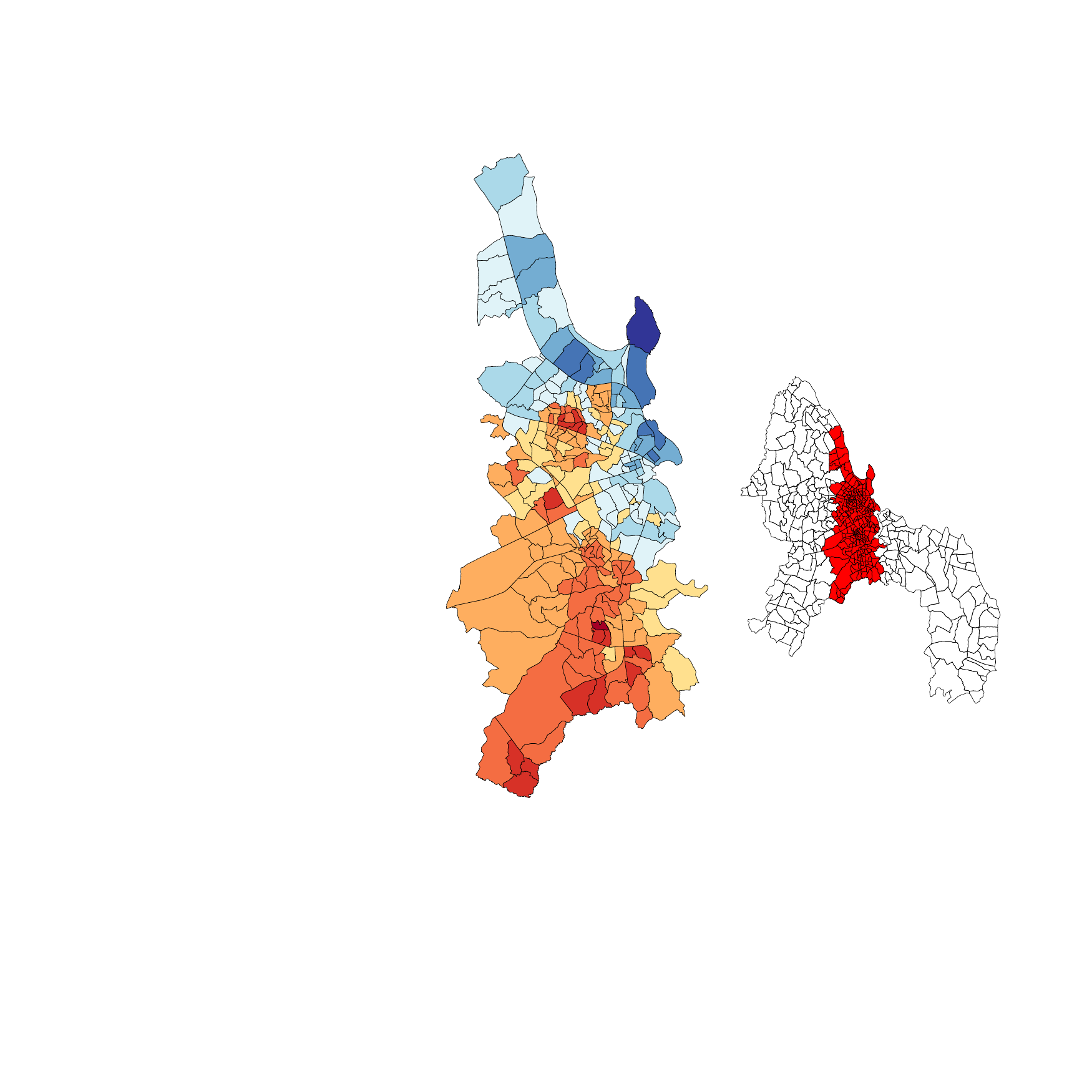}
    \caption{The posterior mean values for the BSBT model applied to the Dar es Salaam data set. The figure on the right shows a magnified section of the centre of the city, together with this central area highlighted on a map of the whole city.}
    \label{fig:dar_results}
\end{figure}

The resulting estimates for the level of deprivation in each subward in the city are shown in Figure \ref{fig:dar_results}. 
We see a north-south trend, whereby the level of deprivation increases further south in the city. We find several sharp changes in deprivation in the city centre, where slums neighbour affluent subwards. The most affluent subward is Masaki, and the ten most affluent areas are all concentrated around the Masaki peninsula directly north of the city centre and home to most of the affluent expatriate communities. The ten most deprived subwards are geographically spread out, with one, Mpakani, being located in the centre of Dar es Salaam and the others spread across the outer regions of the city. Four of the ten most deprived subwards are in the Somangila ward, a coastal ward in the east of the city. 

\begin{figure}[ht] 
    \centering
    \includegraphics[width = 0.49 \textwidth]{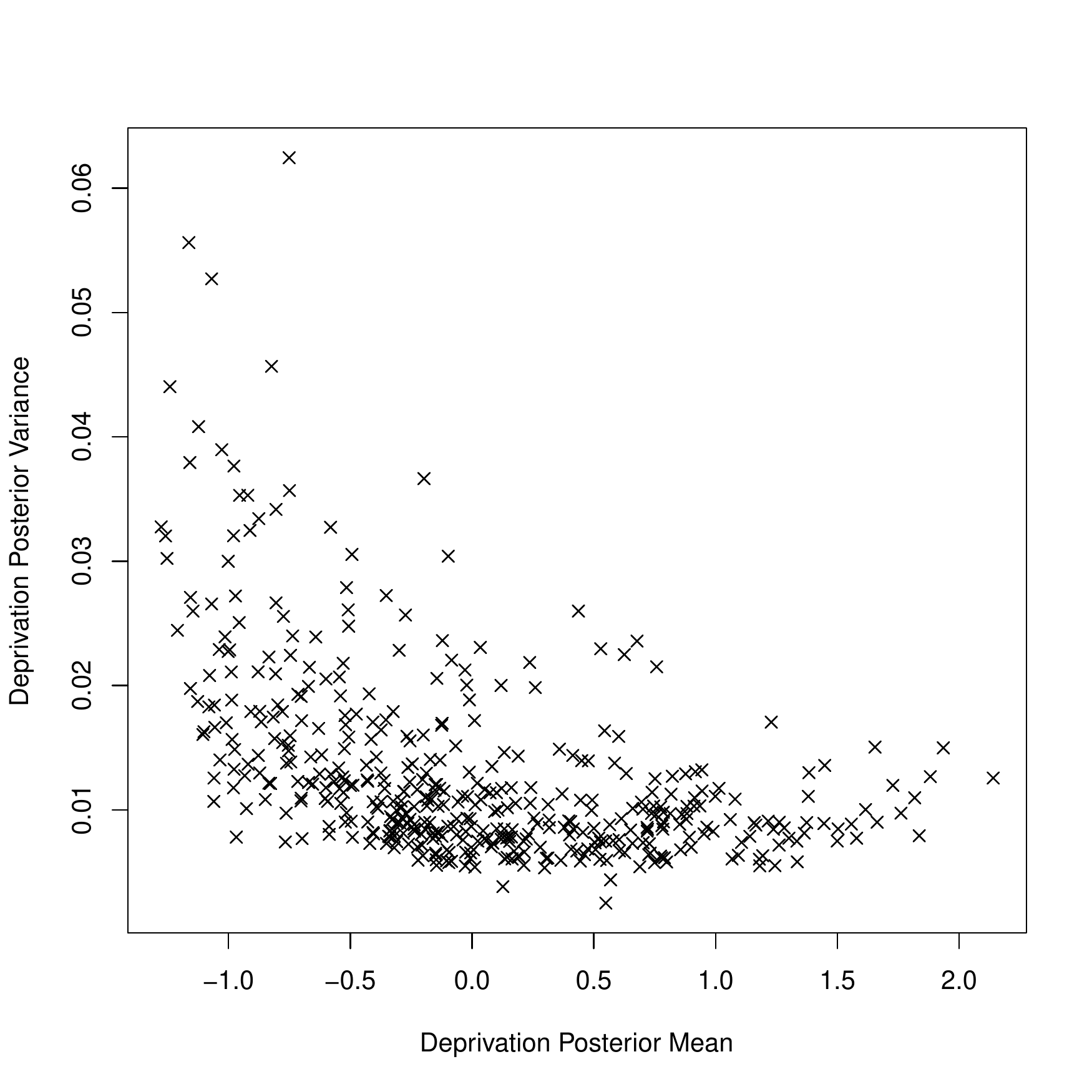}
    \includegraphics[width = 0.49 \textwidth]{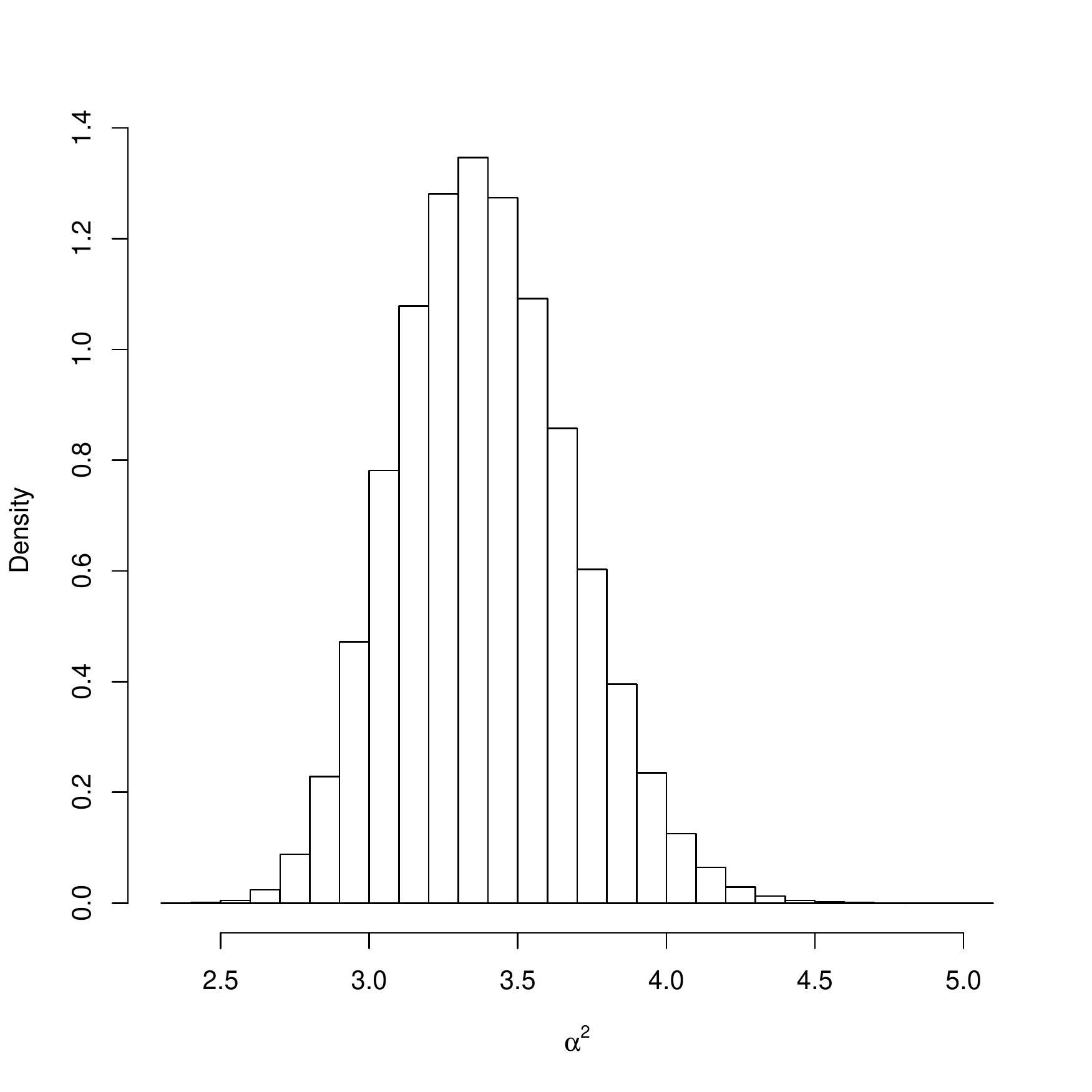}
    \caption{Uncertainty in the fitted BSBT model for the Dar es Salaam data set. Left: The estimated level of deprivation in each subward against the uncertainty in the estimate. Right: The posterior distribution for the variance parameter $\alpha^2_\lambda$.}
    \label{fig: dar uncertainty}
\end{figure}

The uncertainty in the estimates for the level of deprivation in each subward differs considerably, as shown in Figure \ref{fig: dar uncertainty}. We see a correlation between the level of uncertainty in our estimate and the estimated level of deprivation. As the most affluent areas tend to also be well known areas, such as tourist resorts or areas with government buildings, we were able to collect more comparisons involving these subwards and therefore there is less uncertainty in our estimates for the deprivation in these areas. We also estimate the variance parameter $\alpha^2_\lambda$; its posterior mean is 3.378 with 95\% CI (credible interval) (2.868, 3.993) and the posterior distribution is shown in Figure \ref{fig: dar uncertainty}. Section 2 of the supplementary material gives more diagnostic information and a short investigation of judge reliability which concludes that no judges provide judgements which are notably out of line with the fitted model. 

Because approximately 1 in 7 of the comparisons in the data set are tied, which is a substantial proportion, we must take care that our approach to treating ties does not substantially affect the inferred deprivation levels. For the results in this paper, wherever a comparison was tied we randomly allocated a winner. In Section 3.1 of the supplementary material, we carry out a sensitivity analysis of these random allocations, examining 20 data sets generated via different random seeds, and confirm the robustness of our results. In Section 3.2 of the supplementary material, we consider two alternative treatments for the tied comparisons (treating a tie as `half a win' for both subwards involved, and discarding the ties altogether). We found the posterior means were largely unaffected by the treatment of ties. Discarding the ties increases the uncertainty as we are discarding a considerable amount of data, and treating the ties as half a win yields estimates that have the lowest variance of any treatment we considered. 
We have favoured the treatment of allocating winners of the tied comparisons at random. This is on the basis that the results appear insensitive to the specific random allocation used, it makes use of all the available data, and it is conservative in terms of the resulting uncertainty in parameter estimates.

Results for the standard BT and BSBT models are very similar; we see very similar inferred deprivation levels and uncertainties. (See Section 4 of the supplementary material.) However, the data set that we have is quite large, so this is likely a data saturation effect (cf.\ Figure \ref{fig: sim studies MAE}). An important aim of our work is to investigate if many fewer comparative judgements could have been collected, at a much reduced cost, with little loss of information.

\subsection{Efficiency of the BSBT model}
To investigate the effectiveness of the model when we have a smaller number of comparisons, we also fit both the standard BT and BSBT models to the comparisons collected on the first two days of the field work. This subset includes 13,361 comparisons (around 18\% of the original data set). All subwards feature in this partial data set and the number of comparisons each subward was featured in ranged from 2 to 233, with mean 60. Five subwards `lost' every comparison they were featured in, making it difficult to estimate their deprivation level using the standard BT model. We compute the MAE taking the true values to be the inferred deprivation levels using the full data set. Using the BSBT model on this partial data set roughly halves the MAE compared to the standard BT model, reducing it from 0.523 to 0.267.  We are still able to identify sharp changes in deprivation levels, for example where slums neighbour affluent areas in the city centre.

\begin{figure}[ht] 
\centering
       \includegraphics[width = 0.49 \textwidth]{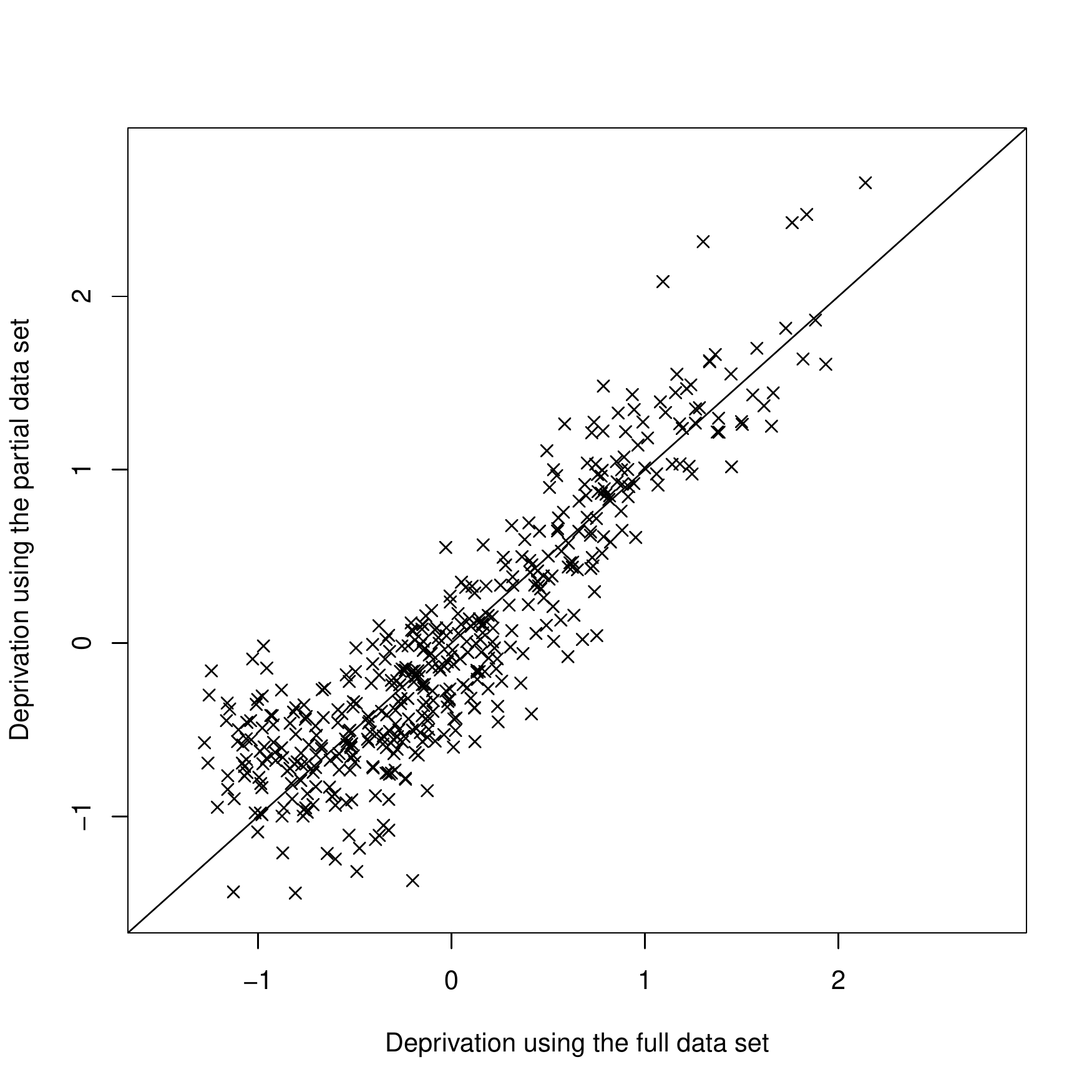} \includegraphics[width = 0.49 \textwidth]{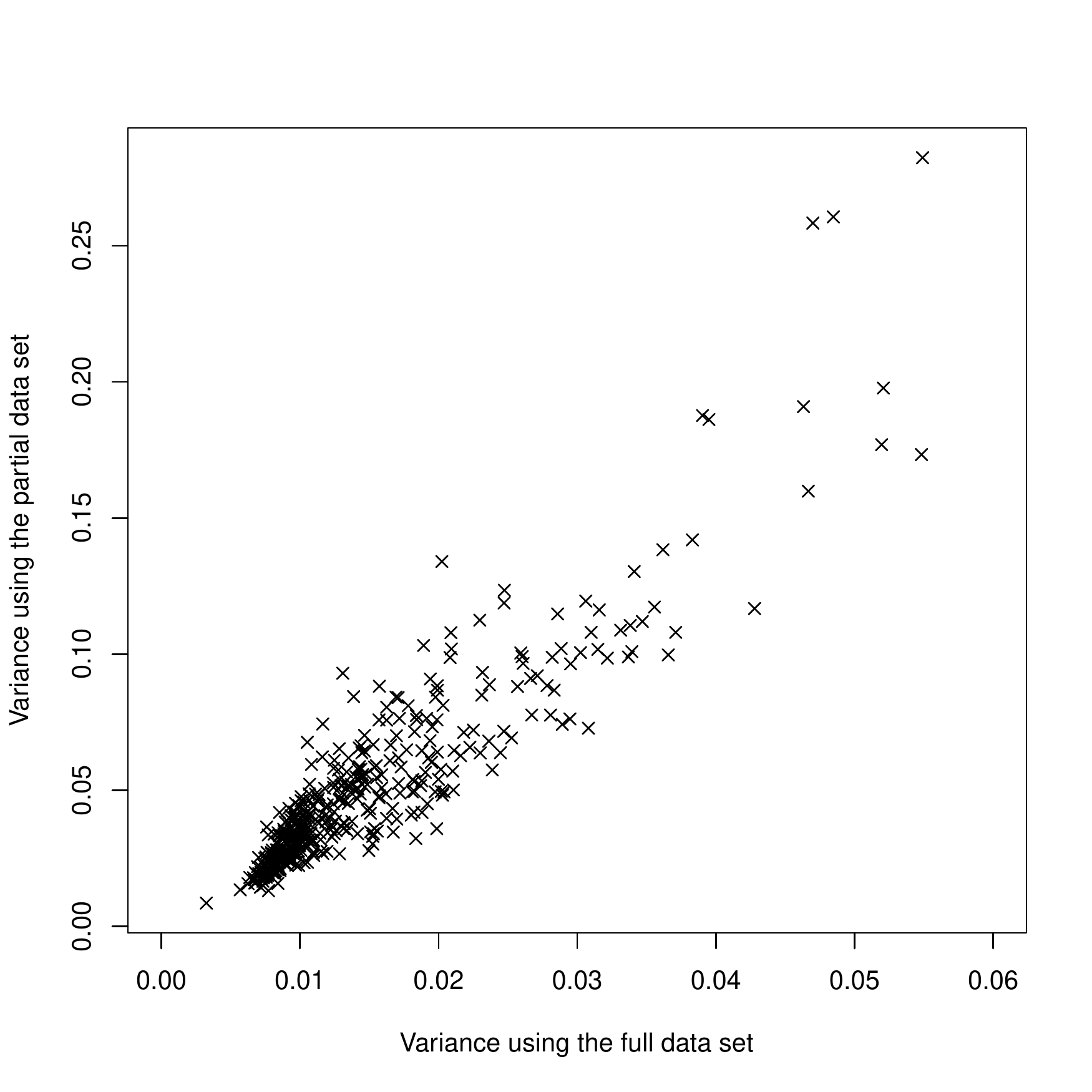}
    \caption{Left: The posterior mean estimates for deprivation using the full and partial data sets. Right: The posterior variance using the full and partial data sets.}
    \label{fig: dar first two days uncertainty}
\end{figure}

In Figure \ref{fig: dar first two days uncertainty}, we report the posterior mean and variance for the deprivation in each subward. There is some shrinkage in the estimates for the most deprived subwards, but no consistent change elsewhere. There is strong linear correlation between the estimated deprivation levels using the full and partial data set ($\rho = 0.832$), showing that in terms of identifying subwards as, for example, somewhat affluent or very deprived, very little is lost by using the partial data set. As is expected, using less data results in higher uncertainty, however the uncertainty is generally small with respect to the deprivation parameter values and the additional uncertainty does not appear to apply to subwards in any systematic way. Alongside the analysis shown in Figure \ref{fig: sim studies MAE}, this shows that by using the BSBT model, in future we can collect far fewer comparisons yet attain similar levels of error in the results. This will reduce the time and cost associated with data collection in similar future fieldwork.

\subsection{Judge information in Dar es Salaam}
\label{sec:dar_judge_gender}
Firstly, we investigate if the men and women in the study perceived subwards differently. For the Dar es Salaam data, there were 91 female judges and 133 male judges. For reasons outlined in the introduction we are interested in determining whether different genders have different perceptions of some parts of the city.  Our first observation is that each male judge did on average 328 comparisons, whereas the average among female judges was 200. This is because the women took longer to carry out individual comparisons than the men. Another difference is that the women tended to be familiar with fewer subwards than the men, perhaps suggesting they are less mobile in the city. We fit the BSBT model with gender effect to the data, here $G = 2$ as we sort comparisons into two groups (men and women) and $P = 2$ as we model the effect of being male or female. We run the MCMC algorithm for 5,000,000 iterations, which took one day on a 2019 iMac with 3 GHz CPU. Diagnostic plots can be found in Section 5 of the supplementary material. We fix $\delta = 0.01$ based on initial runs of the algorithm. We estimate the variance $\alpha^2_\lambda$ (for $\boldlambda$) to be 3.846 (95\% CI: (3.073, 3.694)) and $\alpha^2_1$ (for $\boldsymbol{\beta}_1$) 0.026 (95\% CI: (0.002, 0.034)). Such a small value of $\alpha_1^2$ suggests the men's and women's perceptions are highly correlated. 

\begin{figure}
\centering
\includegraphics[width = 0.49\textwidth]{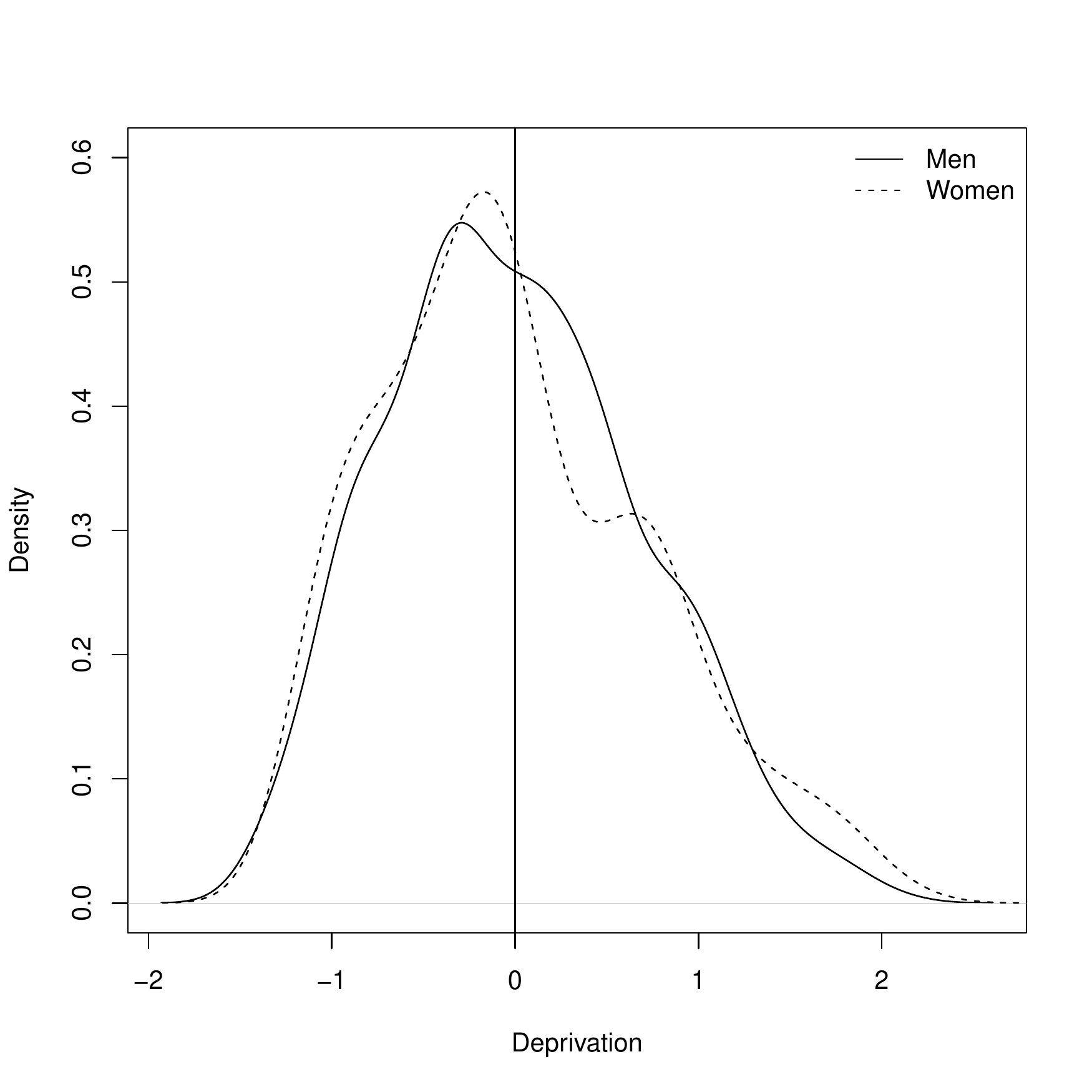}\\
\includegraphics[width = 0.49\textwidth]{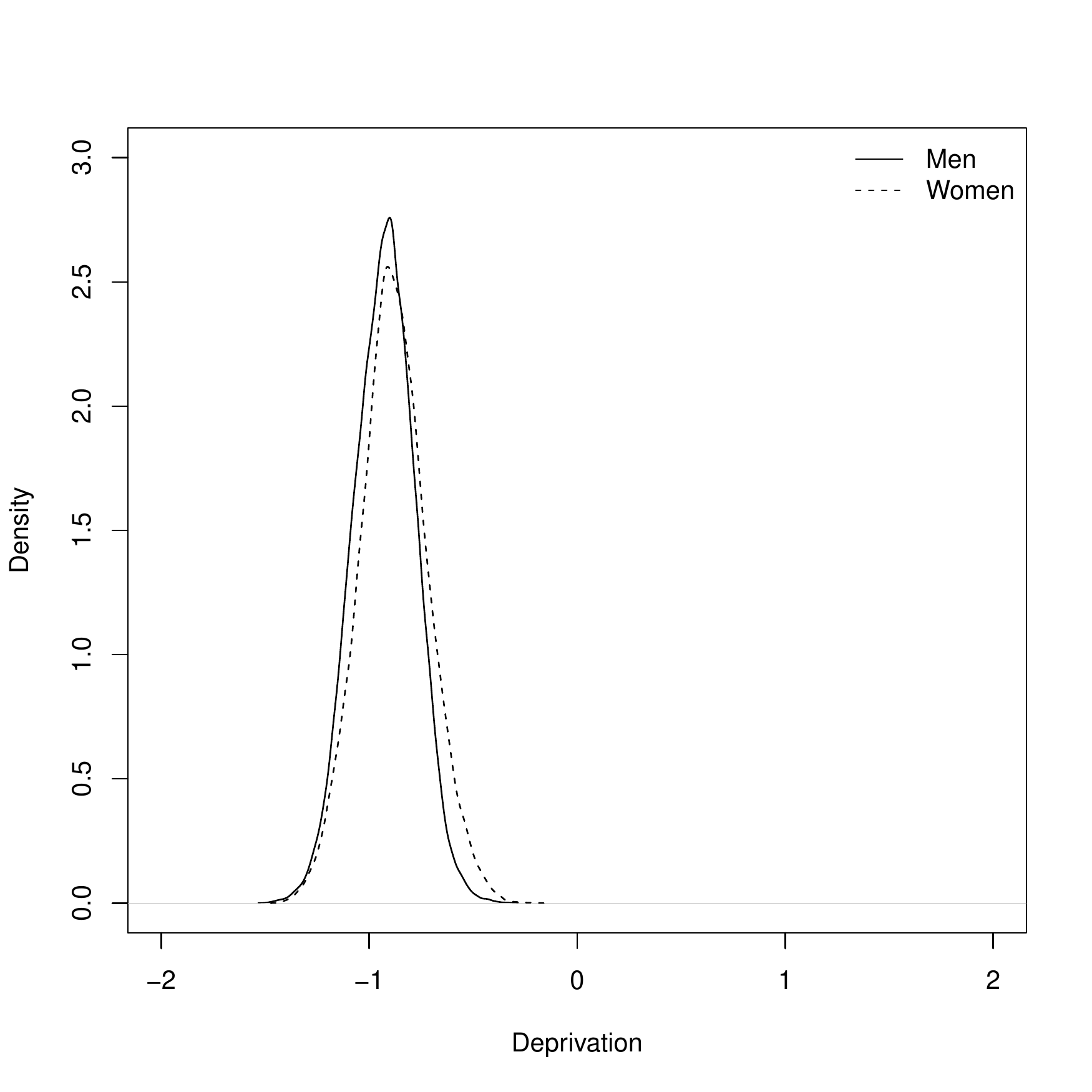}
\includegraphics[width = 0.49\textwidth]{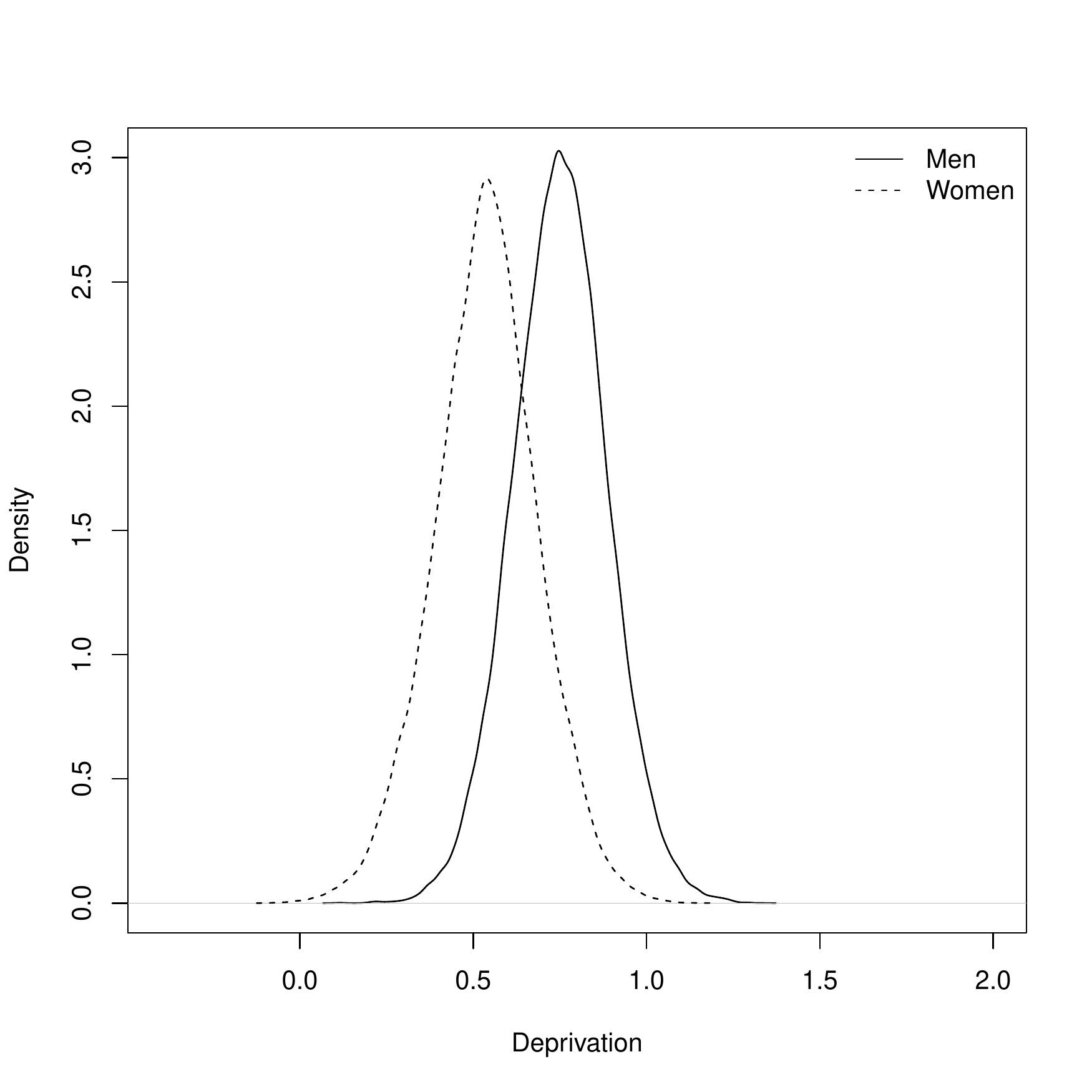}
\caption{Top: Kernel density estimation of posterior means for the levels of deprivation given by men and women. Bottom: The posterior distributions for the men's and women's perceptions of Kibonde Maji A (left) and Hananasif (right). We do not infer a difference between how men and women perceive Kibonde Maji A, but we do for Hananasif.}
\label{fig:dar_gender_results}
\end{figure}

\begin{figure}
    \centering
    \includegraphics[width = 0.49\textwidth]{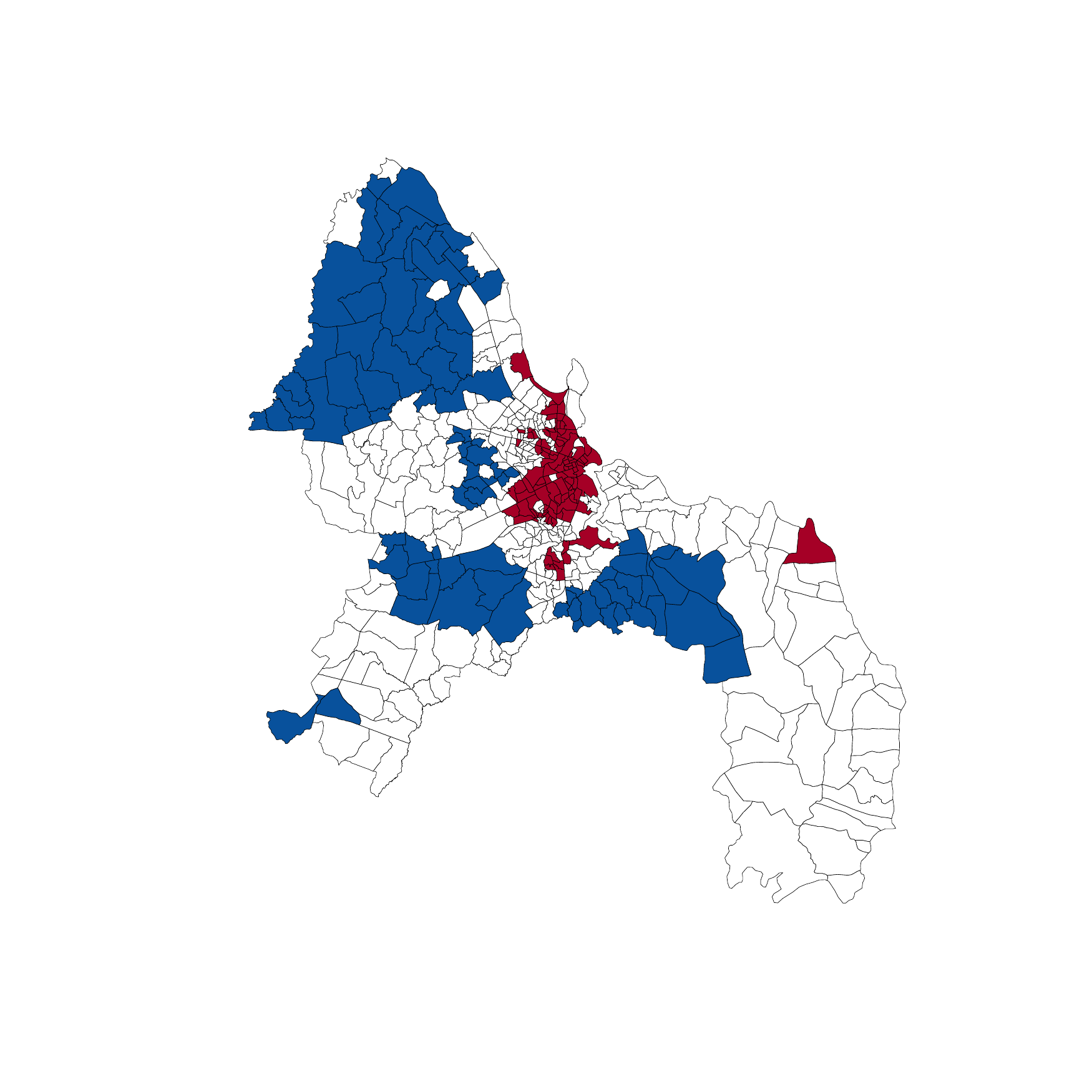}
    \caption{The difference between men and women's perceived deprivation levels, coloured by the 95\% credible interval. Blank subwards have a credible interval which contains 0. Red and blue subwards have a credible interval which does not contain 0, with blue showing positive and red showing negative. }
    \label{fig: gender different}
\end{figure}

Figure \ref{fig:dar_gender_results} shows the distribution of the posterior mean deprivation levels perceived by men and women. We see that the distribution of the levels of deprivation perceived by men and women are largely the same. We also show posterior density estimates for men's and women's perceptions of two subwards. In Kibonde Maji A, a somewhat deprived subward in the south of the city on a trunk road, there is no perceptible difference in how men and women perceive the subward. In Hananasif, an inner city subward near the business district, women perceive the subward to be considerably more deprived than men do. In Figure \ref{fig: gender different} we show the spatial structure in the difference between how men and women view the subwards, based on whether or not CIs for the discrepancies $\beta_{0,i}$ (for each subward $i$) contain zero. The subwards women view as more deprived than men are mostly concentrated in the centre of the city, and the majority of the subwards which women view as less deprived are in the outer regions of the city. We suggest two reasons for the difference in perceptions: the first is personal safety, as women may perhaps feel less safe in the city centre; the second is because the centre is the location of both the central business district and many nightlife venues, which may offer better opportunities to men. 

Secondly, we investigate if students perceived deprivation differently to the other judges. Students made up 37\% of the judges and made 41\% of the comparisons. We fit the BSBT model with $P = 2$, as there are two groups of judges (students and non-students). As in the gender differences model, we run the MCMC algorithm for 5,000,000 iterations.  We run the MCMC algorithm for 5,000,000 iterations, which took one day on a 2019 iMac with 3 GHz CPU. Diagnostic plots can be found in Section 5 of the supplementary material. We find there is no difference between how the students and non-students perceive deprivation in the city; all 95\% CIs for the discrepancy between students and non-students contain 0. The mean absolute discrepancy is 0.016 and the maximum absolute discrepancy is 0.035; indicating very little difference between the two groups. We estimate the variance $\alpha^2_\lambda$ (for $\boldlambda$) to be 4.953 (95\% CI: (3.982, 6.052)) and $\alpha^2_1$ (for $\boldsymbol{\beta}_1$) 0.005 (95\% CI: (0.004, 0.007)). We note that the posterior mean estimate for the variance for the discrepancy parameter is an order of magnitude smaller than in the gender discrepancy results, further suggesting the students and non-students have very highly correlated responses.

\section{Discussion}
\label{sec: discussion}
In this paper we have developed a nonparametric spatial version of the Bradley--Terry model and fitted it to a novel data set to infer deprivation levels in Dar es Salaam, Tanzania. Our methods also allow us to incorporate judge information into the model, e.g.\ judge gender or occupation, to understand the perceptions of different groups of judges.

We analysed a novel data set on deprivation in Dar es Salaam, not only estimating the level of deprivation in the city's 452 subwards, but demonstrating the effectiveness of the BSBT model in significantly reducing data requirements by incorporating spatial correlations in the prior distribution for deprivation levels. As far as we are aware, no estimates for deprivation on such a fine scale are currently available. We were able to identify slums in the centre of the city and estimate the level of deprivation in the peri-urban outer regions of the city. Our findings show that there are several sharp changes in the level of deprivation in the centre of the city where very affluent areas neighbour slums. There is also a difference in how men and women view some areas; specifically we find that women view some parts of the centre of Dar es Salaam as more deprived than men do, but tend to view some parts of the outer regions of the city as less deprived than men do. Our data collection, modelling and analysis provides up-to-date estimates of deprivation levels in Dar es Salaam via the involvement of over 200 of the citizens of the city. 

There is scope for agencies in developing countries to use the BSBT model to design interventions based on a quantitative analysis of social issues. This is advantageous to agencies working in environments where official statistics are low quality or not available. This is not limited to deprivation but any social issue that citizens can compare areas on, for example estimating prevalence of Female Genital Mutilation, or prevalence of black market trading. Similarly, such studies need not be limited to cities, but any context which has a spatial or network component; for example a group of villages spread out across a large area or a network of individuals linked by telecommunications data. 

There are a number of possible directions in which the BSBT model may be fruitfully extended and further explored. The BSBT model has a large computational cost  and there is scope to reduce the computational time required by developing a more efficient MCMC algorithm, for example by adaptive updating of the under-relaxed tuning parameter $\delta$. We could further reduce the amount of data required by optimising the experimental design and identifying pairs of areas which should be asked about or adaptively identifying areas which need to be compared \citep[see, e.g.,][]{Pol12, Pfeiffer2012}.

There is further information to be extracted from the data collected in Dar es Salaam. For example, in addition to our analysis of the effect of gender and occupation, it may be of interest to local agencies to understand whether other covariates (or combinations of covariates) are associated with different perceptions of deprivation. We can also investigate the tied comparisons using a multinomial model, \citep[see, e.g.,][]{Rao67, Dav70}, to investigate the effect of comparing subwards which had similar deprivation levels.  

We have developed new models for efficiently estimating the level of deprivation in urban areas based on comparative judgement data. Existing comparative judgement models require a large amount of data to produce high quality results and collecting such quantities of data is often difficult or infeasible when working in developing countries, where data collection can be prohibitively expensive and time-consuming. Using the Bayesian Spatial Bradley--Terry model, we could have collected considerably fewer comparisons without affecting the quality of our results. When using the data collected only on the first two days on the fieldwork, the error in the BSBT model is small, and substantially smaller than when using the standard model. We achieved this by including a spatial element in the model, where the level of deprivation in one subward is correlated with the level in nearby subwards. We modelled the spatial structure using a multivariate normal prior distribution with a covariance matrix based on the network structure of the city, which avoids making rigid parametric assumptions. We also showed how our method can be used to analyse how different genders perceive the level of deprivation in different areas, and how different their perceptions are. This can help researchers identify areas where one gender may be facing specific problems.

\section{Acknowledgements}
This work was supported by the Engineering and Physical Sciences Research Council [grant reference EP/T003928/1]. We also thank the Humanitarian OpenStreetMap Team (HOT) for their support in data collection. We are grateful to the two reviewers and associate editor for helpful and constructive comments that have improved this article.

\clearpage
\bibliographystyle{rss}
\bibliography{bibliography}
\end{document}


\maketitle

\def\spacingset#1{\renewcommand{\baselinestretch}%
{#1}\small\normalsize} \spacingset{1}

\section{A 1-d Study} \label{sec: 1d sim study}
We construct a one dimensional `city', where the areas which are compared are points on the line. Although this is a simple toy example, it allows us to visualise the city and level of deprivation in each area. We draw the locations for 100 areas from a Laplace distribution with mean 0 and variance 8, as this gives a region near 0 which is densely packed with relatively small areas, and fewer but larger areas in the outlying parts of the city. We specify the level of deprivation in each area by the piecewise function
$$
g(x) = \begin{cases}
\sin(x) + x + \pi &\quad \textrm{if } x < \pi, \\
\sin(4x)&\quad \textrm{if }  \pi \leq x \leq \pi, \\
\sin(x) - x + \pi &\quad \textrm{if } x > \pi. \\
\end{cases}
$$
This gives a level of deprivation which changes quickly in the city centre compared to the outskirts. 

We simulate the comparisons according to the model in equation (2), choosing pairs of areas uniformly at random to compare. We simulate data sets of various sizes to mimic real data collection. The sizes of simulated data sets used in this paper are shown in Table \ref{tab:simStudySizes}. These are the same as in the 2-d study in Section 3 of the main text, but with the larger sizes excluded since here there are many fewer areas and we find that both models reach data saturation by 9,000 comparisons. 

\begin{table}
\caption{Data set sizes used in the simulations studies, using 180 comparison per judge hour.\newline}
\label{tab:simStudySizes}
\centering
\begin{tabular}{c|*{10}{c}}
Judge hours & 1 & 2 & 5 & 10 & 25 & 50 \\
\hline
Comparisons & 180 & 360 & 900 & 1,800 & 4,500 & 9,000 
\end{tabular}
\end{table}

\begin{figure}[htbp]
\centering
\includegraphics[width = 0.49\textwidth]{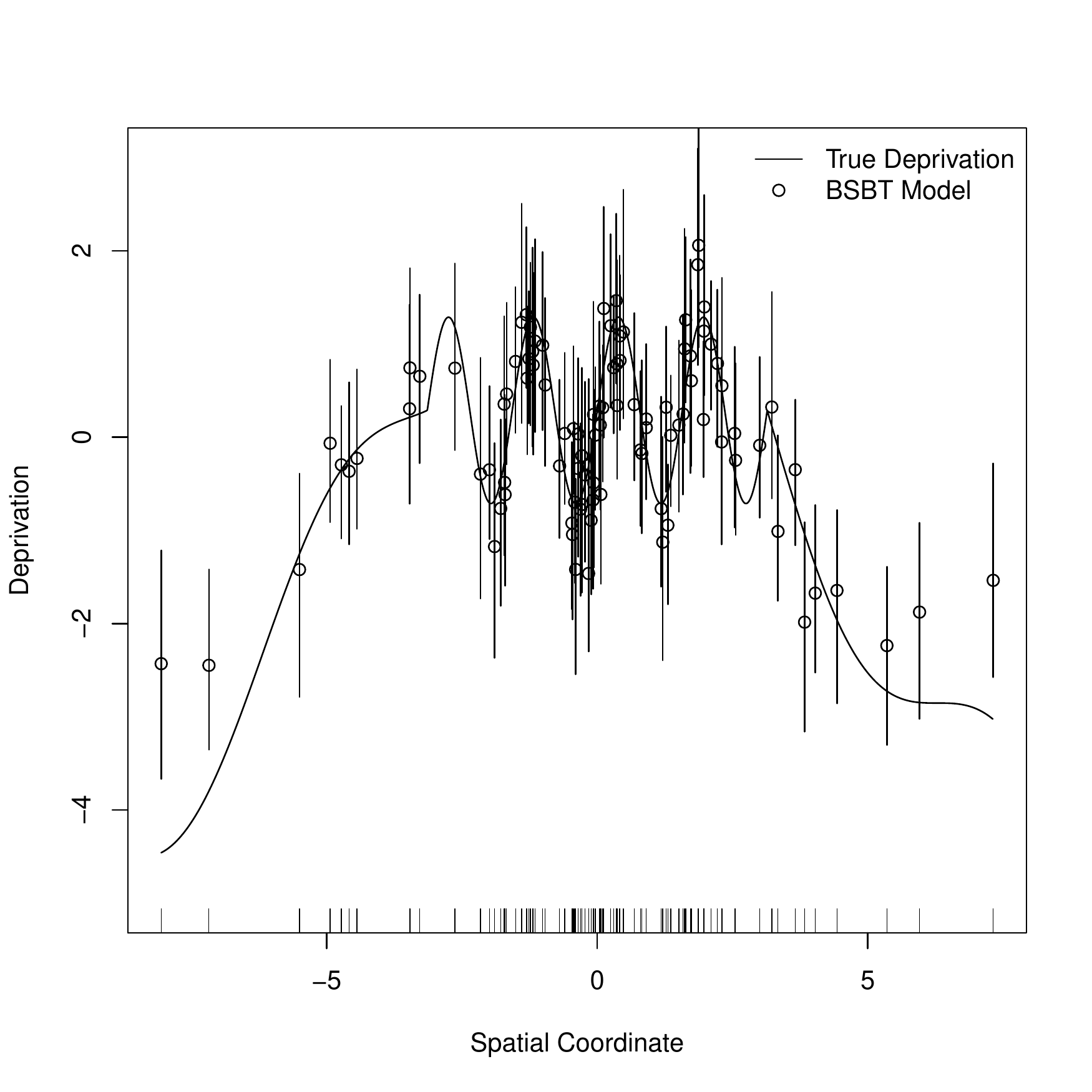}
\includegraphics[width = 0.49\textwidth]{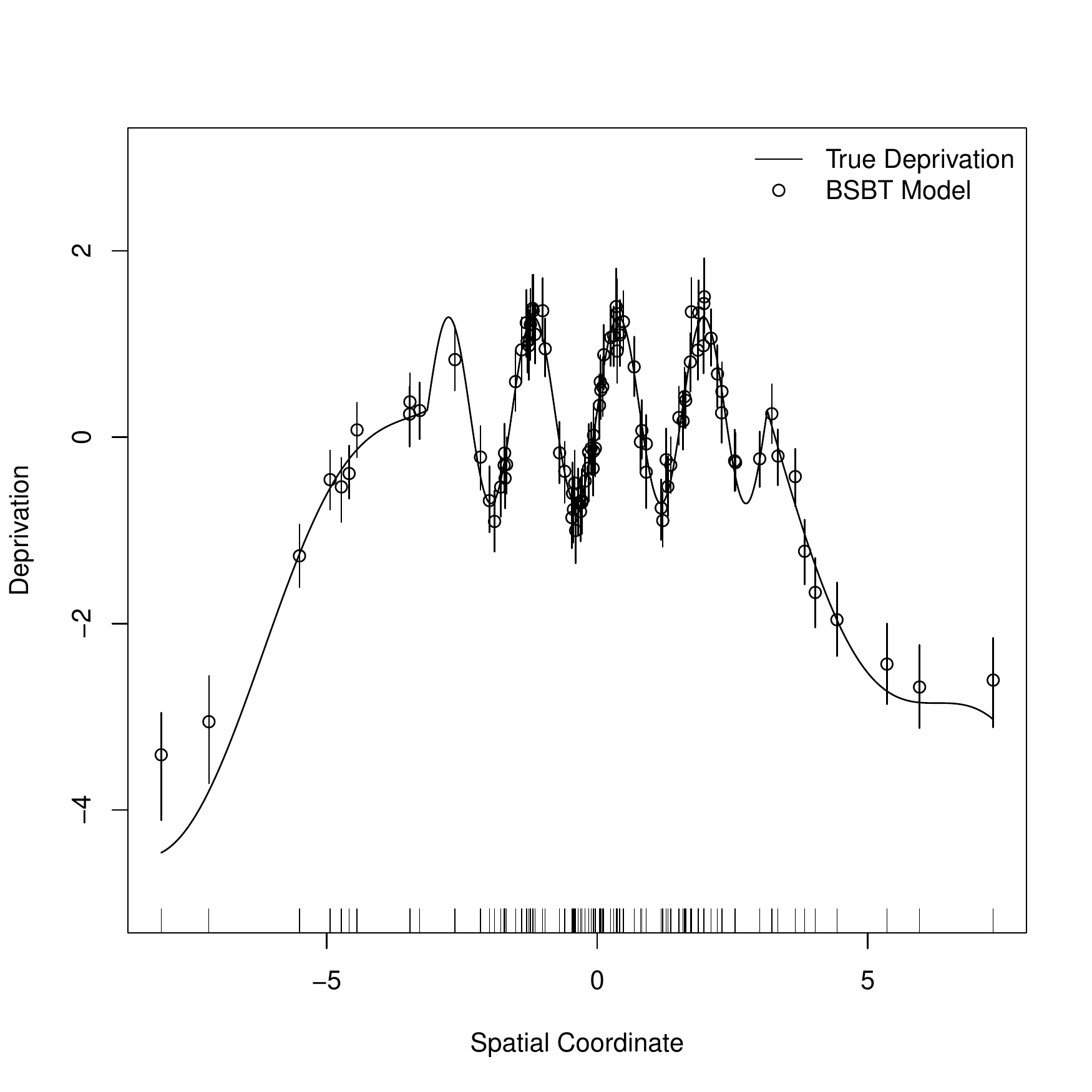} \\
\includegraphics[width = 0.49\textwidth]{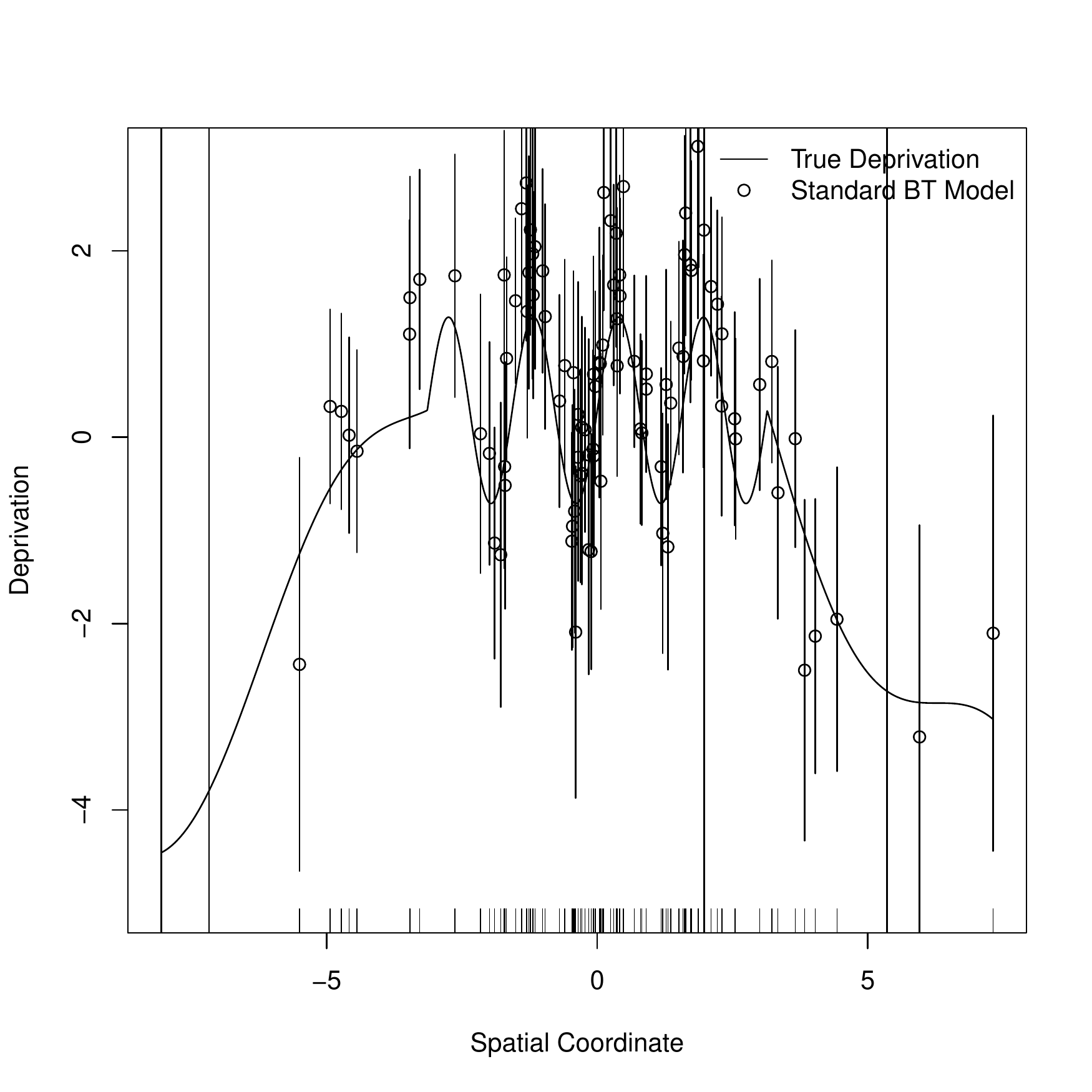}
\includegraphics[width = 0.49\textwidth]{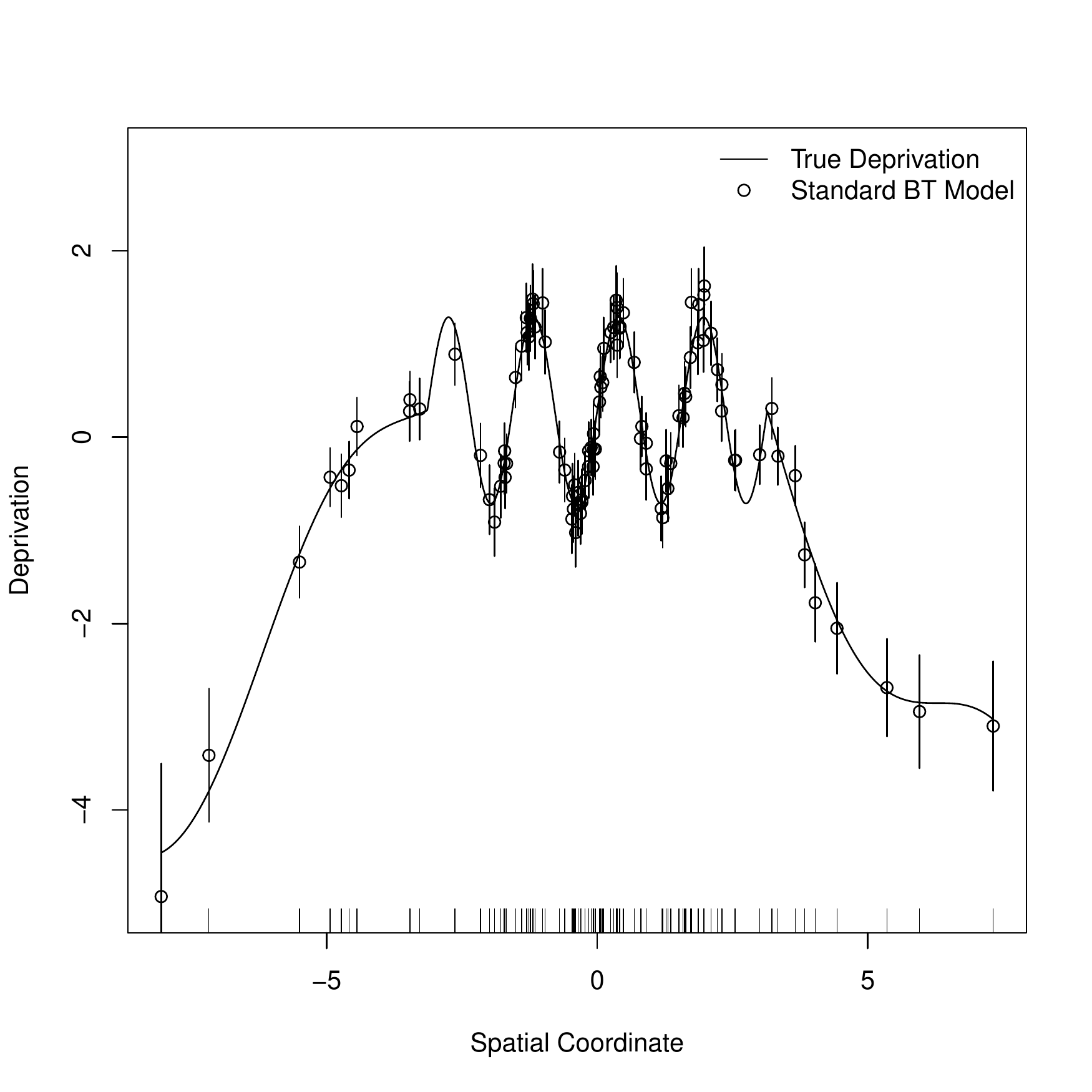}
\caption{Results for the 1-d simulation study using the BSBT and standard BT models. The top row shows the results of the BSBT model and the bottom row shows the results of the standard BT model. The left column shows the models fitted with 900 comparisons and the right column with 9,000 comparisons. The same scales are used on all plots for ease of comparison, but this means that in the bottom row some intervals are larger than can be displayed. Tick marks on the horizontal axes show the locations of the areas in the 1-d city.
} 
\label{fig: 1D study result}
\end{figure}

We fit the BSBT model and run the MCMC algorithm for 500,000 iterations, removing the first 100,000 as a burn-in period.  We estimate the quality of each area and learn plausible values of the variance hyperparameter $\alpha^2_\lambda$. As in the 2-d study, we fix the tuning parameter $\delta = 0.01$, based on initial runs of the algorithm, and fix $\chi = \omega = 0.1$ in the prior distribution on $\alpha^2_\lambda$. The top row of Figure \ref{fig: 1D study result} gives results for the BSBT model using 900 and 9,000 comparisons; it shows the true deprivation levels, the location of the areas and the posterior median deprivation with a 95\% credible interval for each area. We see two main effects from increasing the number of comparisons. The first is increasing accuracy of inference, with better estimates and less uncertainty when using 9,000 comparisons. The second is the model's ability to deal with extreme levels of deprivation, either very deprived or very affluent areas. The areas on the outskirts of the synthetic city make this a challenging data set for the BSBT model, since they are extreme both spatially and in terms of deprivation. When using smaller data sets, inferred deprivation levels in the BSBT model are pulled towards 0 by the prior distribution. As we do not have sufficient data to estimate the extent of the deprivation in the most extreme areas, we also underestimate the variance parameter. Although we do not accurately estimate the extent of the extremes, we do successfully identify which areas have extremes of deprivation. 

Corresponding results from fitting the standard BT model to the same data sets are given in the bottom row of Figure \ref{fig: 1D study result}. These plots show the same detail of the true deprivation and locations of areas, but here show MLEs and corresponding intervals based on quasi-variances. In the smallest data set there are some areas which feature in very few comparisons, so estimates for their level of deprivation are highly uncertain. This shows in the lower-left plot as intervals spanning all shown deprivation values and/or the point estimate not being visible on the scale used. Many estimates are also quite poor compared to the BSBT results.  Figure \ref{fig: 1D study result} also shows that when using the 9,000 comparisons the standard and BSBT models perform fairly similarly. This is expected since the data set is large enough to estimate the model parameters well using either model. 

\begin{figure}[ht]
    \centering
\includegraphics[width=0.49\linewidth]{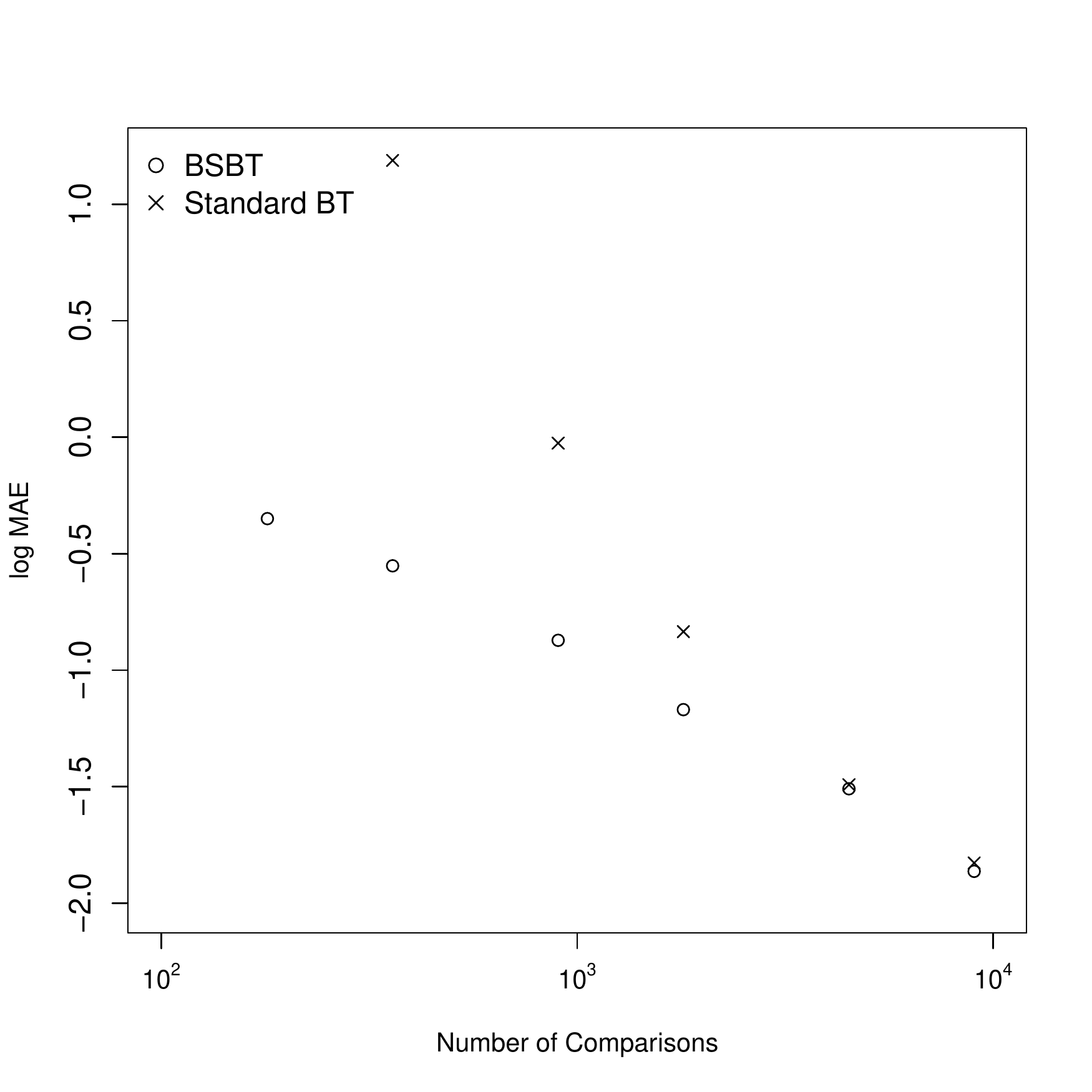} 
    \caption{Log MAE for the simulation study comparing performance of the standard BT and the BSBT models in terms of error as a function of the number of comparisons.}
    \label{fig: sim studies 1D}
\end{figure}

To assess the model fit, we compute the Mean Absolute Error (MAE) for the result of each set of comparisons, which is given by 
$$
\textrm{MAE} = \frac{1}{N}\sum_{i=1}^N |\lambda_i - \hat{\lambda}_i|,
$$
where $\hat{\lambda}_i$ is the estimate corresponding to the MLE or posterior median for area $i$. 
Figure \ref{fig: sim studies 1D} shows the log MAE for each data set. The BSBT model performs better than the standard BT model in all cases, though as discussed above we approach data saturation for the largest data sets. Figure \ref{fig: sim studies 1D} also  shows that when we fix the number of comparisons, the error in the BSBT model is always lower than in the standard model, and especially for smaller data sets offers a notable improvement. For example, when using 900 comparisons, MAE in the BSBT model (0.418) is less than half that in the standard model (0.975). The level of error in the standard model with 1,800 comparisons is of the same order as the BSBT model with 900 comparisons, demonstrating that with the BSBT model we can collect appreciably less data without compromising the quality of the estimates. In small data sets we cannot compute the MLE for areas which are not featured in any comparisons. As such, when using 180 comparisons we are unable to compute the MAE for the standard BT model. The BSBT model does not suffer from this issue since it uses a correlated prior distribution, so we still get an estimate of deprivation for these areas, albeit with large uncertainty. 

\section{BSBT diagnostics for the Dar es Salaam data set}
\begin{figure}
    \centering
    \includegraphics[width = 0.3 \textwidth]{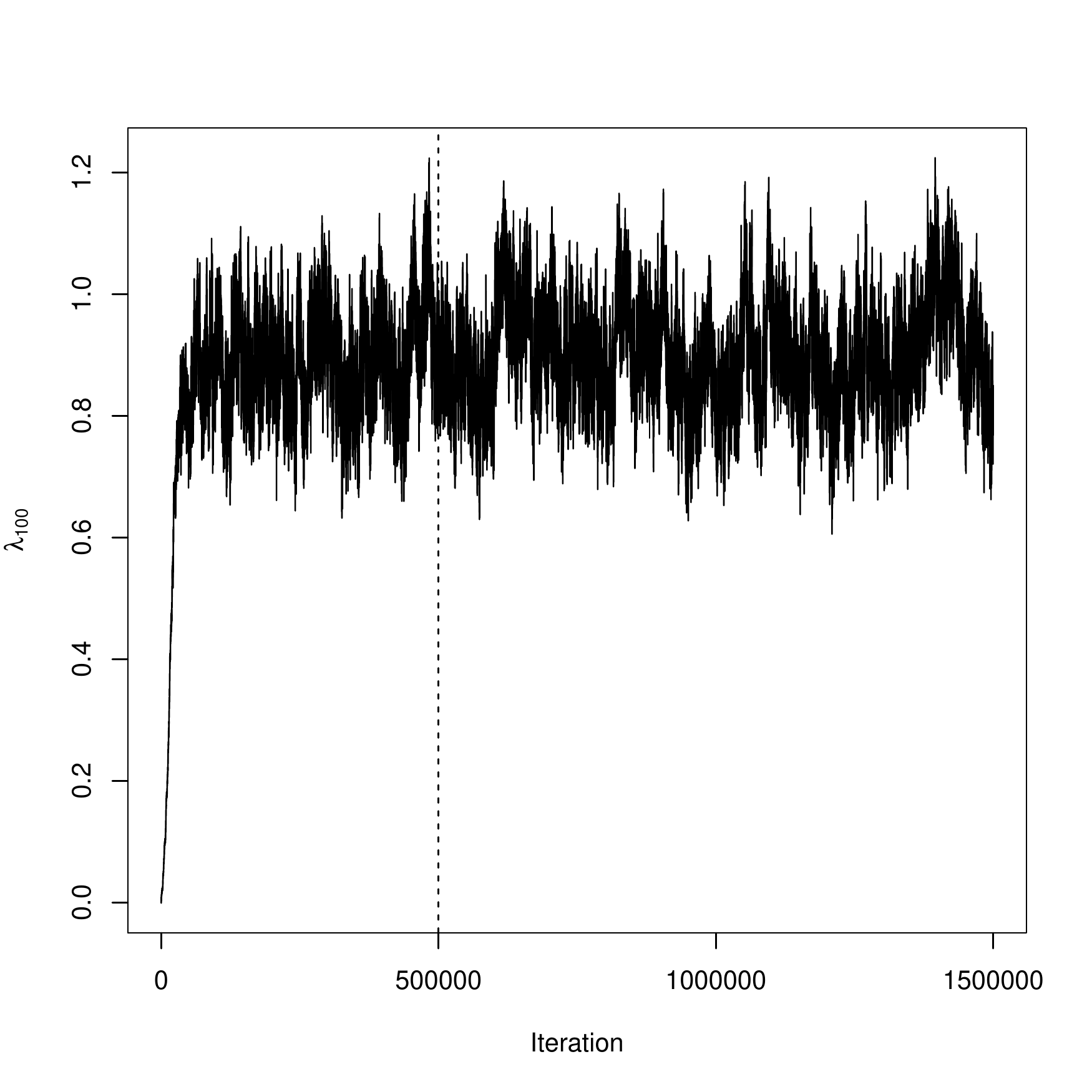}
    \includegraphics[width = 0.3 \textwidth]{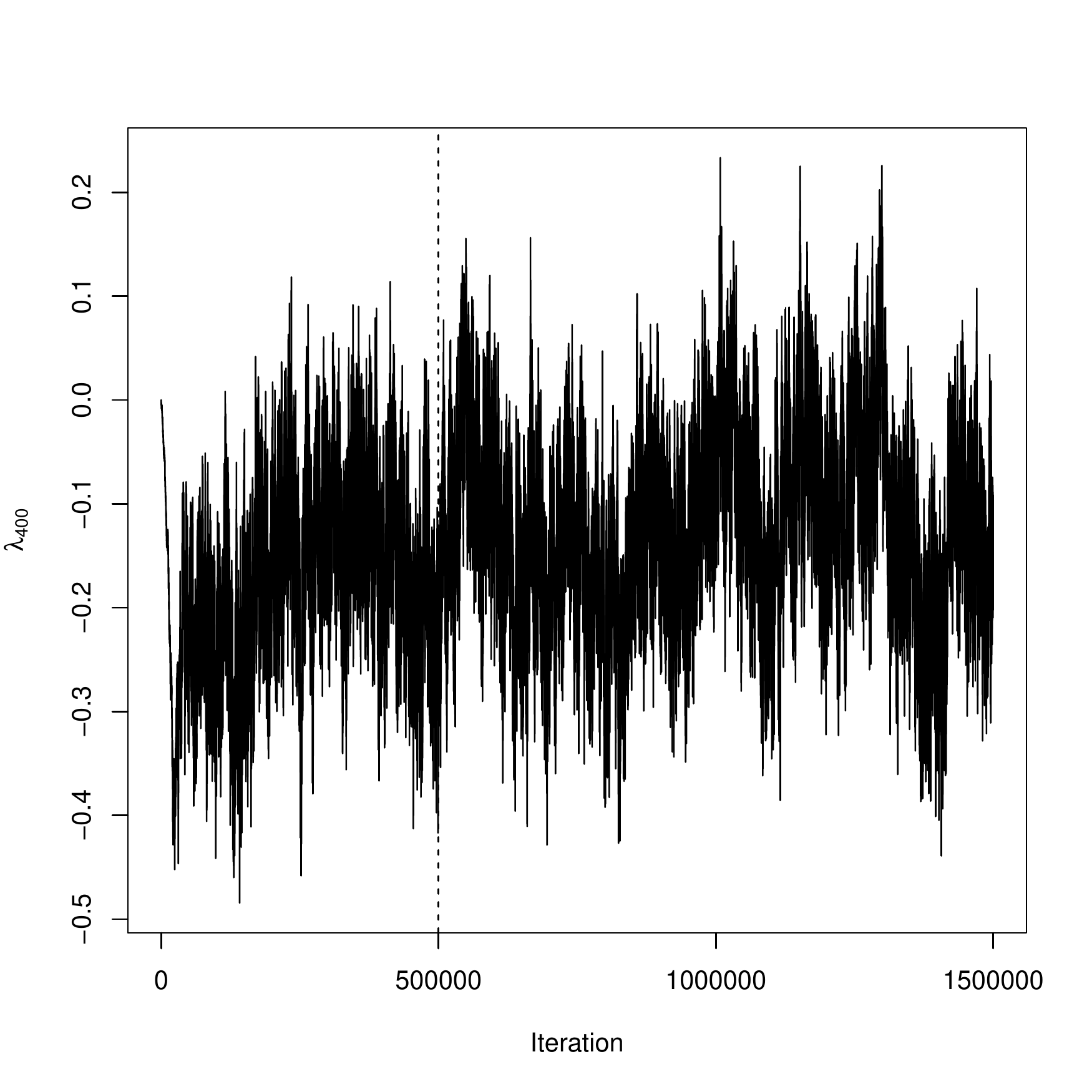}
    \includegraphics[width = 0.3 \textwidth]{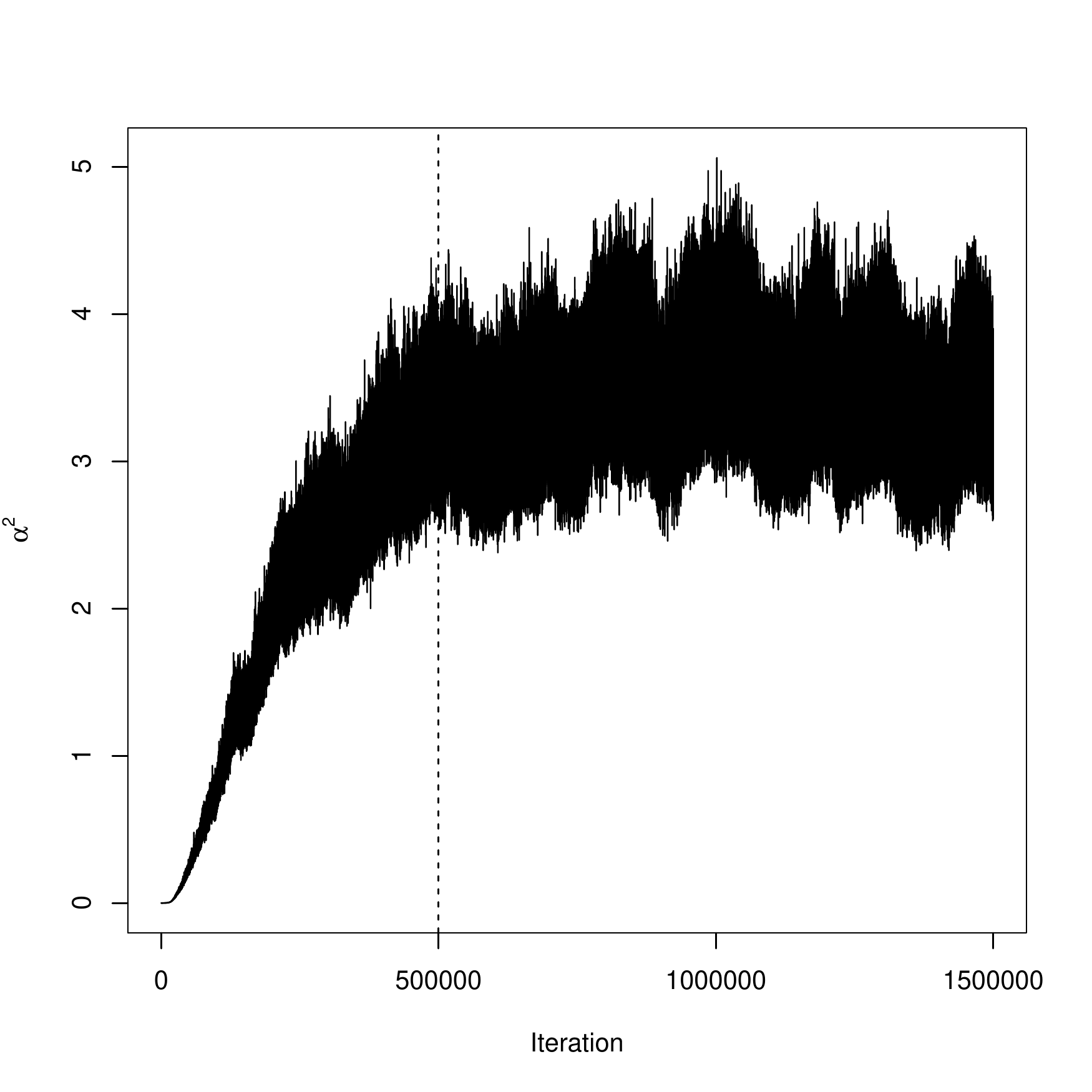}
    \caption{Trace plots for $\lambda_{100}$ (left), $\lambda_{400}$ (middle) and $\alpha^2_\lambda$ (right). The dashed line is at 500,000 iterations and marks the end of the burn-in period. }
    \label{fig: BSBT trace plots}
\end{figure}

We fit the BSBT model to the Dar es Salaam data set and produce the results shown in Section 4.1 of the main text. Trace plots for $\lambda_{100}$ and $\lambda_{400}$ and $\alpha^2_\lambda$ are shown in Figure \ref{fig: BSBT trace plots}. These show that the deprivation parameters converge quickly, but the variance hyperparameter is slower to converge. Based on the diagnostic plots, we consider the first 500,000 iterations as a burn-in period, and compute the posterior distributions for the model parameters from the remaining 1,000,0000 iterations. The trace plots show the Markov chain is mixing well. 

\begin{figure}
    \centering
    \includegraphics[width = 0.5\textwidth]{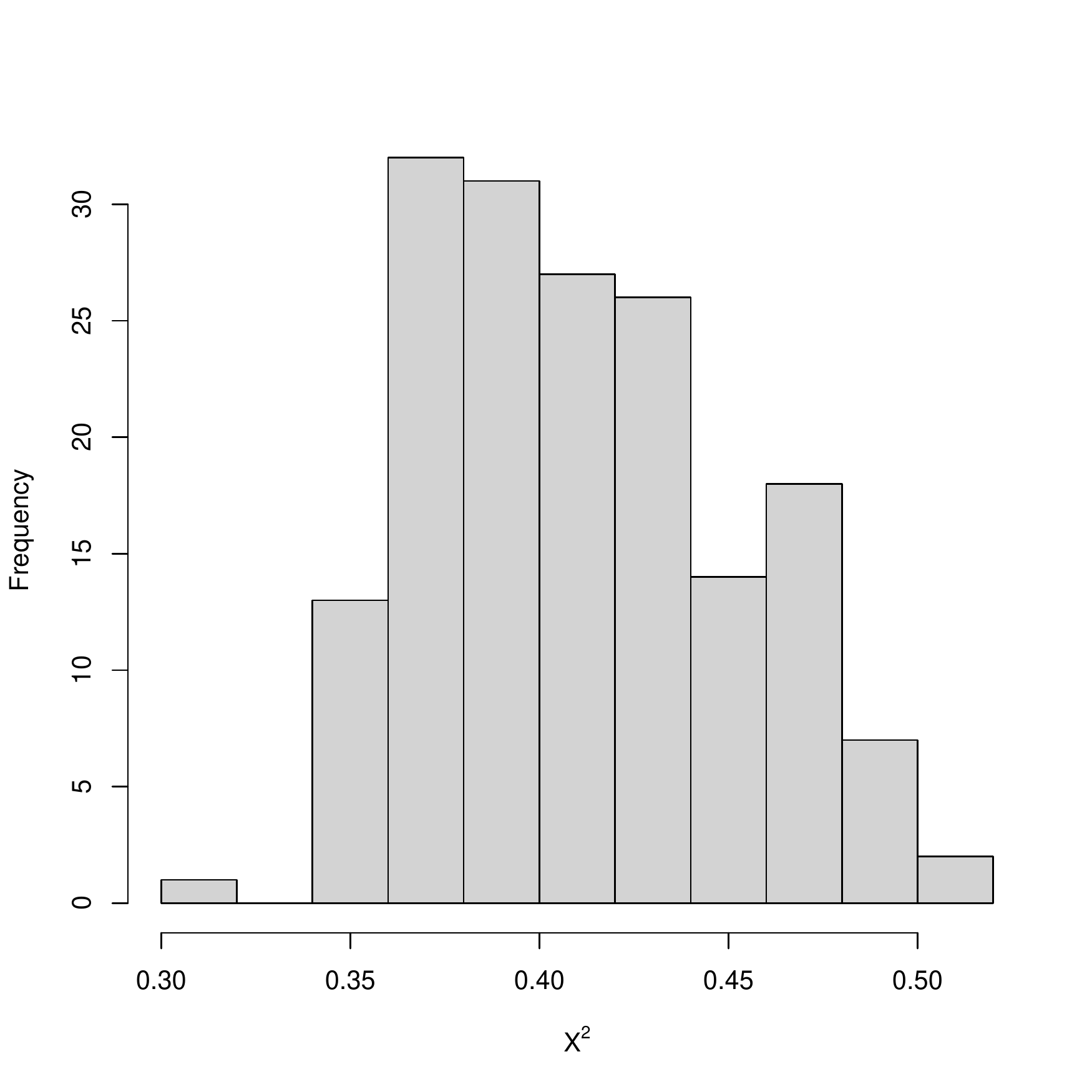}
    \caption{A histogram showing the judge reliability heuristic for each of the judges in the study.}
    \label{fig: judge chi sq}
\end{figure}

To determine the reliability of the judges and if they were carrying out the comparisons faithfully, we use a $\chi^2$ style heuristic. For each judge, we see how the observed comparisons differ from the expected comparisons based on the fitted model. The value of the heuristic for each judge ($j$) is given by 
$$
X^2_j = \frac{1}{N_j}\sum_{i = 1}^{N_j} \frac{(O_{j,i} - E_{j,i})^2}{E_{j,i}},
$$
where $N_j$ is the number of comparisons judge $j$ made, and $O_{j,i}$ and $E_{j,i}$ are the observed and expected outcomes of judge $j$'s $i$-th comparison. The latter is calculated using equation (1) of the main text and posterior mean estimates of $\boldsymbol{\lambda}$. The values are shown in Figure \ref{fig: judge chi sq}, which shows that the judges are largely homogeneous by this measure. We investigate the comparisons for the worst five judges (by this measure) and find that they are nonetheless largely consistent with the consensus.

\section{Treatment of ties for the Dar es Salaam data set}
Around one in seven of the comparisons are tied comparisons and their treatment affects the results. Throughout, we have considered a treatment where the winner of each tied comparison was randomly allocated. We carry out two sensitivity analyses to justify our choice of treatment. The first concerns the random random allocation of winners for the tied comparison and the second investigates other treatments of these ties. 

\subsection{Randomly allocating a winner}
\begin{figure}
    \centering
     \includegraphics[width = 0.45\textwidth]{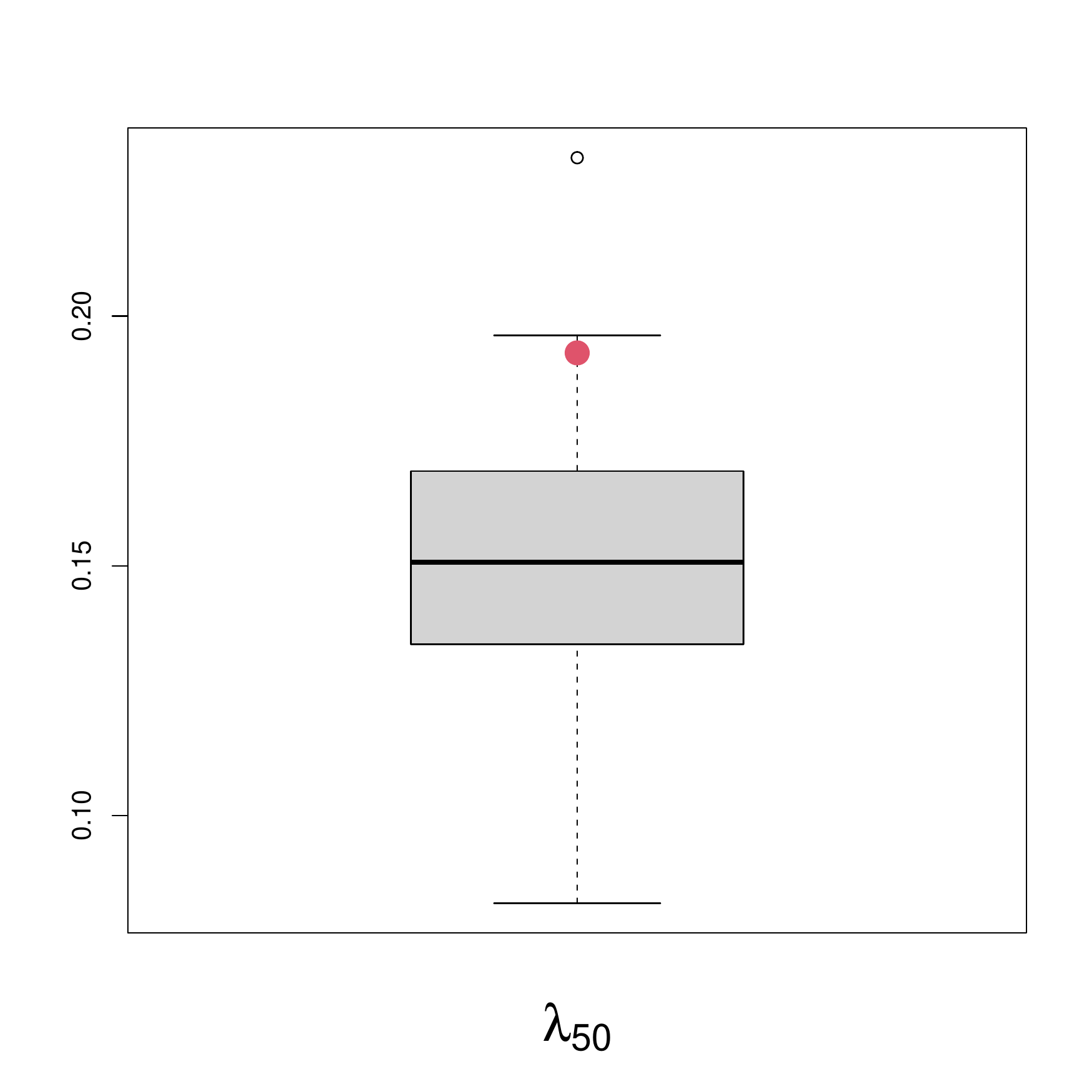}
      \includegraphics[width = 0.45\textwidth]{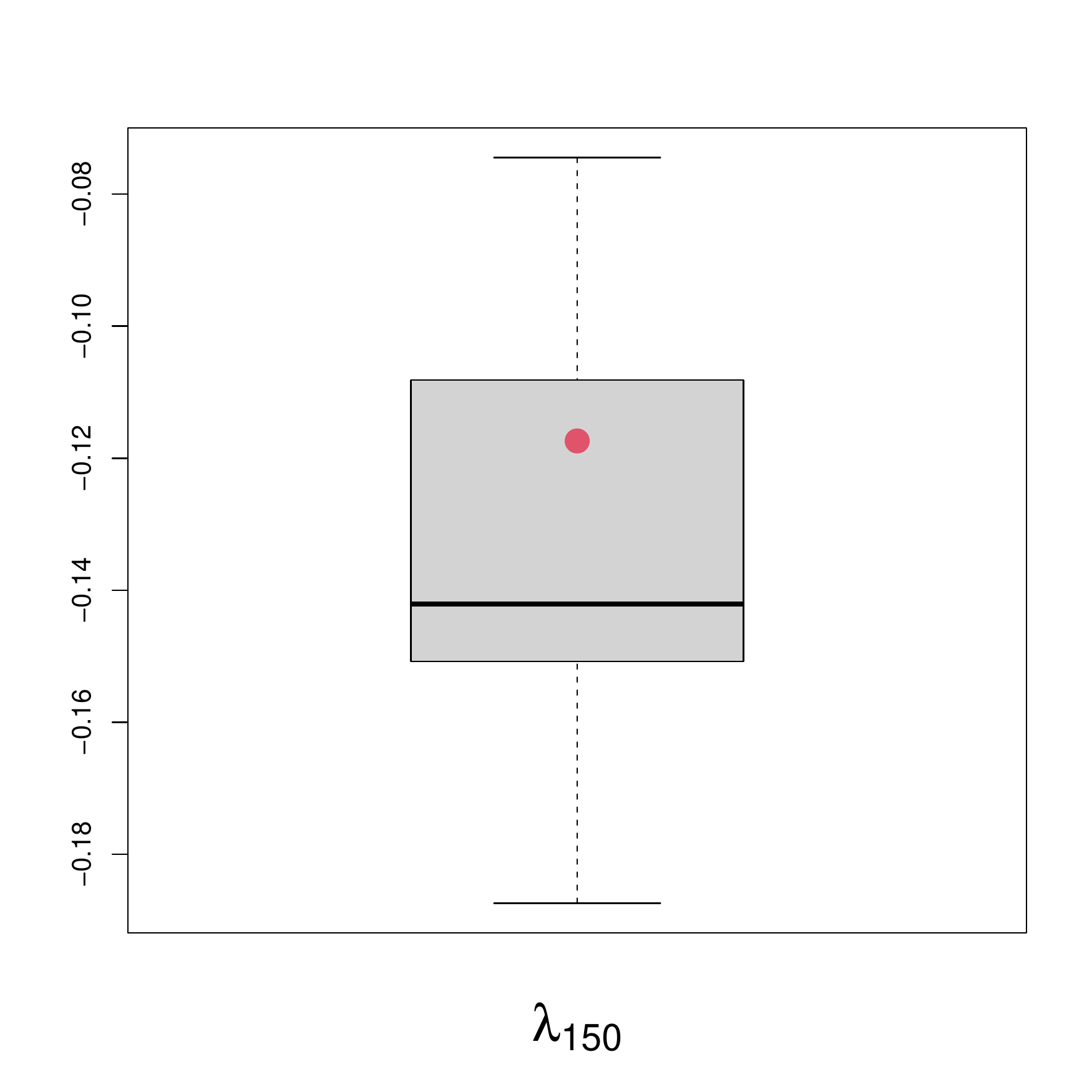}
     \includegraphics[width = 0.45\textwidth]{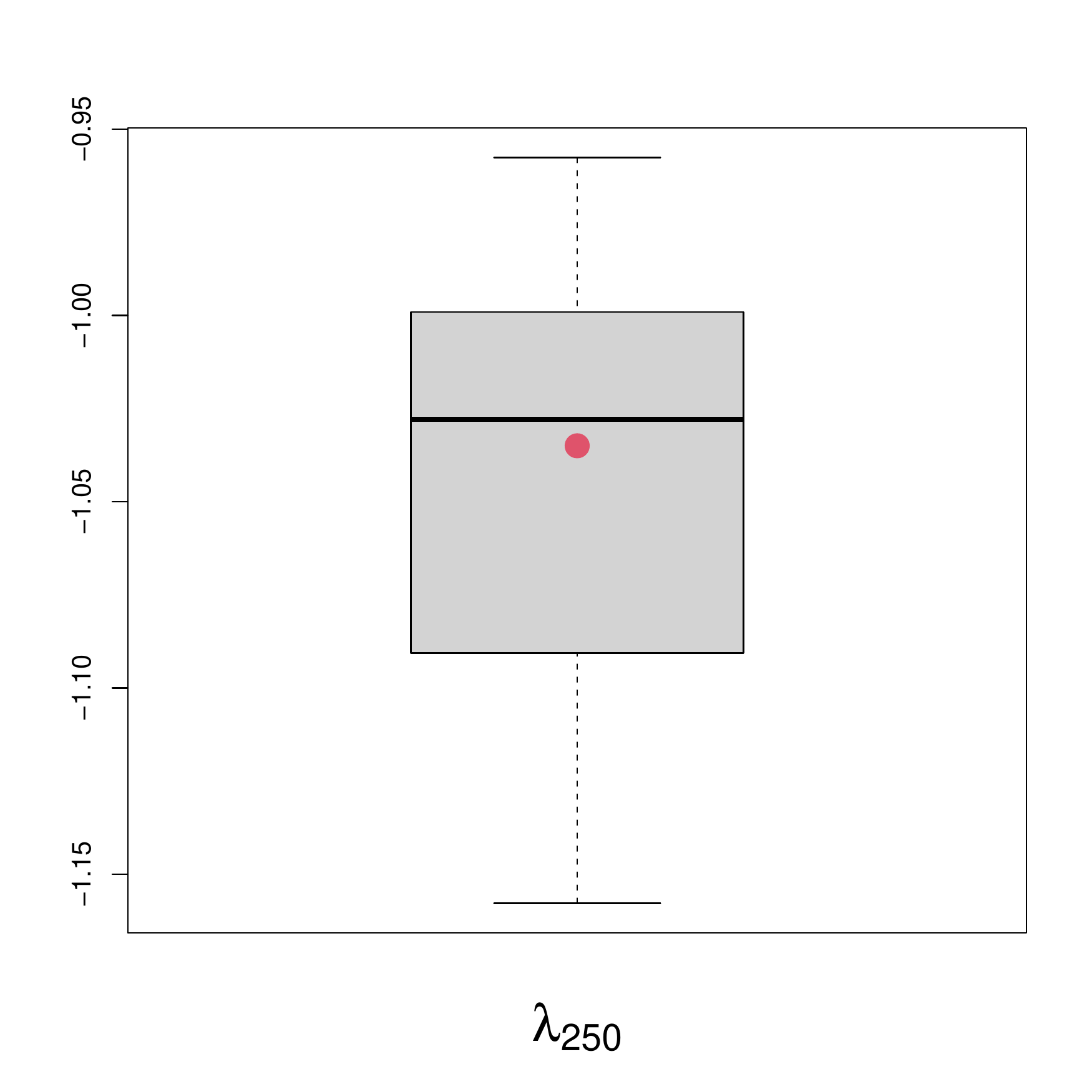}
      \includegraphics[width = 0.45\textwidth]{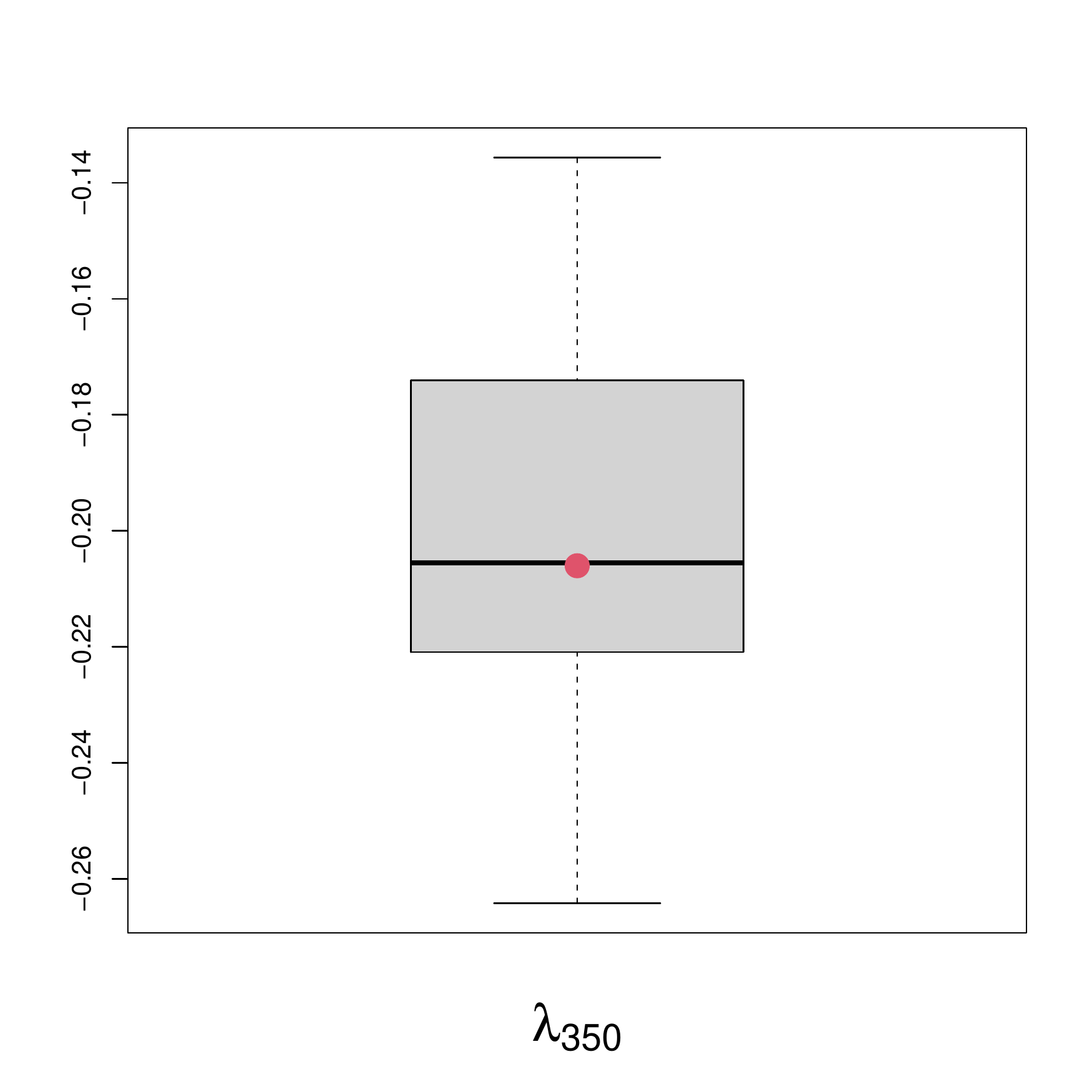}
    \caption{Box plots showing the posterior means for the 20 data sets using random allocations. The solid point shows the estimate reported in the main text. }
    \label{fig: box plots}
\end{figure}

We investigate the effect of the random allocation of winners of each tied comparison on the results. We generate 20 new data sets, in each allocating a winner of each tied comparisons according to a different random seed. We fit the standard BSBT model to each data set using the same fitting procedure used in the main text (1,500,000 iterations with the first 500,000 removed as a burn-in period). We compute the posterior mean deprivation levels for each data set and compare these estimates to those reported in the main text using Spearman's rank correlation coefficient. This is given by
$$
r_s = 1 - \frac{6\sum_{i=1}^N d_i^2}{N(N^2-1)},
$$
where $d_i$ is the difference in ranks of $\lambda_i$ stemming from the two data sets. Over all 20 data sets, the lowest correlation coefficient is 0.993 and the mean value of the correlation coefficients is 0.994, suggesting that the random allocation has little effect on the results. Figure \ref{fig: box plots} shows the values of the estimates for $\lambda_{50}$, $\lambda_{150}$, $\lambda_{250}$ and $\lambda_{350}$, and in all four cases the range of estimates is small and the value reported in the main text is a representative estimate. 

\subsection{Other treatments}
We consider two further treatments of the tied comparisons, which are:
\begin{itemize}
    \item treating ties as half a win for both subwards, as described in \cite{Glickman1999}, effectively replacing $y_{ij}$ in equation (2) of the main text by $y_{ij} + t_{ij}/2$, where $t_{ij}$ is the number of tied comparisons of subwards $i$ and $j$; and
    \item discarding the tied comparisons altogether.
\end{itemize}
We fit the BSBT model to the data using each of the treatments, running the MCMC algorithm for 1,500,000 iterations and removing the first 500,000 iterations as a burn-in period. This is the same model fitting procedure as used in the main text. 

\begin{figure}
    \centering
     \includegraphics[width = 0.45\textwidth]{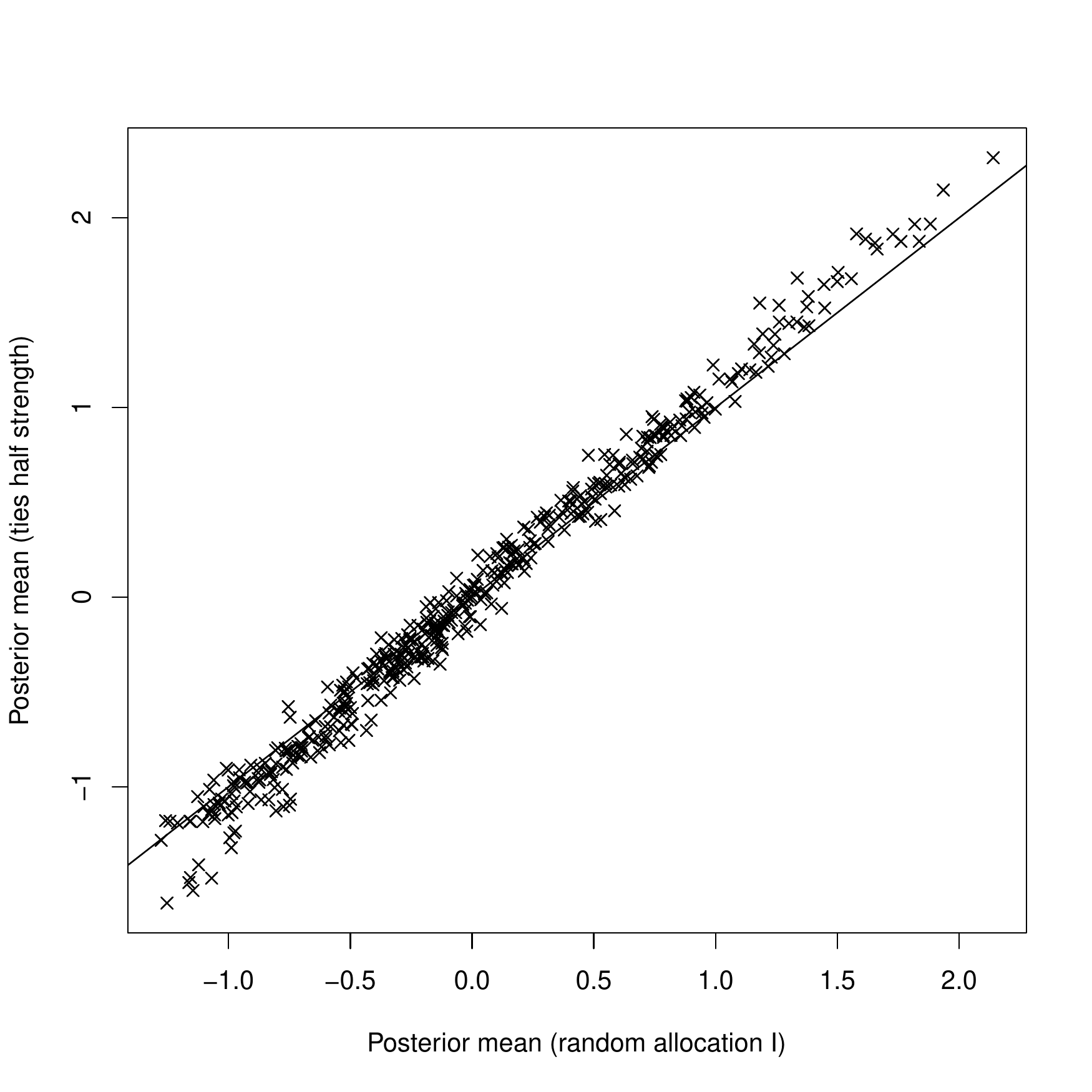}
      \includegraphics[width = 0.45\textwidth]{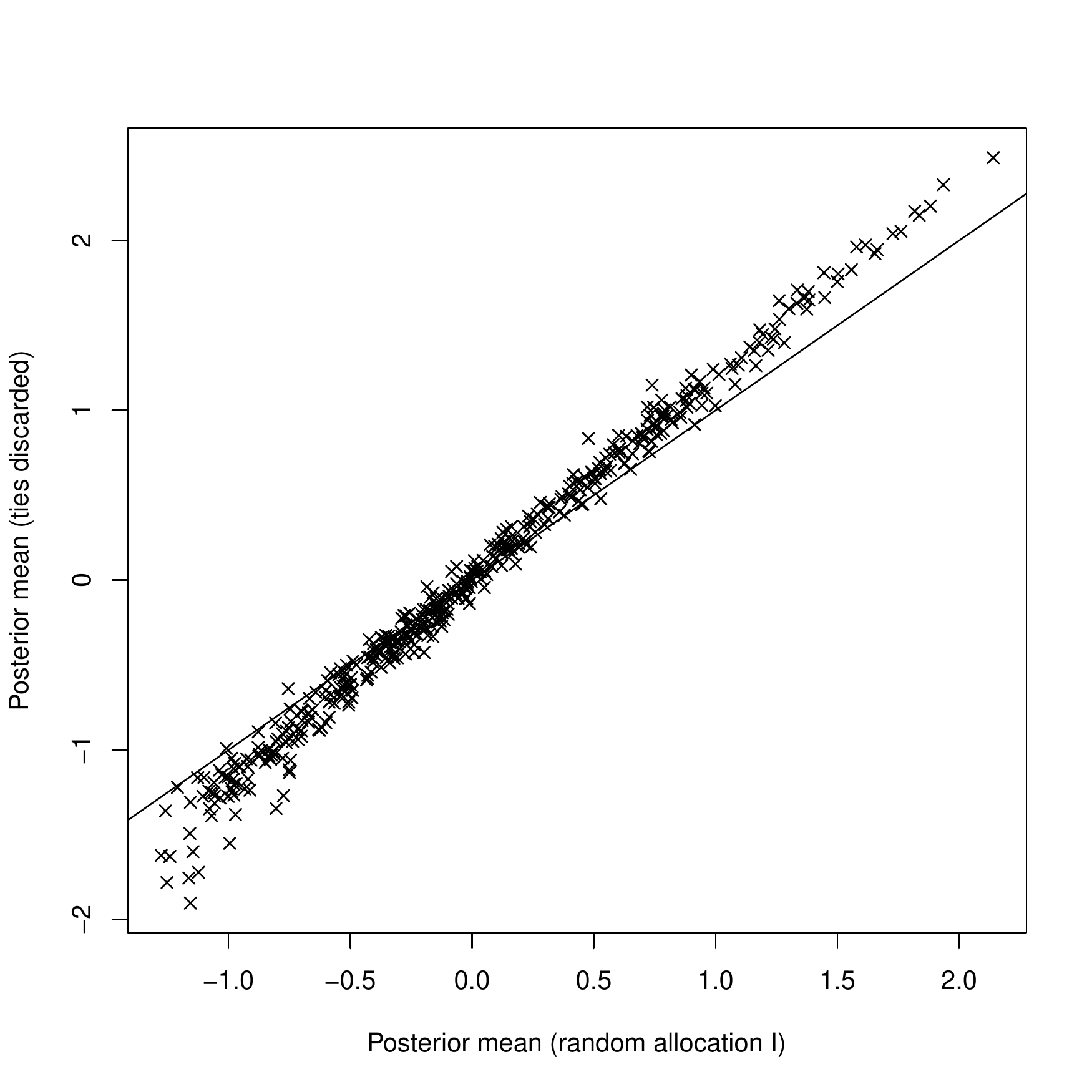}
     \includegraphics[width = 0.45\textwidth]{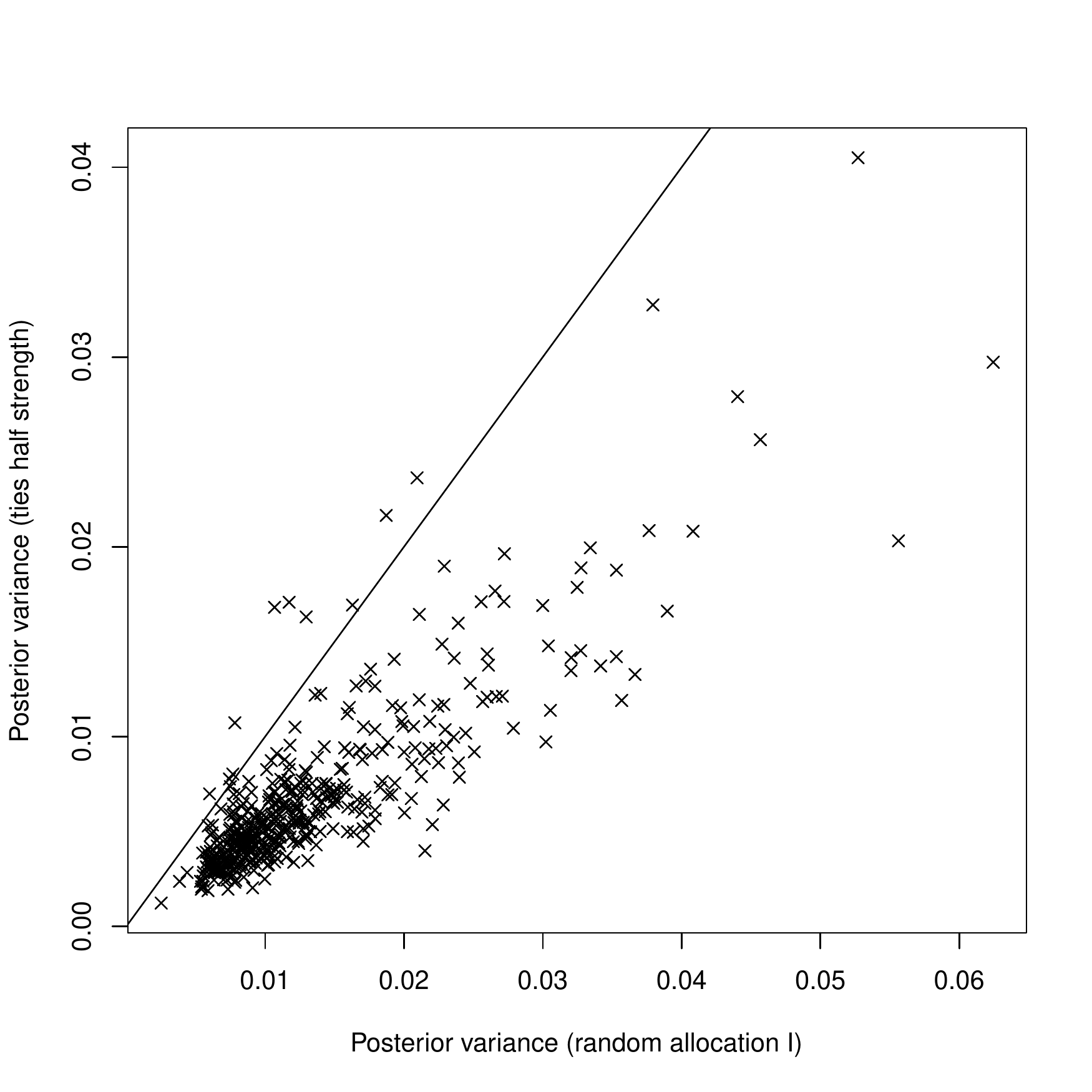}
      \includegraphics[width = 0.45\textwidth]{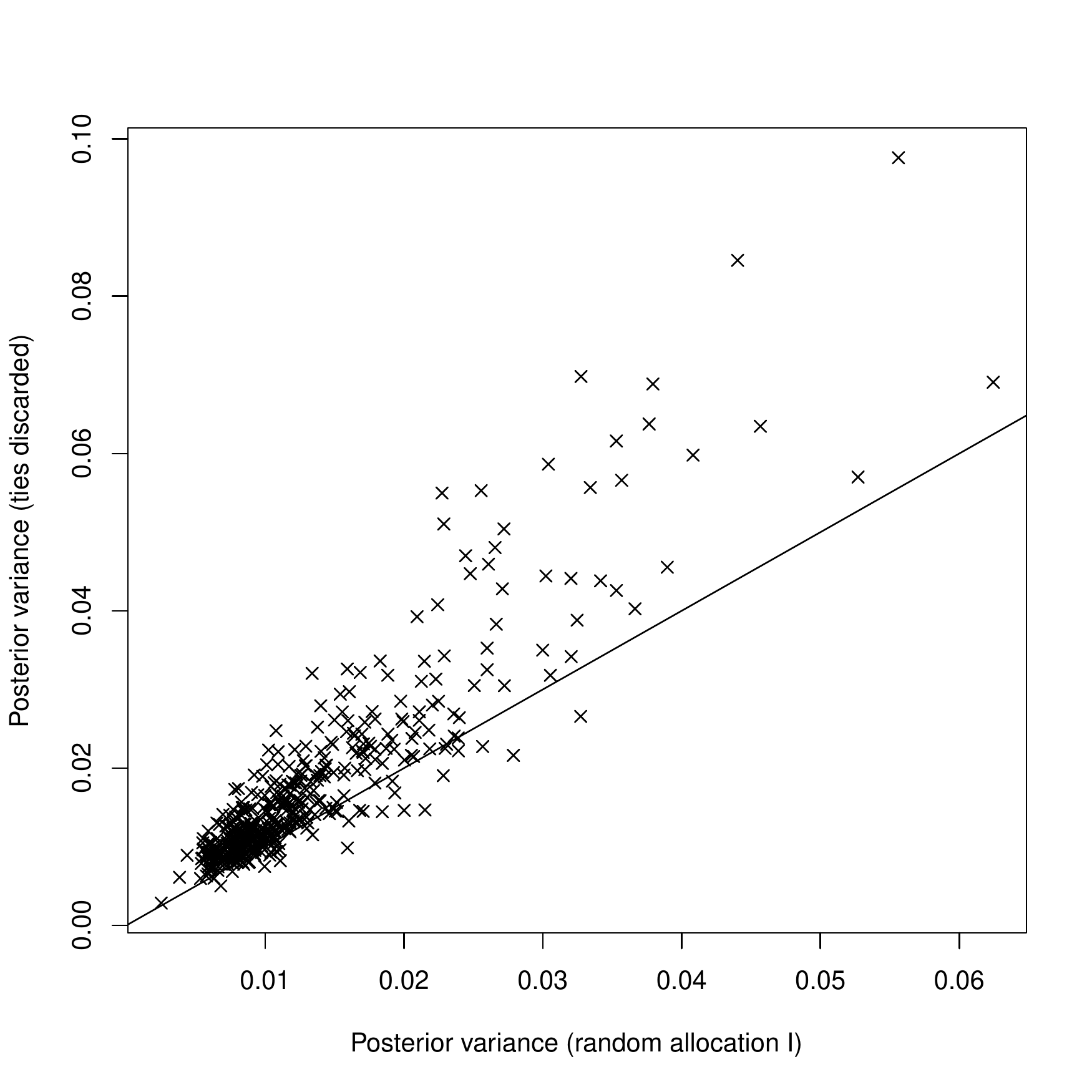}
    \caption{Posterior means and variances for the deprivation in each subward for the two treatments of ties, compared to the treatment used in the main text. }
    \label{fig: sensitivity}
\end{figure}

Figure \ref{fig: sensitivity} shows the posterior means (top row) and variances (bottom row) for the deprivation in each subward using the two treatments, compared to the results reported in the main text using a random allocation. When using either treatment, the rankings of the subwards are largely unchanged; the Spearman's rank correlation coefficient between randomly allocating a winner and treating ties as half a win is 0.993, and when discarding the tied comparisons is 0.995. Treating a tied comparison as half a win produces deprivation estimates that are broadly similar to allocating a winner at random, but suggests there is some slight shrinkage when using a random allocation. The variances using this treatment, however, are considerably smaller than using a random allocation. Finally, discarding the tied comparisons is the least attractive of the three treatments, but provides an upper bound for the level of uncertainty. Discarding the tied comparisons leads to the estimates of the deprivation levels in the most affluent and deprived subwards to be more extreme, as we are discarding data which will tend to make estimated affluences more similar to each other.

Overall we conclude that our results, using random allocation of a winner to break the ties, are robust to the random allocation used and are most conservative in terms of uncertainty of estimates amongst methods which use the ties in some way.

\section{The standard BT model for the Dar es Salaam data set}
\begin{figure}
    \centering
    \includegraphics[width = 0.45\textwidth]{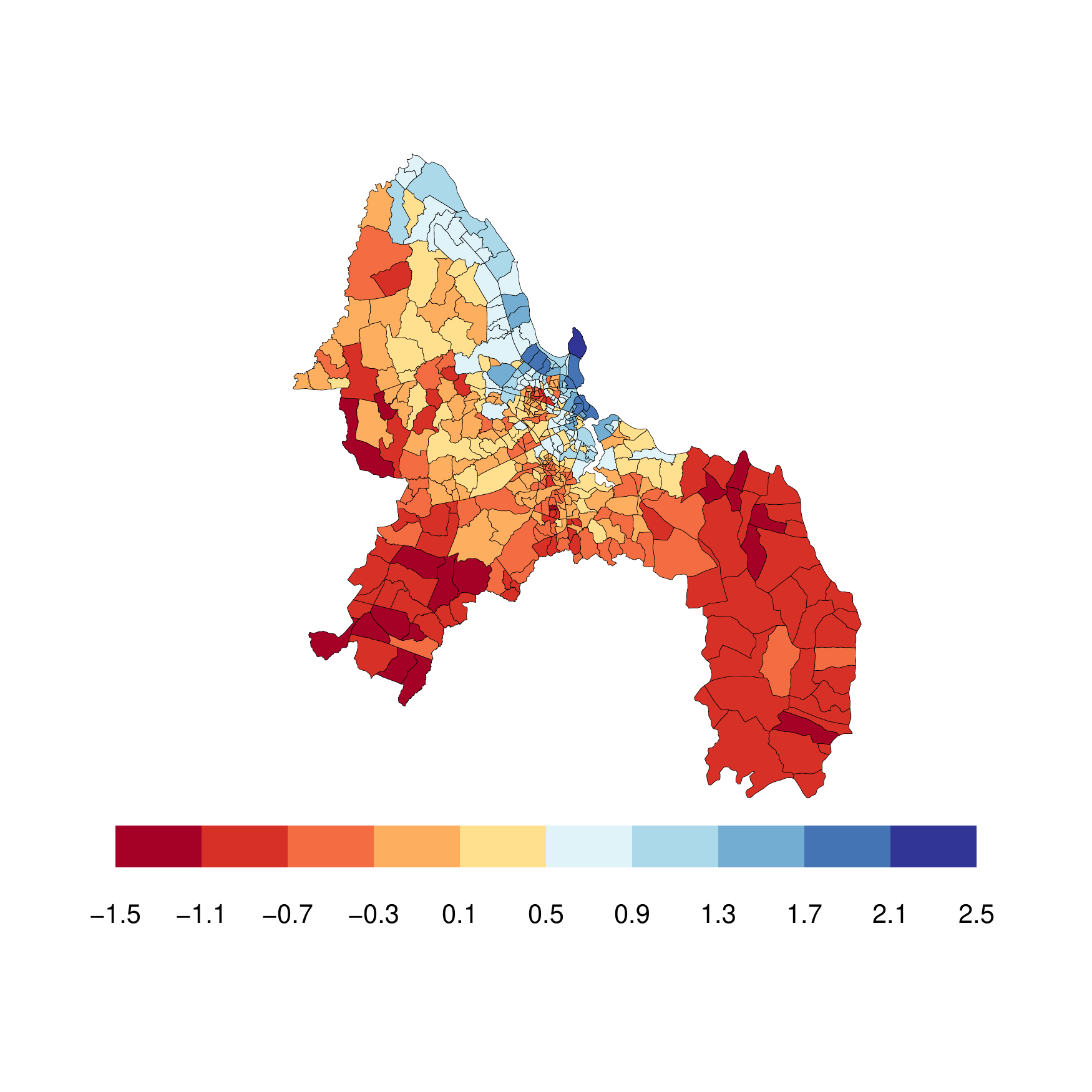}
    \caption{The results of the standard BT model applied to the full Dar es Salaam data set. }
    \label{fig: dar standard model}
\end{figure}

\begin{figure}
    \centering
    \includegraphics[width = 0.45\textwidth]{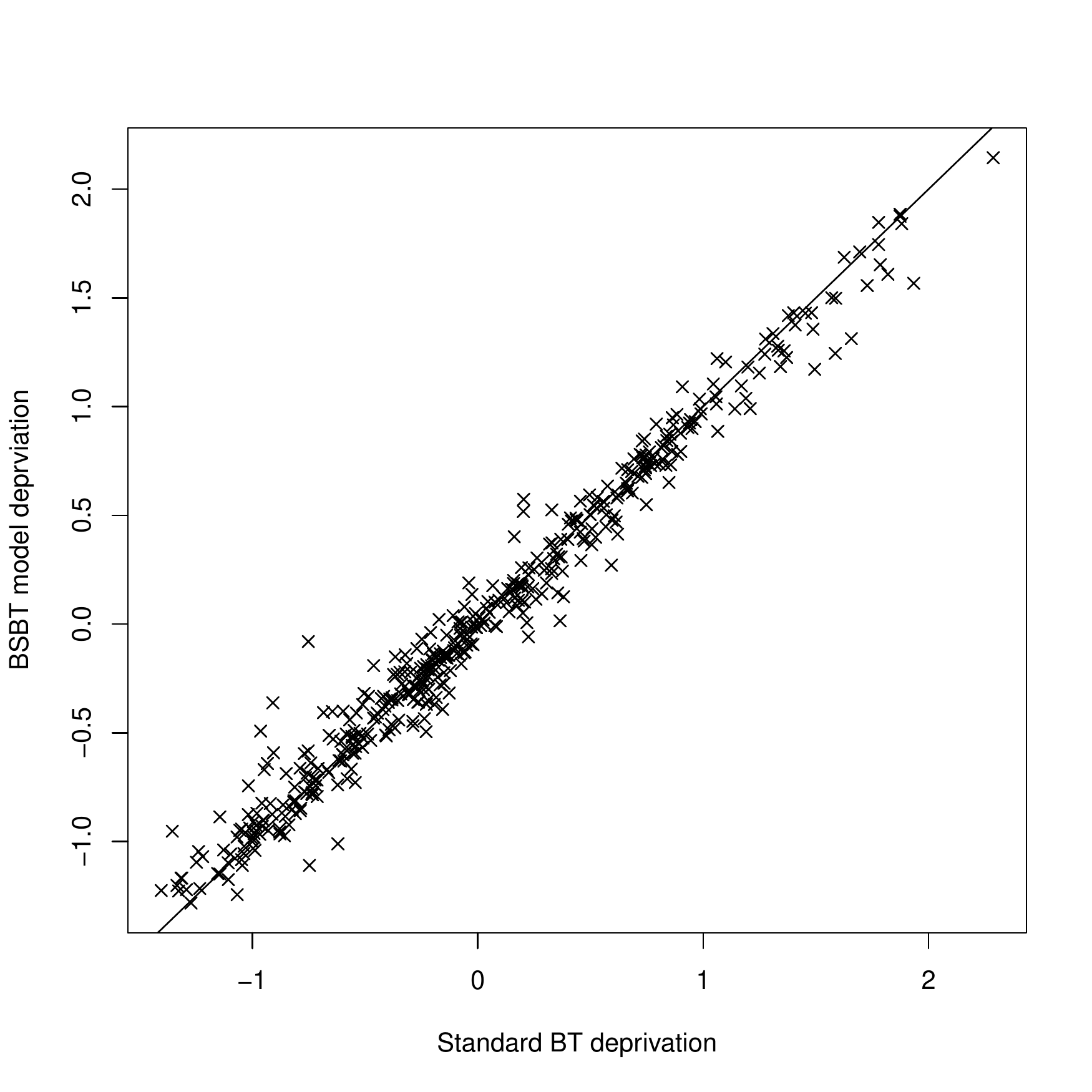}
    \includegraphics[width = 0.45\textwidth]{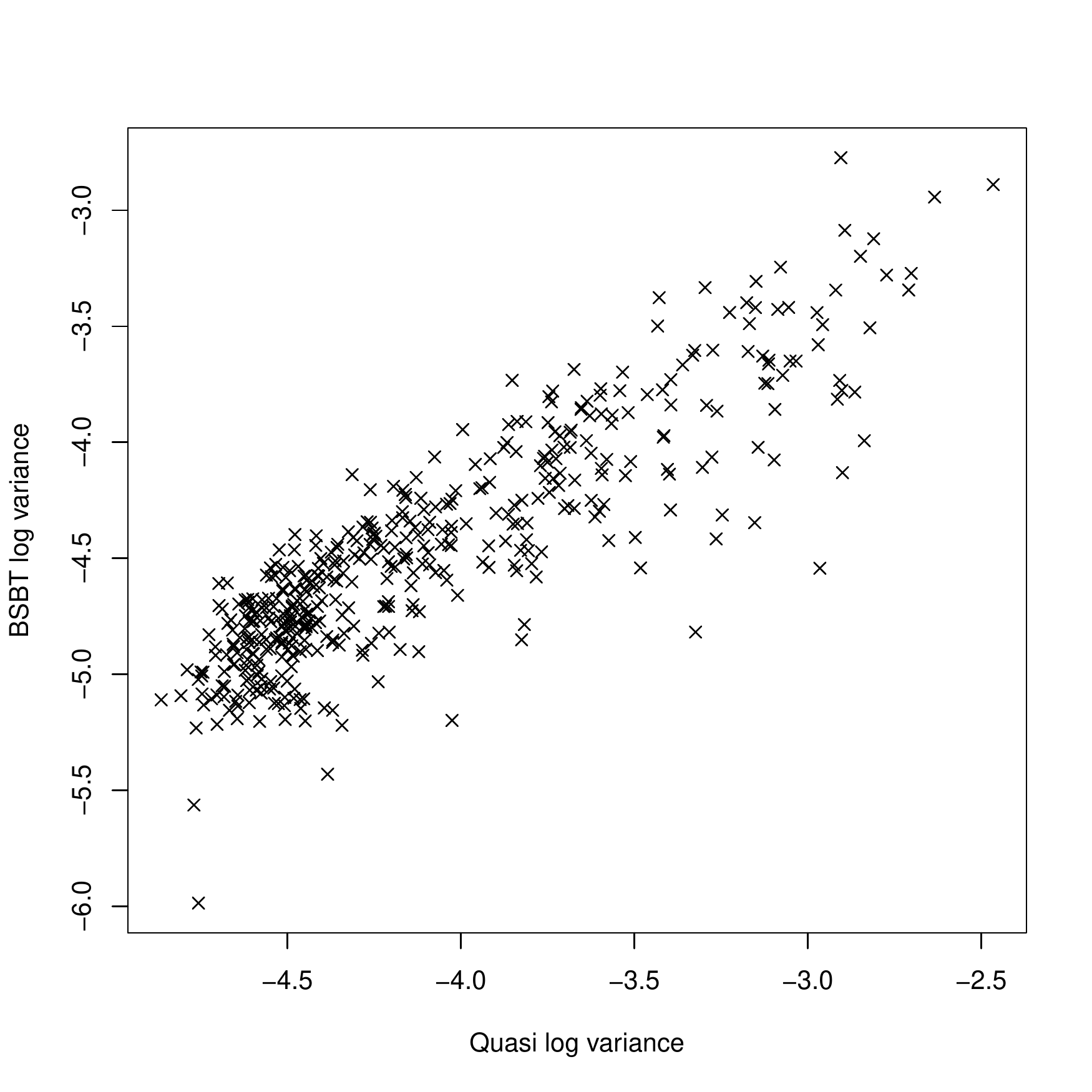}
    \caption{Left: The BSBT estimates plotted against the standard BT estimates. Right: The quasi variances for the BT model plotted against the variances of the posterior distributions for the BSBT model. Both are plotted on a log scale.}
    \label{fig: dar comparisons}
\end{figure}

We fit the standard BT model to the Dar es Salaam data set. To fit the model we use the \texttt{BradleyTerry2} R package. This took around 5 minutes on a 2019 iMac with a 3 GHz CPU. A map of the inferred deprivation parameters from the standard BT model are shown in Figure \ref{fig: dar standard model}. This is broadly very similar to the map shown in the main text (Figure 5) of the results for the BSBT model, though the smoothing effect of the prior is apparent in a few places (mainly near the most affluent and most deprived areas). We directly compare the results of the two models in Figure \ref{fig: dar comparisons}. There is little difference between the results from the two models. As the uncertainty in the two models is measured in different ways, they are difficult to compare directly, however the uncertainties are of the same order and there is a clear pattern of subwards with relatively high/low uncertainty under one model also having relatively high/low uncertainty under the other model.

\section{BSBT diagnostics for the Dar es Salaam data set with judge information}
\begin{figure}
    \centering
        \includegraphics[width = 0.3 \textwidth]{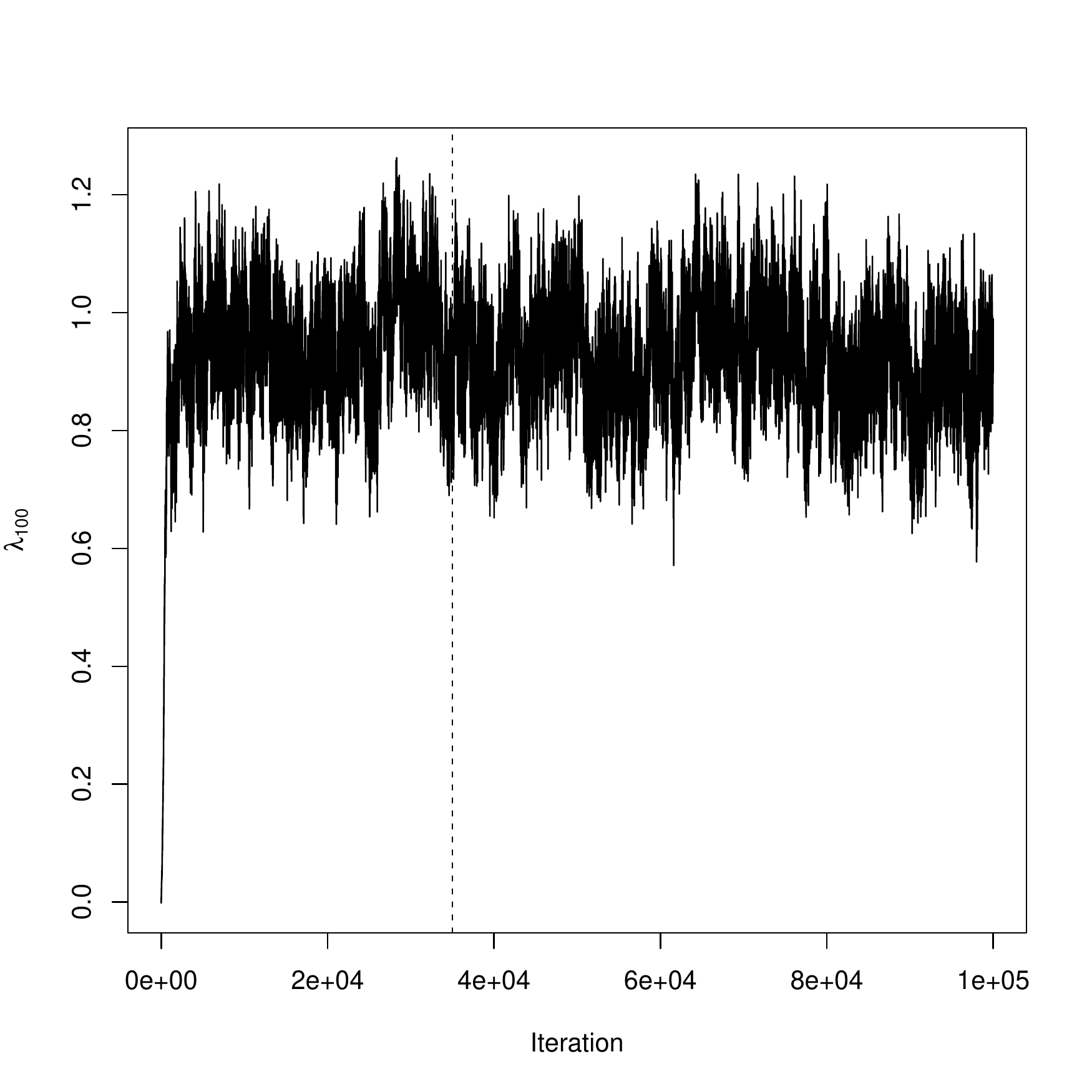} 
    \includegraphics[width = 0.3 \textwidth]{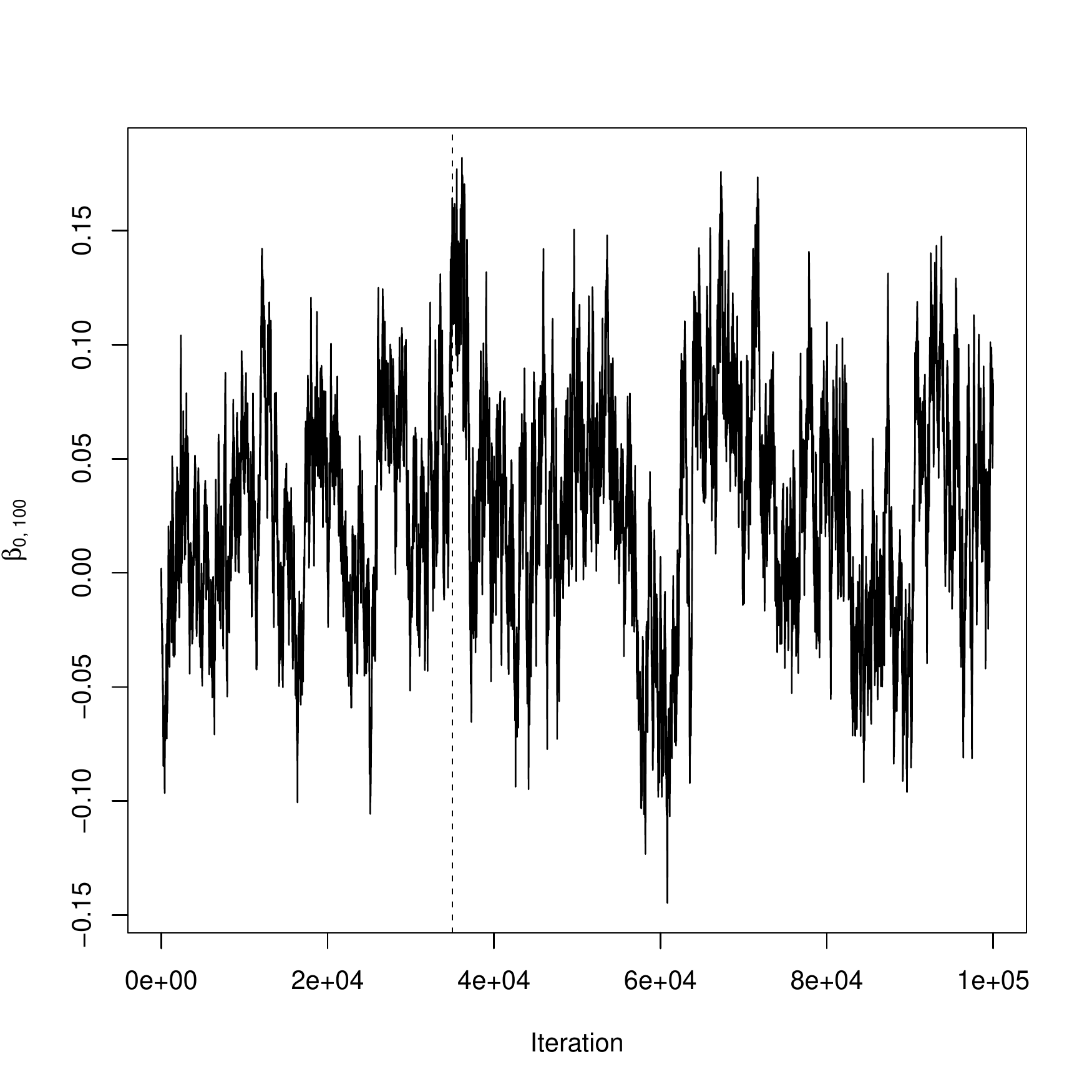}\\
    \includegraphics[width = 0.3 \textwidth]{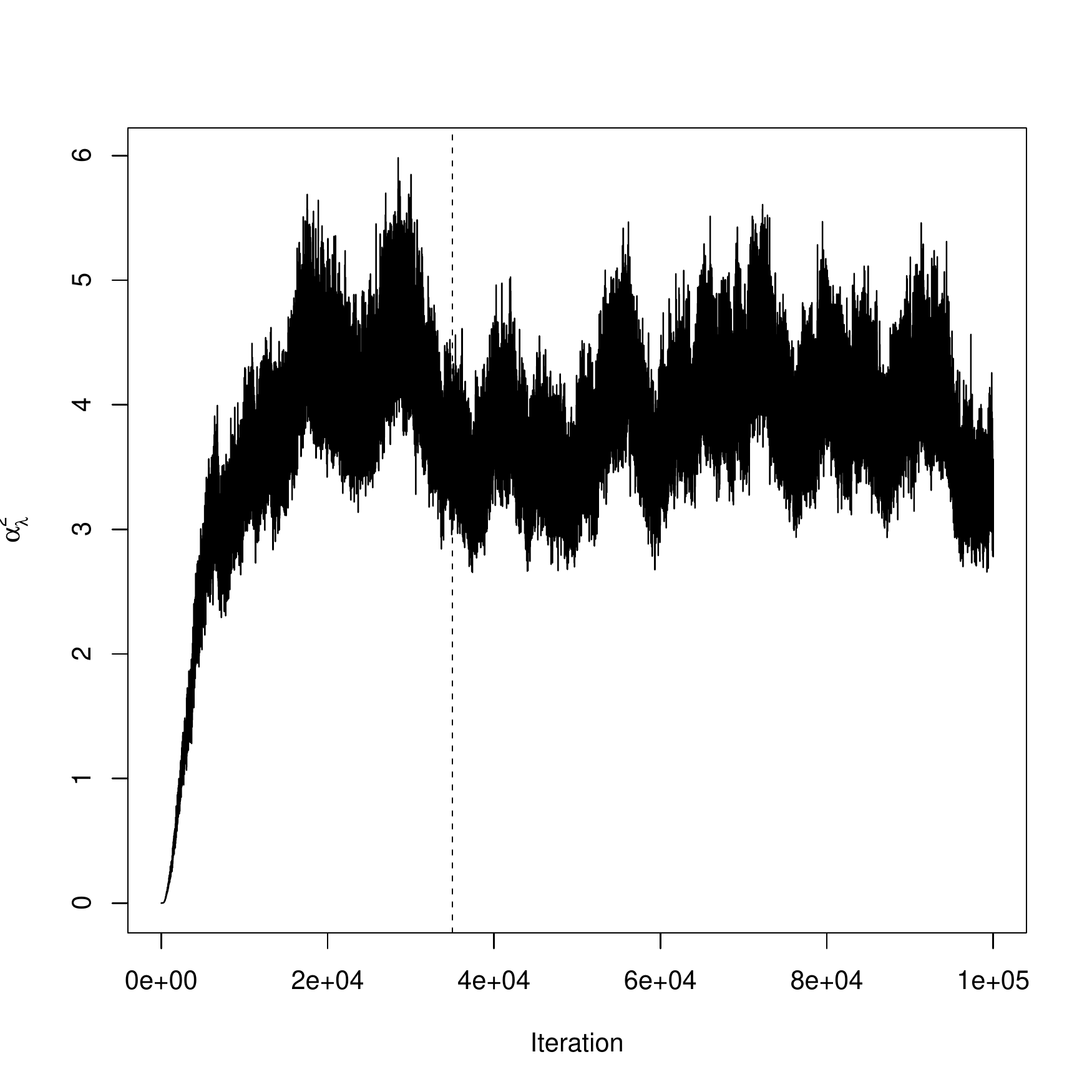}
    \includegraphics[width = 0.3 \textwidth]{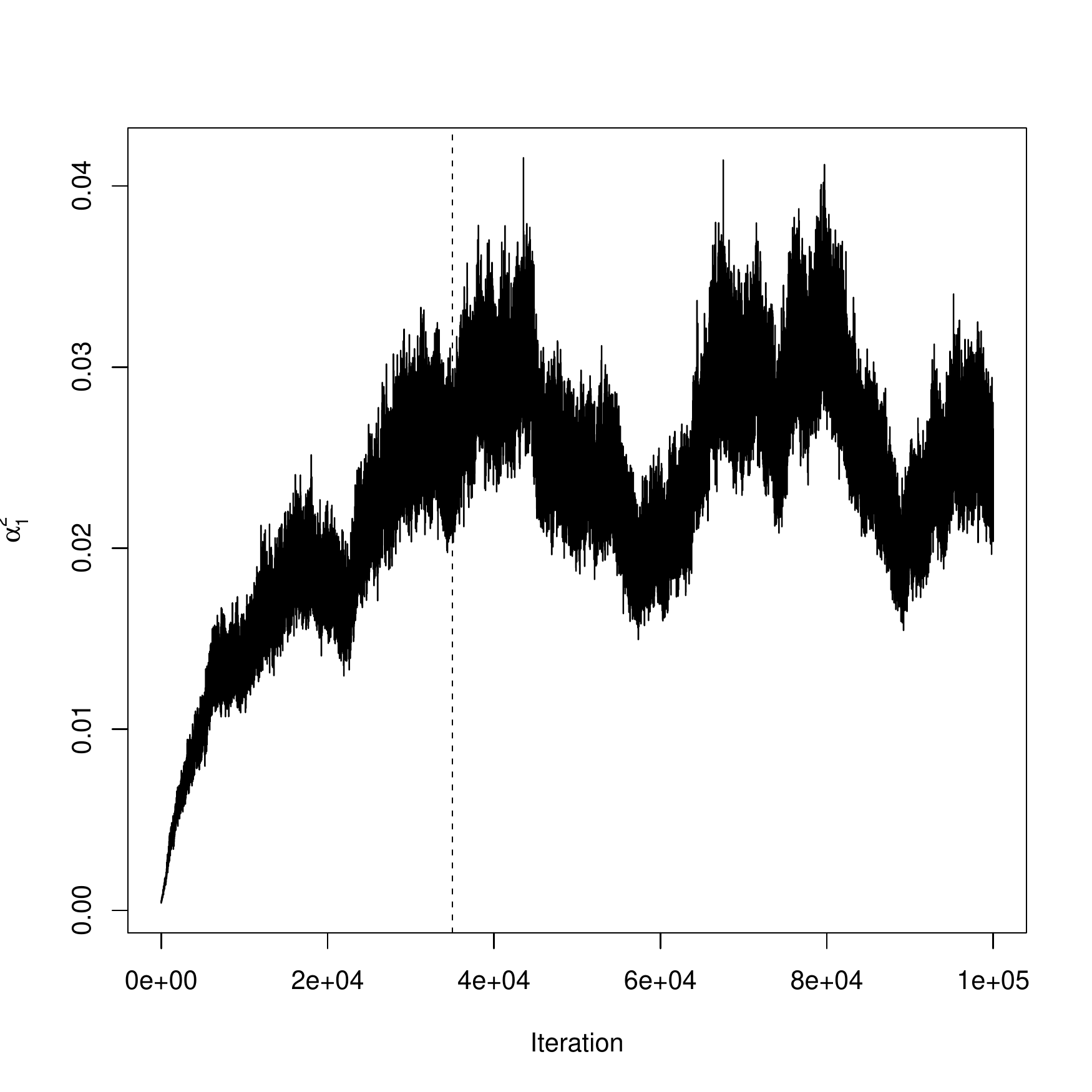}
    \caption{Men and women study. Trace plots for $\lambda_{100}$ (top left), $\beta_{0, 100}$ (top right), $\alpha^2_\lambda$ (bottom left) and $\alpha^2_1$ (bottom right). The 5 million iterations have been thinned to 100,000 and the dashed line at 35,000 iterations marks the end of the burn-in period.}
    \label{fig: BSBT gender trace plots}
\end{figure}

\begin{figure}
    \centering
        \includegraphics[width = 0.3 \textwidth]{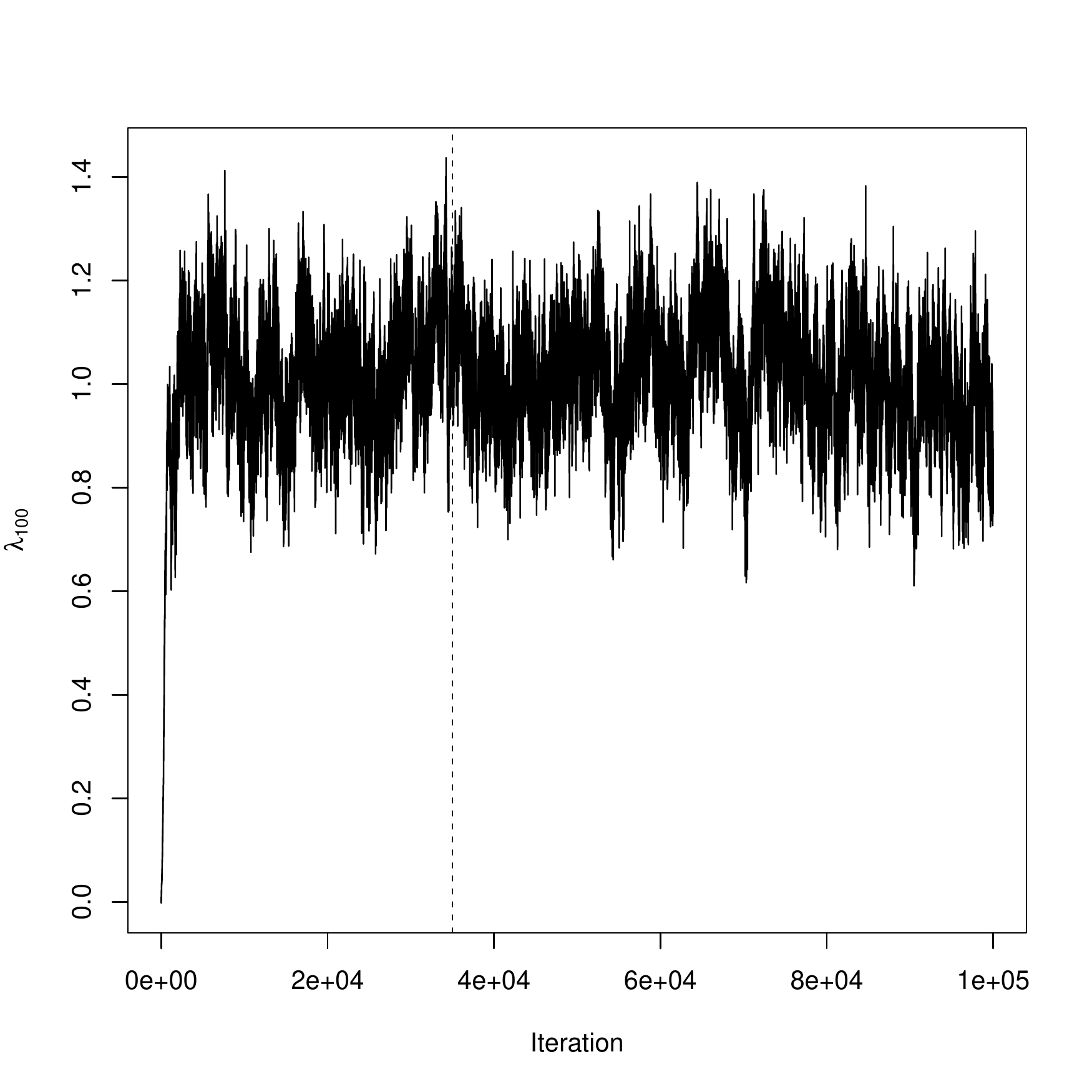} 
    \includegraphics[width = 0.3 \textwidth]{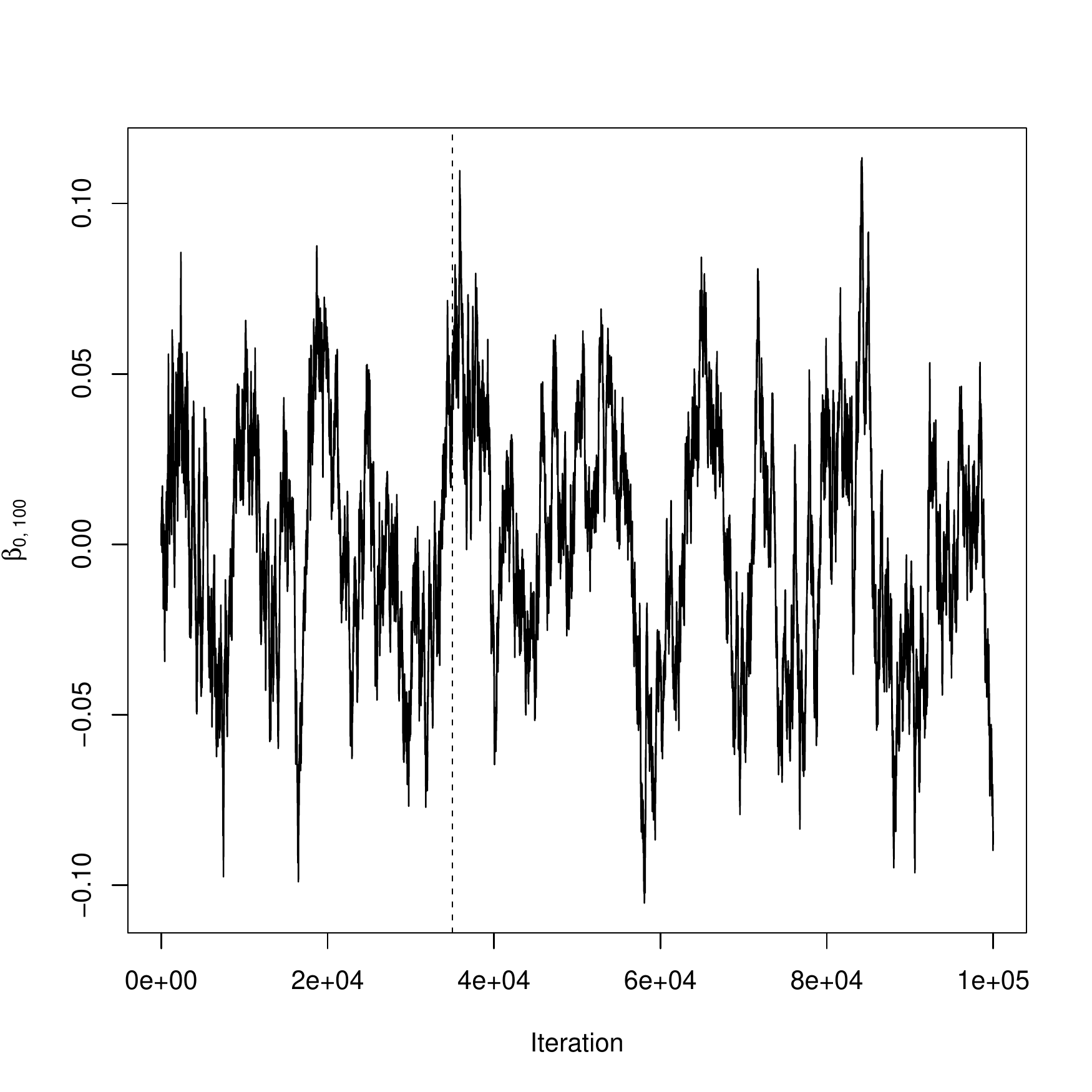}\\
    \includegraphics[width = 0.3 \textwidth]{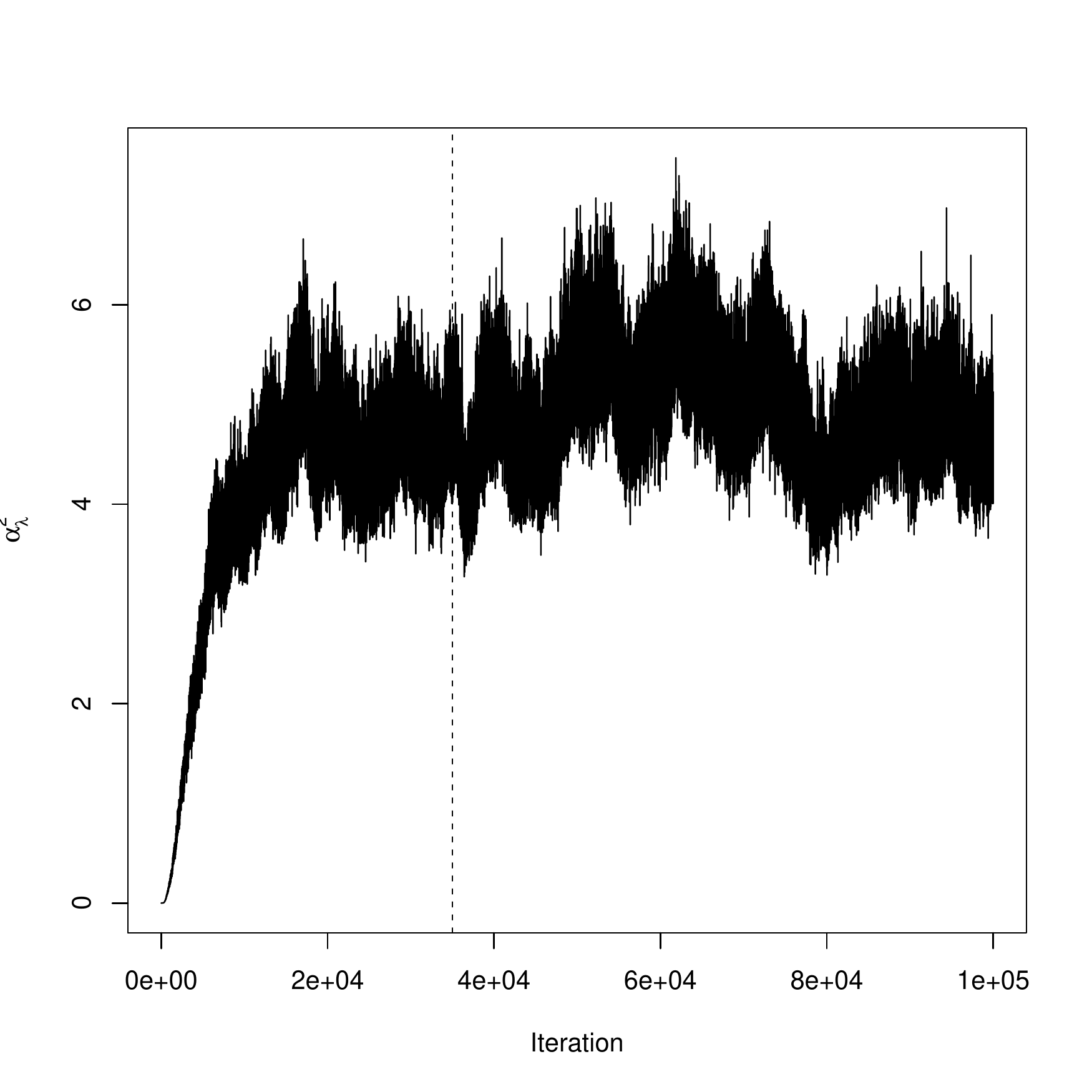}
    \includegraphics[width = 0.3 \textwidth]{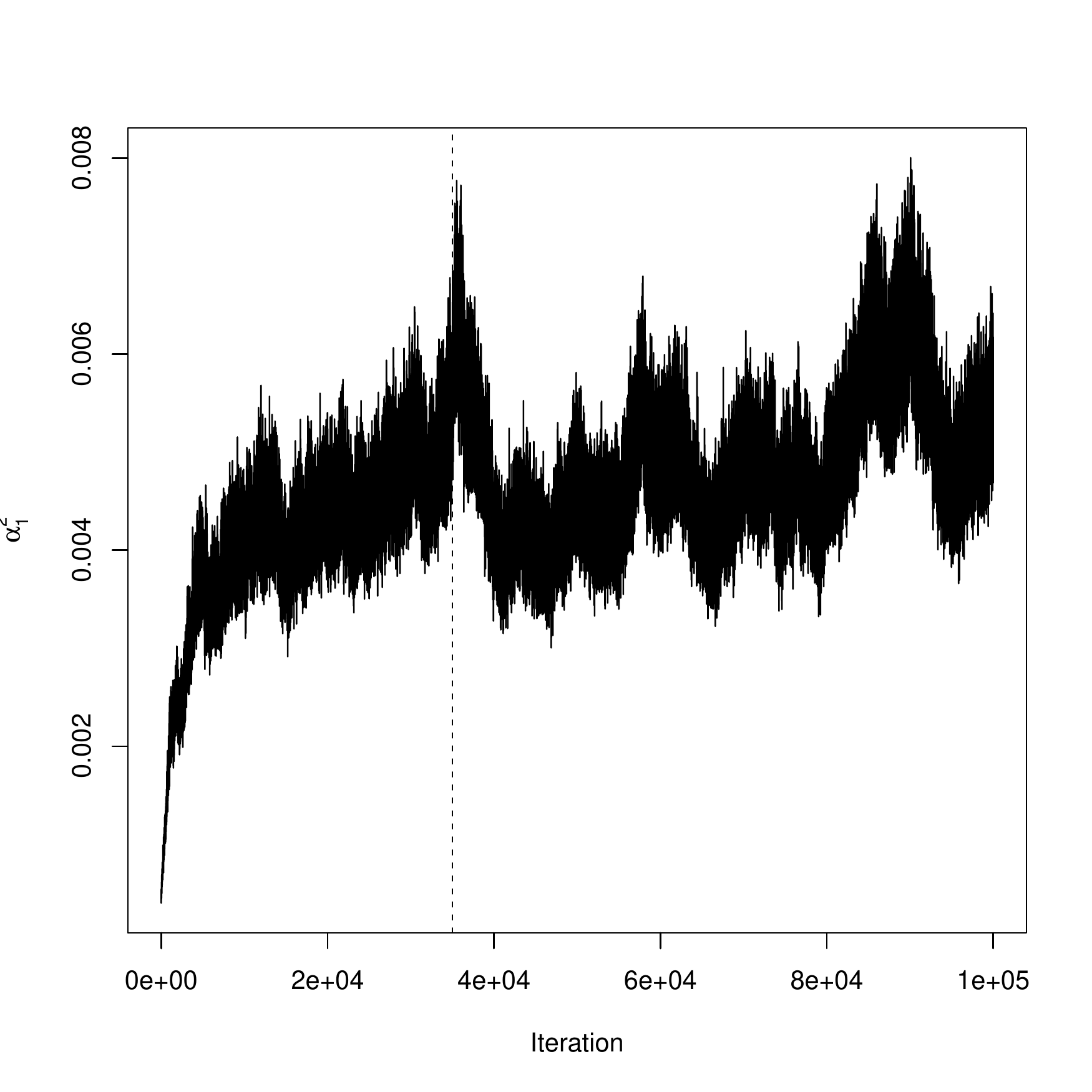}
    \caption{Student and non-student study. Trace plots for $\lambda_{100}$ (top left), $\beta_{0, 100}$ (top right), $\alpha^2_\lambda$ (bottom left) and $\alpha^2_1$ (bottom right). The 5 million iterations have been thinned to 100,000 and the dashed line at 35,000 iterations marks the end of the burn-in period.}
    \label{fig: BSBT student trace plots}
\end{figure}

We fit the BSBT model with judge information to the Dar es Salaam data set and produce the results shown in Section 4.2. Trace plots for $\lambda_{100}$, the grand mean for subward 100, $\beta_{0, 100}$, the difference between men and women's judgements in subward 100, $\alpha^2_\lambda$ and $\alpha^2_1$ are shown in Figure \ref{fig: BSBT gender trace plots}. The same plots for students and non-students are shown in Figure \ref{fig: BSBT student trace plots}. As we ran the MCMC algorithm for 5,000,000 iterations in both studies, we thin the results to reduce the required memory and store every 50th iteration. We take the 35,000th thinned iteration to be the end of the burn-in period. The mixing could be improved, particularly for $\alpha^2_1$, which could be achieved by using adaptive MCMC and adapting the underrelaxed tuning parameter, $\delta$. 

\bibliographystyle{rss}
\bibliography{bibliography}